\pdfoutput=1
\documentclass[aps,prd,superscriptaddress,notitlepage,11pt,final,longbibliography]{revtex4-1}

\usepackage[pdftex]{graphicx}
\usepackage[pdftex]{xcolor}
\usepackage[pdftex]{hyperref}

\usepackage{float}

\usepackage{amsmath,amssymb,mathrsfs}

\usepackage{subfigure}
\usepackage{epsf}

\usepackage{bbm}
\usepackage{color}
\usepackage{comment}
\usepackage{cleveref}
\usepackage{comment}
\usepackage{natbib}
\usepackage{appendix}
\renewcommand{\bm}[1]{{\mbox{\boldmath $#1$}}}

\graphicspath{{./fig/}}

\begin{document}
\title{Inelastic Scattering, Emergent Interactions of Solitons in the Zakharov-Kuznetsov Equation
through Conservative and non-Conservative Physics-Informed Neural Networks}
	
\author{A. Nakamula~}
\email{nakamula@sci.kitasato-u.ac.jp}
\affiliation{Department of Physics, School of Science, Kitasato University, Sagamihara, Kanagawa 252-0373, Japan}

\author{K. Obuse~}
\email{obuse@okayama-u.ac.jp}
\affiliation{Graduate School of Environmental and Life Science, Okayama University, Okayama 700-8530, Japan}

\author{N. Sawado~}
\email{sawadoph@rs.tus.ac.jp}
\affiliation{Department of Physics and Astronomy, Tokyo University of Science, Noda, Chiba 278-8510, Japan}

\author{K. Shimasaki~}
\email{shimasakitus@gmail.com}
\affiliation{Department of Physics and Astronomy, Tokyo University of Science, Noda, Chiba 278-8510, Japan}

\author{Y. Shimazaki~}
\email{shimazakitus@gmail.com}
\affiliation{Department of Physics and Astronomy, Tokyo University of Science, Noda, Chiba 278-8510, Japan}

\author{Y. Suzuki~}
\email{ytszkyuta@gmail.com}
\affiliation{Department of Physics and Astronomy, Tokyo University of Science, Noda, Chiba 278-8510, Japan}

\author{K. Toda~}
\email{kouichi@yukawa.kyoto-u.ac.jp}
\affiliation{Department of Mathematical Physics, Toyama Prefectural University, Imizu, Toyama 939-0398, Japan}
\affiliation{Research and Education Center for Natural Sciences, Keio University, Hiyoshi 4-1-1, Yokohama, Kanagawa 223-8521, Japan}

\begin{abstract}

The Zakharov-Kuznetsov equation, originally a three dimensional mathematical model 
of plasma with a uniform magnetic field, 
is a direct extension of the KdV equation into higher dimensions and 
is a typical quasi-integrable system.  
Physics-Informed Neural Networks (PINNs) are used to study the collision of soliton solutions in the 2+1 dimensional Zakharov-Kuznetsov equation. 
PINNs are able to successfully solve the equations in the forward process, and the solutions 
are obtained using a mesh-free approach and automatic differentiation, 
taking into account conservation laws. 
In the inverse process, the proper form of the equation can be successfully derived from a given training data. 
However, the situation becomes intractable in the collision process. 
The forward analysis result no longer adheres to the laws of conservation, and 
is better described as a dynamically incompatible field configuration (DIFC) than a 
solution to the system. 
Conservative PINNs have thus been introduced for this purpose, and in this paper 
we succeed in obtaining solutions that satisfy conservation laws. 
The inverse analysis suggests a different equation in which the coefficients exhibit significant changes, 
implying an emergence of temporary interactions. 
With these modulated coefficients, we recalculate the equation and confirm that the 
adherence to the laws of conservation has unquestionably improved. 

\end{abstract}

\maketitle

\section{Introduction}

Many natural phenomena people have long wished to understand still lack suitable mathematical models. 
Even for systems for which mathematical models have been proposed
and research has made progress, analytical understanding is often challenging because of factors such 
as the nonlinearity of the equations. Numerical analysis is effective, especially for nonlinear systems. 
However, despite continuing advances in computer performance, it remains difficult to comprehend the properties 
of nonlinear systems or obtain accurate solutions through numerical calculations. 
In addition, there are often cases where it is uncertain whether the obtained numerical solution is indeed a correct 
approximation of the intended equation. 
When there is sufficient interest in breakthroughs in the numerical analysis of nonlinear systems,  
the application of deep learning methods---specifically, deep neural networks---to nonlinear partial 
differential equations (PDEs) has become an attractive solution.

Numerous complex scientific problems, including those in fluid and solid mechanics~\cite{Raissi2020,Brunton2020,Kadeethum2020,cai2021,Kashinath2021,Jin2021,
LINGHU2025116223,ALHUBAIL2024104797,ESHKOFTI2024118485,SUN2024106421}, 
cyber-physical systems~\cite{9064519}, 
biological systems~\cite{Wu2017,KISSAS2020112623,Ruizherrera2021,SEL2023},
can efficiently be solved by Physics-Informed Neural Networks (PINNs) ~\cite{RaissarxivI,RaissarxivII,RAISSI2019686}.
One remarkable feature of PINNs is that in addition to efficiently solving PDEs (forward analysis), 
they may also provide a precise estimate of the equation based on the governing data for
the physical issues of our concern (inverse analysis). 
PINNs have much greater extrapolation power than conventional deep-learning techniques, making them appropriate for analyses involving limited learning data~\cite{Fang2020,SAHLICOSTABAL2024107324,WU2025106750}. 

Notably, PINNs are completely different scheme from traditional numerical algorithms~\cite{JAGTAP2020113028,MISHRA2021107705,Chen2022UsingPI,YANG2023109656,Sedykh_2024}. 
Many numerical studies have been conducted with the finite difference and finite element methods, 
in which the governing PDEs are ultimately discretized over the computational domain. 
In contract,  PINNs have the unique feature of using a mesh-free 
methodology, as automatic differentiation approximates the differential operators 
in the governing PDEs. 
The grid independence of PINNs is undoubtedly efficient, particularly 
for solving high-dimensional problems and inverse analysis. 
With PINNs, inverse analysis can be used to verify the validity of a 
solution derived using the conventional finite difference method.  
This capability of PINNs is our main focus in this paper. 
An additional benefit of PINNs is their ability to integrate 
a system's conservation laws
into the analysis by incorporating conditions into the loss function. 
Conservative-Physics-Informed Neural Networks~(cPINNs) can be used to 
further improve accuracy of the analysis
~\cite{JAGTAP2020113028,LIN2022111053,FANG2022112118,WU2022112143,cardosobihlo2025}. 
Ref.\cite{cardosobihlo2025} employs the standard method of projecting geometric numerical
integration into PINNs and claims theirs is an exact, or \textit{hard-conservative} method. 
To the best of our knowledge, no similar technique has been developed for nonlinear-wave analysis. 
In Ref.\cite{Nakamula:2024cmx}, we presented a compact cPINN technique 
(without the \textit{hard} method), 
in which we included a weight function $C(\mathcal{E})$ that allows us to regulate 
the convergence of the loss function. 

There have been numerous studies of PINNs for integrable PDEs, such as the 
Bergers eq.~\cite{RaissarxivI,RaissarxivII,RAISSI2019686,LIN2022111053,JAGTAP2020113028}, 
the Korteweg--de Vries (KdV) and the modified KdV equations
~\cite{Li_2020,JAGTAP2020113028,FANG2022112118,Junkai2024,Zhou2024,LIN2023133629}, 
the nonlinear Schr\"odinger eq.~\cite{ZHOU2021127010,Pu2021}, 
a coupled Schr\"odinger-KdV eq.~\cite{ZHANG2024108229} and many other variations. 
In 2+1 dimensions, the Kadomtsev--Petviashvili (KP) eq. and the spin-nonlinear Schr\"odinger eq. 
have been studied using PINNs~\cite{Zhengwu2022,ZHOU2023164,WANG202317}. 
Quasi-integrable deformations of the KdV eq.
have also been analyzed~\cite{Bai2021,Zijian2023}.
We have previously studied two quasi-integrable equations in 2+1-dimensions and thoroughly discussed how 
inverse PINNs identified them~\cite{Nakamula:2024cmx}. 

In the present paper, we use PINNs to analyze the properties of several configurations of the Zakharov-Kuznetsov (ZK) eq. 
The ZK eq.~\cite{Zakharov74,IWASAKI1990293} is a direct extension of the KdV eq. into higher dimensions. 
To the best of our knowledge, past studies~\cite{PetYan82, IWASAKI1990293, Klein21} 
have only focused on the two-dimensional version of the model. 
The equation possesses stable isolated vortex-type soliton solutions 
that exhibit typical solitonic properties, i.e., they move with a constant speed without 
dissipation and have several conserved quantities.
As a result, they display a certain longevity on a standard computational time scale,    
and thus may be candidates for a vortex in flow fields, e.g., 
the Great Red Spot (GRS) on Jupiter~\cite{Koike:2022gfq}.
However, inelastic properties emerge when the heights of solitons 
significantly differ. In a collision process, the taller soliton gains more height 
whereas the shorter one tends to wane with the radiation~\cite{IWASAKI1990293}. 
These unusual inelastic solitons, which are frequently found in several quasi-integrable systems ~\cite{ABDULLOEV1976427,COURTENAYLEWIS1979275,terBraak:2017jpe,KAWAHARA199279}, 
are caused by insufficient conserved quantities. 
Many integrable and the quasi-integrable systems have their origin in 
fluid mechanics or plasma physics. 
Although the two systems are almost on the same footing physically, few studies
have explored the latter's mathematical nature as an initial value problem. 
The rare exception was for the 1+1-dimensional regularized long-wave eq.~\cite{Benjamin72}. 
A crucial question has been long overlooked as to whether linear superimposed 
configurations in these quasi-integrable equations could be solutions. 
In this paper, we will address this issue. 
To distinguish these anomalous objects from the known, well-behaved soliton solution, 
we introduce the concept of the dynamically incompatible field configuration (DIFC). 
Therefore, the main objective of the present paper is to apply forward/inverse PINNs/cPINNs 
to some DIFCs in the collision process of the 2+1 ZK eq.

The remainder of the paper is organized as follows. In Section \ref{sec:2}, we give a brief introduction 
to the ZK equation 
and present an overview of the PINNs as well as the 
basic formalism of our cPINNs.   
In Section \ref{sec:3}, we present our numerical results, including 
a discussion of 
the coefficient modulations in the inverse PINNs. 
This section also discusses the coincidence or disagreement 
between the forward and inverse analyses. 
Conclusions and final remarks are presented in the last section.

\section{\label{sec:2}Zakharov-Kuznetsov equation}

The ZK equation originally was originally a three dimensional model of plasma 
with a uniform magnetic field~~\cite{Zakharov74}. 
It can simply be regarded as a direct extension of the KdV equation into higher dimensions and 
is a typical quasi-integrable system.  
Although the term ``quasi-integrability'' is frequently employed in the literature, 
mathematically it is still somewhat obscure. 
A straightforward definition of a quasi-integrable system is 
an equation with a finite number of exactly conserved quantities. 
From this perspective, we investigate the ZK equation, 
the solutions of which resemble solitons in the integrable system and have just 
four conserved quantities. 
At the moment, most existing research~\cite{PetYan82, IWASAKI1990293, Klein21} 
has focused on the two-dimensional version of the model. 
One prominent aspect of the solution to the equation lies 
in the irregular inelastic features of the collision process. 
When the heights of solitons coincide, the collision appears similar to
the integrable cases: however, when they differ by a large margin, 
the taller soliton gains more height during the collision process 
while the shorter tends to wane with radiation~\cite{IWASAKI1990293}.
Furthermore, in the case of offset scattering,  the shorter soliton tends to 
disappear after the impact of the collision.  

The ZK equation is given as
\begin{equation}
\frac{\partial u}{\partial t}+2u\frac{\partial u}{\partial x}+\frac{\partial}{\partial x}\left(\nabla^2 u\right)=0 
\label{ZK}
\end{equation}
where the Laplacian operator is $\nabla^2=\partial_{\tilde{x}}^2+\partial_y^2$. 
Eq.(\ref{ZK}) possesses meta-stable isolated vortex-type solutions 
that exhibit solitonic properties. 
Solutions to the ZK equation propagate in specific directions with uniform speeds. 
Here, we describe propagation in the positive $x$ direction as having velocity $c$: 
that is, we assume $u=U(\tilde{x}:=x-ct,y)$. 
Plugging this into (\ref{ZK}), we obtain
\begin{align}
\nabla^2U=cU-U^2\,,
\label{ZK circ}
\end{align}
where $\nabla^2=\partial_{\tilde{x}}^2+\partial_y^2$. 
A steady progressive exact wave solution is of the form
\begin{align}
U_{\rm 1d}=\frac{3c}{2}{\rm sech}^2\biggl[\frac{\sqrt{c}}{2}(\tilde{x}\cos\theta+y\sin\theta)\biggr],
\label{ZK wave}
\end{align}
where $\theta$ is a given inclined angle of the solution. 
This indicates that this solution is simply a trivial embedding of the KdV soliton into two spatial dimensions. 
In addition to the solution in \eqref{ZK wave}, 
\eqref{ZK circ} possesses another solution that maintains circular symmetry.
To obtain this solution,  we introduce cylindrical coordinates and rewrite the equation as \begin{align}
\frac{1}{r}\frac{d}{dr}\biggl(r\frac{dU(r)}{dr}\biggr)=cU(r)-{U}(r)^2\,,
\label{eq:ZKcylind}
\end{align}
where $r:=\sqrt{\tilde{x}^2+y^2}$. 
We can numerically determine the solutions with the boundary condition $U\to 0$ 
as $r\to \infty$ and solutions from specific parameter families of $c$ such as $U(r):=cF(\sqrt{c}r)$. 

The solutions of \eqref{ZK} exhibit solitonic behavior; however, they are not like known integrable solitons, 
such as those of the KdV equation, in that the stability of the solutions may be supported by the infinite number 
of conserved quantities in the equation.  
Eq.\eqref{ZK} admits only the four integrals of motion~\cite{KUZNETSOV1986103}, which are
\begin{align}
&I_1:=\int i_1(y)dy= \int u dx dy,~~\textrm{with}~~ i_1(y):=\int u dx,
\label{ZKCQ1}
\\
&I_2:=\int\frac{1}{2}u^2dxdy\,,
\label{ZKCQ2}
\\
&I_3:=\int\biggl[\frac{1}{2}(\nabla u)^2-\frac{1}{3}u^3\biggr]dxdy\,,
\label{ZKCQ3}
\\
&\bm{I}_4:=\int \bm{r}u dxdy-t\bm{e}_x\int u^2dxdy\,,
\label{ZKCQ4}
\end{align}
where $\boldsymbol{r}$ and $\boldsymbol{e}_x$ are the two-dimensional position vector and the unit vector in the $x$-direction, respectively.
Here, $I_1$ is interpreted as the ``mass" of the solution, and $i_1(y)$ itself is conserved similarly to the KdV equation.
Additionally, $I_2, I_3, I_4$ correspond to the momentum, energy, and laws for the center of mass, respectively.  
We therefore conclude that the ZK equation is not an integrable system in the manner of ordinary soliton equations.

The algorithms used in this work consist of two parts. 
The first is the well-known neural network part, 
which is constructed with 4 hidden layers and of 20 nodes
each to produce an output $u$ given the temporospatial input coordinates $(t,x,y)$. 
However, the output of this network has no physical meaning. 
In the second segment, the output derivatives and a loss function for network optimization are estimated 
(see Secs.3.2 and 3.3 for details).

The PINNs can be applied to both forward and inverse problems.   
Through the forward analysis, 
solutions to a governing equation can be found without
high computational demand or sophisticated numerical algorithms.  
Let us consider the ZK equation of the form
\begin{align}
\mathcal{F}:=u_t+\mathcal{N}(u,u_x,u_{xxx},u_{xyy})=0,~~
\mathcal{N}(u,u_x,u_{xxx},u_{xyy}):=2uu_x+\left(\nabla^2 u\right)_x \label{zk}\,,
\end{align}
We focus on the soliton solutions moving in the positive $x$ direction.  
We define the rectangular mesh space 
\begin{align}
&x\in [-L_x,L_x],~N_x\textrm{~grid~points};~~~~y\in [-L_y,L_y],~N_y\textrm{~grid~points}\,:
\nonumber \\
&t\in [T_0,T_1]\,.
\end{align}
For optimization of the networks, we set the loss function as the mean-squared error 
$\mathit{MSE}$ 
which measures the discrepancy between the predicted and true values. 
The loss function can be defined such that
\begin{align}
&\mathit{MSE}:=\mathit{MSE}_\textrm{init}+\textit{MSE}_\textrm{eq}+\textit{MSE}_\textrm{bc}\,,
\label{MSEB}\\
&\textit{MSE}_\textrm{init}:=\frac{1}{N_u}\sum_{i=1}^{N_u}|u_{\mathrm{pred}}^{0}(x^i,y^i,0)-u_\mathrm{correct}^{0}(x^i,y^i,0)|^2\,,
\\
&\textit{MSE}_\textrm{eq}:=\frac{1}{N_F}\sum_{i=1}^{N_F}|\mathcal{F}(x^i,y^i,t^i)|^2\,,
\end{align}
where $u^0_{\mathrm{pred}}$ and $u^0_{\mathrm{correct}}$ are the predicted and 
true initial profile and $u^0_{\mathrm{correct}}$, respectively. 
$\{x^i,y^i,0\}_{i=1}^{N_u}$ is $i$th set of $N_u$ random residual points at $t=0$ 
and $\{x^i,y^i,t^i\}_{i=1}^{N_F}$ 
is the $i$th set of $N_F$ random points for the PINNs $\mathcal{F}(x,y,t)$. 
Finally, $\textit{MSE}_\textrm{bc}$ represents the mean squared error caused 
by the boundary conditions with the $N_{\mathrm{b}}$ random points. 
Here, doubly periodic boundary conditions are employed as
\begin{align}
\textit{MSE}_\textrm{bc} 
:=\frac{1}{N_\textrm{b}}\Bigl[\sum_{i=1}^{N_{\mathrm{b},x}}|u_{\mathrm{pred}}(x^i, L_y,t^i)-u_{\mathrm{pred}}(x^i,-L_y,t^i)|^2
\nonumber \\
~~+\sum_{j=1}^{N_{\mathrm{b},y}}|u_{\mathrm{pred}}(L_x,y^j,t^j)-u_{\mathrm{pred}}(-L_x,y^j,t^j)|^2\Bigr].
\end{align}
where $N_{\mathrm{b}}\equiv N_{\mathrm{b},x}+N_{\mathrm{b},y}$. 

Taking into account the conservation laws $I_i~(i=1,\cdots,4)$~\eqref{ZKCQ1}--\eqref{ZKCQ4}, 
which position the framework as a cPINN,  
we add to $\textit{MSE}$~\eqref{MSEB} the following conservational contribution:
\begin{align}
\textit{MSE}_\textrm{C}:=C(\mathcal{E})\frac{1}{N_\mathrm{p}}\sum_{i=1}^4\sum_{a=1}^{N_\mathrm{p}}
|I_i^\mathrm{pred,(a)}-I_i^\mathrm{correct,(a)}|^2\,.
\label{loss_conv}
\end{align}  
The weight function $C(\mathcal{E})$ is 
the key component of the term and it is evaluated using the conventional MSE 
($\mathcal{E}\equiv \textit{MSE}_\textrm{B}$) as
\begin{align}
C(\mathcal{E})=\left\{
\begin{aligned}
&\exp(-\gamma \mathcal{E}_\textrm{crit})~~~\mathcal{E}> \mathcal{E}_\textrm{crit}  \\
& \exp(-\gamma \mathcal{E})~~~~~~\mathcal{E}\le \mathcal{E}_\textrm{crit} \,.
\end{aligned}
\right.
\end{align}
This describes the schedule of the network optimization of the simulation. 
In the early stages, 
the conventional MSE, $\mathcal{E}$ is the main concern. 
The weight function then starts to rely on the value of $\mathcal{E}$ at 
a specific critical point $\mathcal{E}\sim \mathcal{E}_\textrm{crit}$,
after which the MSE$_\textrm{C}$ term takes precedence.  
Here, we introduce two parameters $\gamma$ and $\mathcal{E}_\textrm{crit}$ that define 
the rate of change of the weight function $C(\mathcal{E})$,
and the following values were selected heuristically for the present study
\begin{align}
\gamma=3000,~~\mathcal{E}_\textrm{crit}=5.0\times 10^{-3}\,.
\nonumber 
\end{align} 
We establish the reference number $N_\textrm{p}$ of time, steps over which 
the numerical integration by the conventional Simpson method is carried out, for the estimation of the conserved quantities.
Because it produces the most notable effect of the four laws, we employ only the second conservation 
law $I_2$ in the present analysis.

Another more prominent application of PINNs is the generation of the PDE 
from given data, known as inverse analysis. 
Now, we introduce a slightly modified PINN:
\begin{align}
\tilde{\mathcal{F}}:=u_t+\tilde{\mathcal{N}}(u,u_x,u_{xxx},u_{xyy},\bm{\lambda})=0,
\label{inv}
\end{align}
where the unknown parameters $\bm{\lambda}$ are included in the equations. 
For the inverse analysis, we define the MSE as
\begin{equation}
\mathit{MSE}_\textrm{inv}:=\frac{1}{N_u}\sum_{i=1}^{N_u}|u_{\mathrm{pred}}(x^i,y^i,t^i)-u_\mathrm{correct}(x^i,y^i,t^i)|^2
+\frac{1}{N_F}\sum_{i=1}^{N_F}|\tilde{\mathcal{F}}(x^i,y^i,t^i)|^2, 
\label{loss_inv}
\end{equation}
where $u_{\mathrm{pred}}$ and $u_\mathrm{correct}$ are the predicted and true profiles including the boundary data. 
The networks are optimized by varying the parameters of the neural networks $\left(w^{(j)}_{lk} , b^{(j)}_{k}\right)$ 
along with the parameters $\bm{\lambda}$. 
This yields optimal parameter values $\bm{\lambda}$ and the idealized equations
for the corresponding training data. 
For inverse analysis with cPINNs, 
we add the conservation term~\eqref{loss_conv} to
the inverse MSE~\eqref{loss_inv}. 
PINNs are tuned by optimizing the MSE. 
For this, we employ the limited-memory Broyden--Fletcher--Goldfarb--Shanno (L-BFGS) optimizer~\cite{LIU1989}. 
Convergence is attained the norm of the \textit{MSE} gradient is less than the machine epsilon:
\begin{align}
|\nabla \mathit{MSE}| < \varepsilon,~~\varepsilon=2.220\times 10^{-16}\,.
\end{align}

\begin{figure}[t]
	\centering 
	\includegraphics[width=1.0\linewidth]{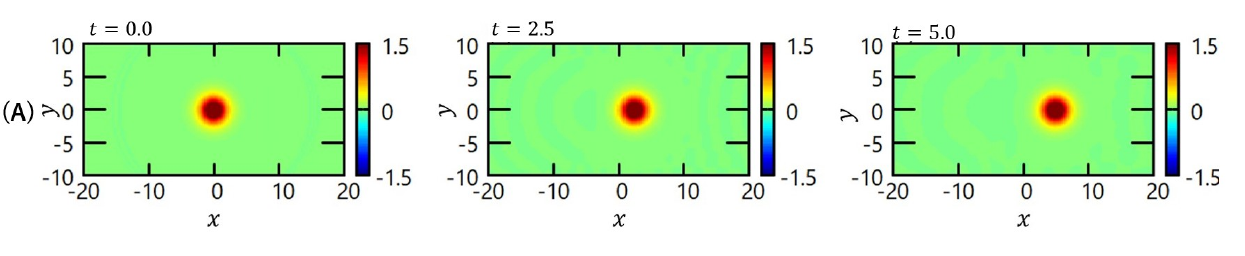}
	\includegraphics[width=1.0\linewidth]{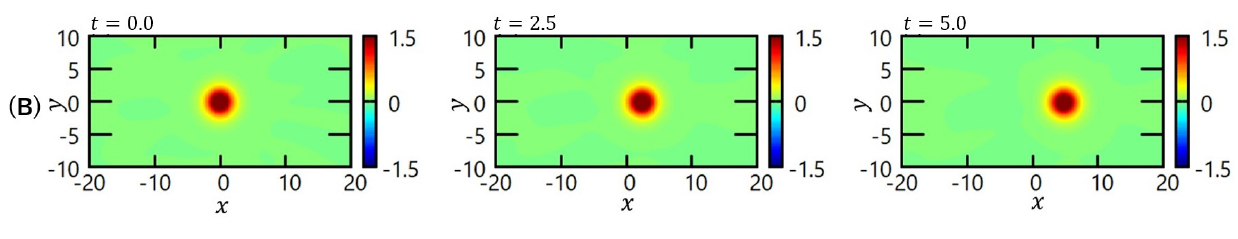}
	\includegraphics[width=1.0\linewidth]{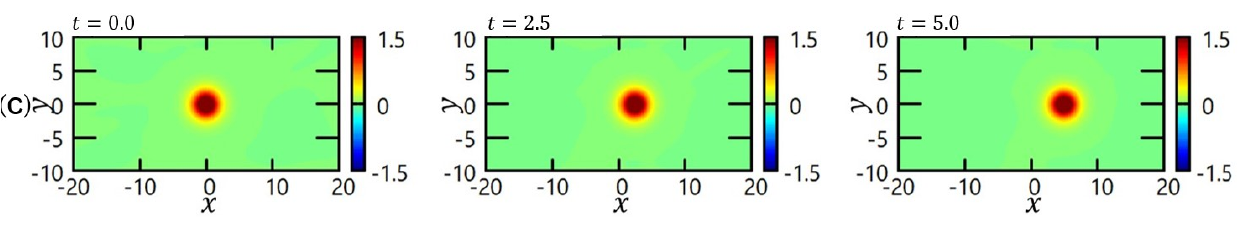}
	\caption{\label{fwd1}1-soliton solutions obtained with (A) the exact numerical Runge-Kutta method, (B) a PINN, 
	and (C) a cPINN at the times $t=0.0,2.5,5.0$.}

\end{figure}

We then further improved the analysis to address the computational demands of long-term simulation. 
In general, PINNs need a lot of training data to handle long-term phenomena, 
in addition to a significant amount of processing resources. 
Some researchers have attempted to increase the training time interval
based on a study of the splitting of the integration intervals~\cite{Bihlo2022,Krishnapriyan2024}. 
After many time steps, the PINN output frequently converges to a trivial solution. 
In this paper, we also employ a multi-time step algorithm, but we aim 
to focus on events that may occur during a short time window.
We divide the total time $T$ into $n$ segments as
\begin{align}
t_s\in [T_0+(s-1)\Delta t,T_0+s\Delta t],~~s=1,\cdots,n;~~\Delta t:=\frac{T_1-T_0}{n}\,.
\end{align} 
We initialize the PINN or cPINN in the first segment $s=1$ with the initial condition 
$\{u(x^i,y^i,t=T_0)\}$.
Once we obtain the solution $\{u(x^i,y^i,t^i_{s=1})\}$,  
the PINN/cPINN for the next segment $s=2$ can be defined 
with the initial profile $u(x^i,y^i,\Delta t)$. 
We iterate the procedure by incrementing $s$ until $s=n$. 
We perform our analysis for each segment with 
\begin{align}
N_u=20000,~~N_F=50000,~~N_\textrm{b}=5000,~~N_\textrm{p}=11\,.
\end{align}
We will see in the next section that the inverse analysis results in 
fluctuations of the equation's coefficients. 
For the inverse analysis, we thus use $I_2$ in \eqref{loss_conv} 
because it is invariant in the presence of such changes.

\begin{figure}[t]
	\centering 
	\includegraphics[width=0.5\linewidth]{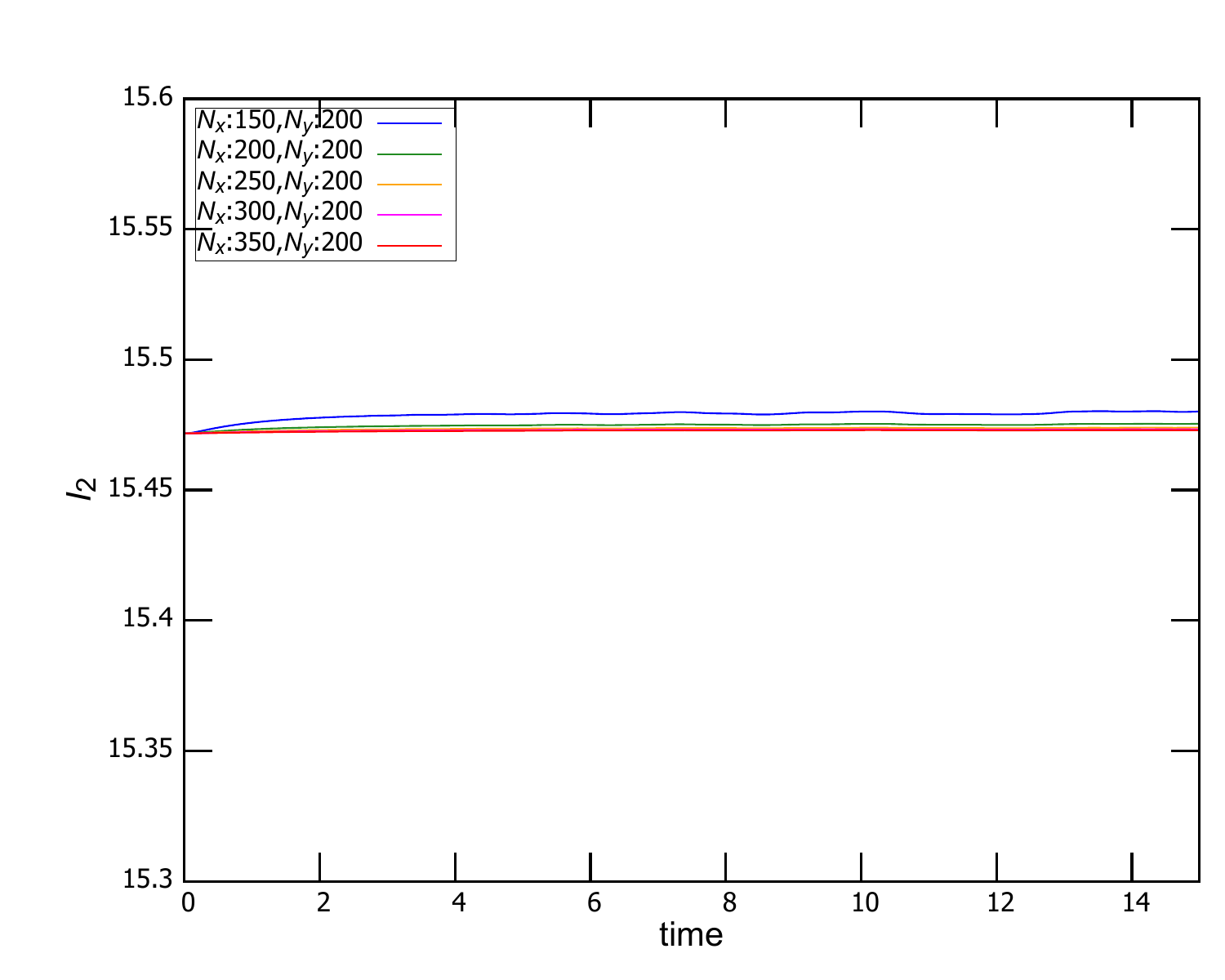}\hspace{-0.5cm}
	\includegraphics[width=0.5\linewidth]{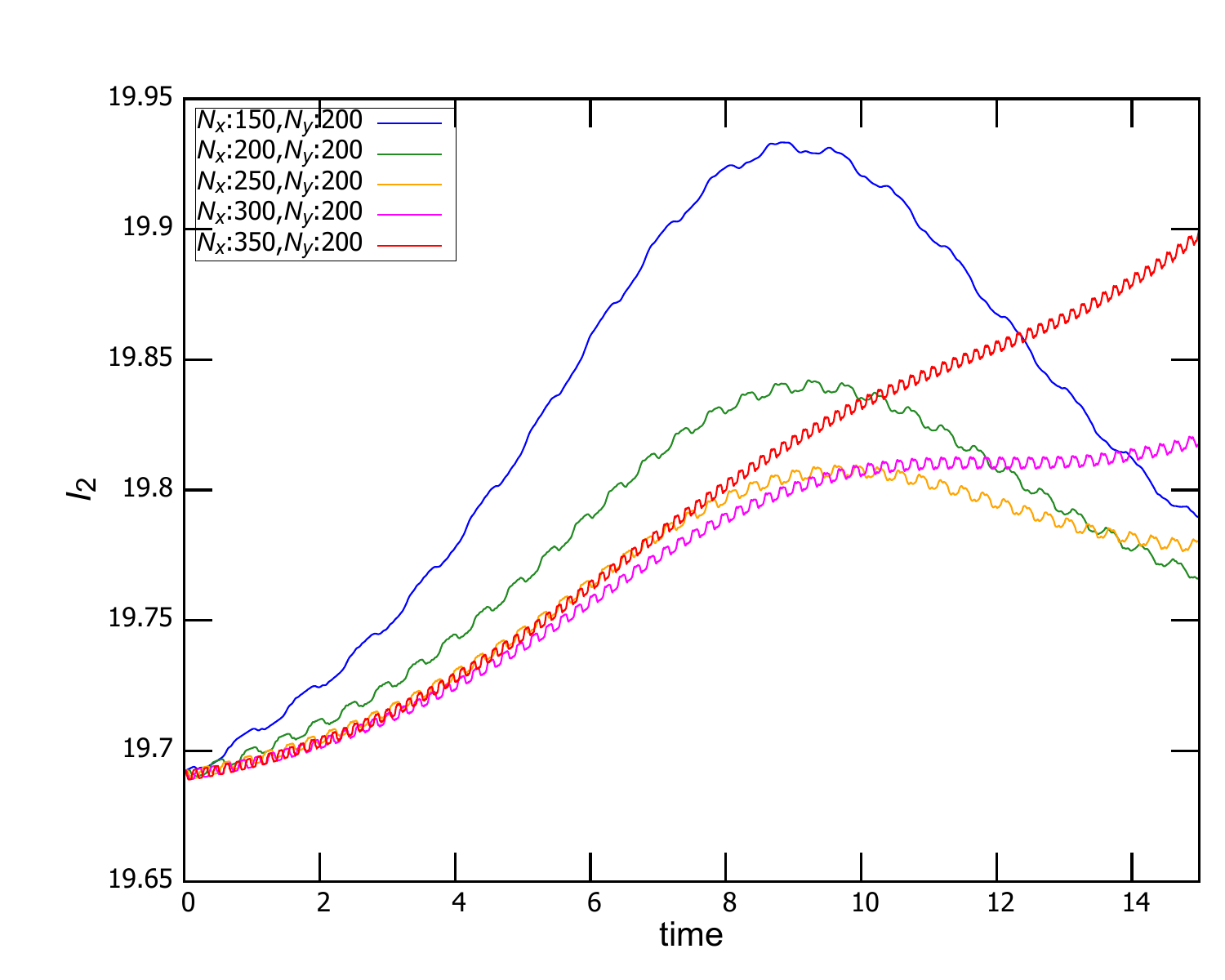}
	\caption{\label{cons1}Conserved quantity $I_2$ \eqref{ZKCQ2} of the 1-soliton (\textit{left})
	and the 2-soliton (\textit{right}) solutions. The evolution equation was solved by the Runge-Kutta 
	method. }

\end{figure}

\section{\label{sec:3}Numerical results}

In this section, we compare the results of several 1- and 2-soliton configurations 
in PINNs or cPINNs obtained using the standard Runge-Kutta method. 
For the inverse analysis, we 
focus particularly on data from short-period segments, allowing us to estimate 
changes in the coefficients of the equation, 
which is our current area of interest.  
 
 \begin{figure}[t]
\centering 
	\includegraphics[width=1.0\linewidth]{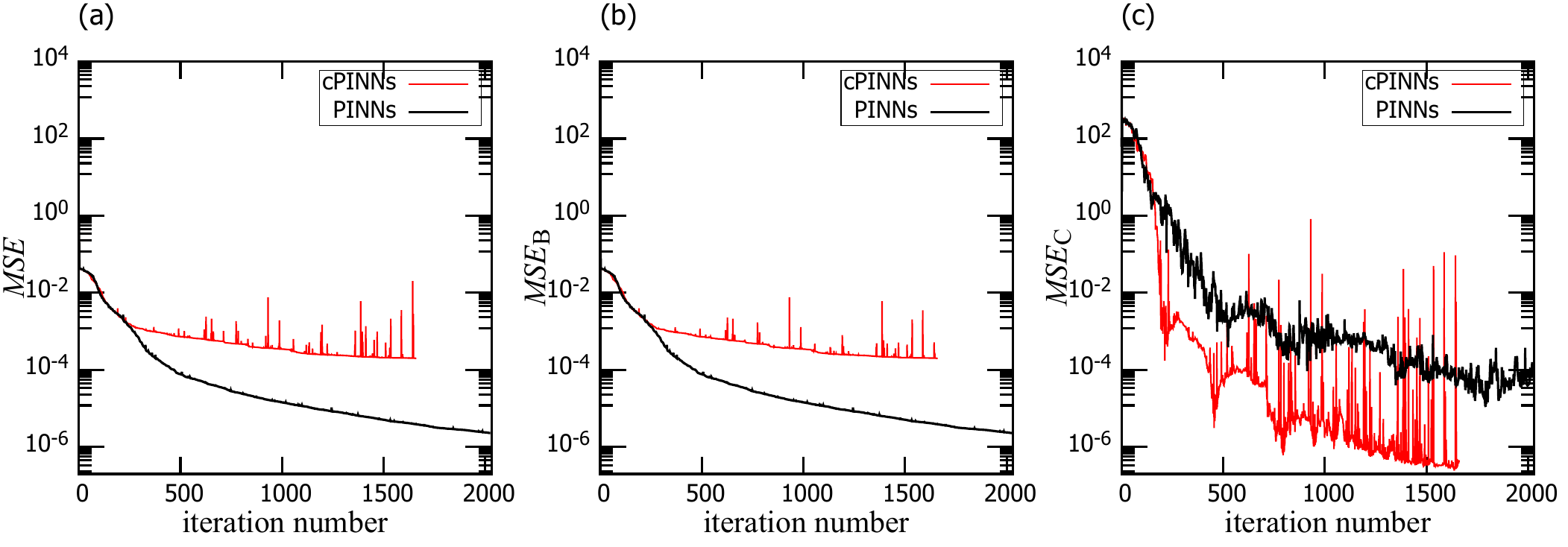}\hspace{-1.1cm}
	\caption{\label{loss} MSEs of the PINN and cPINN solutions with $c=1.0, 0.25$.
	(a)~Total \textit{MSE}, (b)~$\mathit{MSE}_\textrm{B}$, (c)~$\mathit{MSE}_\textrm{C}$.}
\end{figure}

\subsection{Forward analysis}

Fig.\ref{fwd1} shows the 1-soliton solutions obtained using different method at times $t=0.0, 2.5, 5.0$. 
The Runge-Kutta method, PINN, and cPINN all yielded similar results. 
It has been noted in \cite{ABDULLOEV1976427,COURTENAYLEWIS1979275,Goda1980,Makino81,IWASAKI1990293,KAWAHARA199279,
Ferreira:2013nda,terBraak:2017jpe} that 
the system becomes untenable in the process of the 2-solitons collision process. 
The nonlinearity in the equation has a detrimental effect on how the solutions 
evolve. This difficulty can be readily identified in the conservation laws, 
which, as we have shown, are supported by the equation itself. 
Fig.\ref{cons1} illustrates the time dependence of the conserved quantity $I_2$ \eqref{ZKCQ2}
of the 1- and the 2-soliton configurations obtained using the Runge-Kutta method
with different mesh granularity $N_x=150,200,250,300,350$ and $N_y=200$.   
The behaviors in the 1-soliton solution are nearly constant over time, within the numerical uncertainty,  
and seem to converge as the mesh becomes more fine-grained. 
It is natural to conclude that the 1-soliton solution of the equation exists in the limit of an 
infinite number of mesh points.  
However, we do not observe such convergence for the 2-soliton: they exhibit both a modest, 
rapid fluctuation over time and considerable divergence with increased granularity.
This latter feature is more noteworthy because it subtly 
demonstrates that the 2-soliton configuration is not a solution to the equation, 
whereas the former feature might suggest some chaotic aspect of the system. 
The $I_2$ quantity is not unique in this regard: the other conserved quantities 
$I_1,I_3$, and $I_4$ also exhibit similar behavior. 
Although some sophisticated numerical techniques like the implicit method might improve the conservation, 
one cannot resolve the significant discrepancy seen in Fig.\ref{cons1}.
Consequently, we must draw the conclusion that the numerous collision process configurations 
of the quasi-integrable equations found in earlier research~\cite{IWASAKI1990293} 
violated the laws of conservation of the equations. 
Further, the laws of conservation are likely to be broken for other multi-soliton configurations as well.  
The 2-soliton configuration is therefore considered as a DIFC, 
as defined in the introduction. 
Using cPINNs ---that is, PINNs that maintain conserved quantities--- enables us to seriously consider 
solutions that obeys the laws of conservation.  

\begin{figure}[t]
	\centering 
	\includegraphics[width=0.7\linewidth]{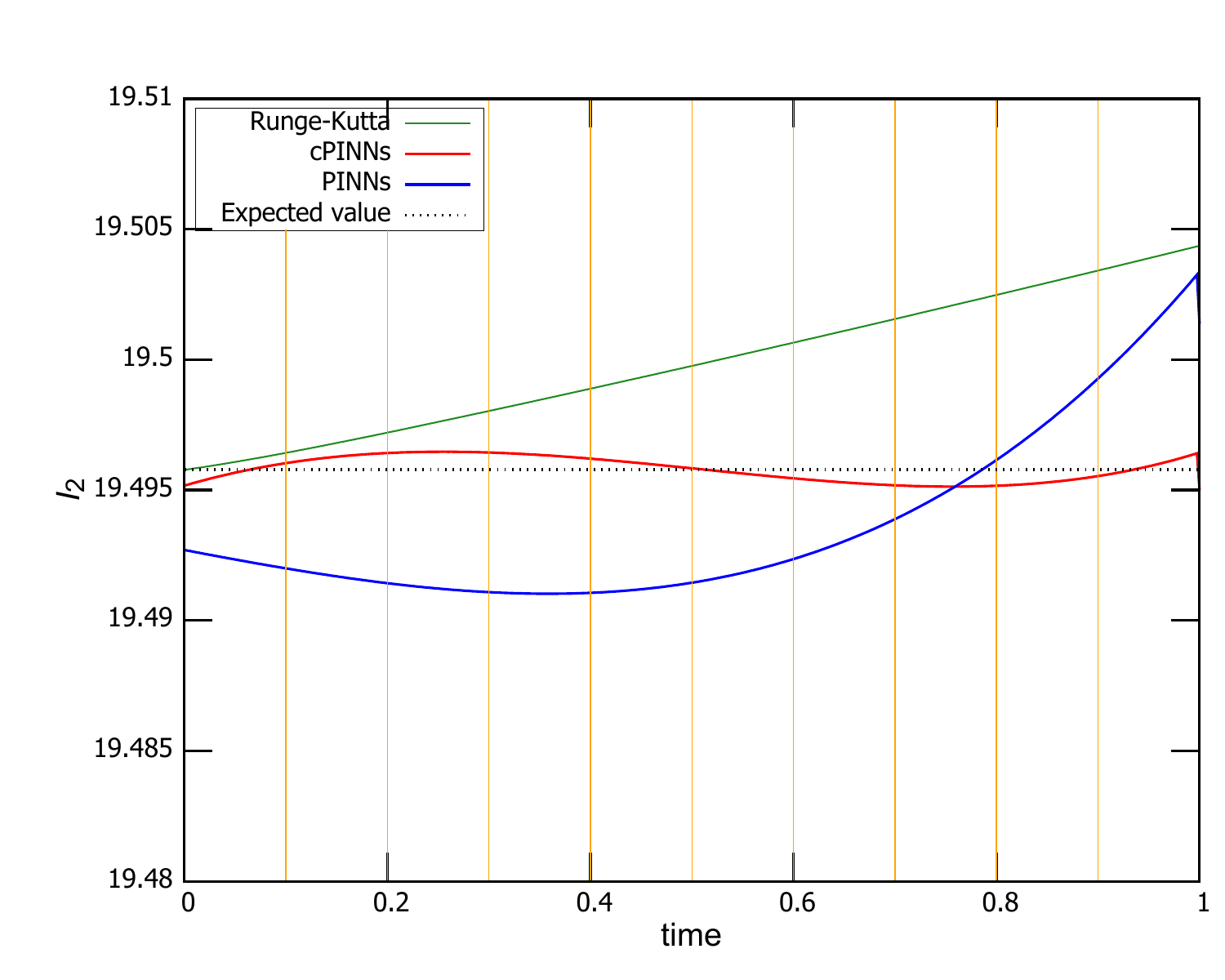}
	\caption{\label{cons0-1}Conserved quantity $I_2$ 
    for the configuration of the onset collision with $c=1.0, 0.25$ 
    from the Runge-Kutta, PINN and cPINN solutions, on the interval $0\leqq t\leqq 1$.}

\end{figure}

\begin{figure}[htbp]
	\centering 
	\includegraphics[width=1.0\linewidth]{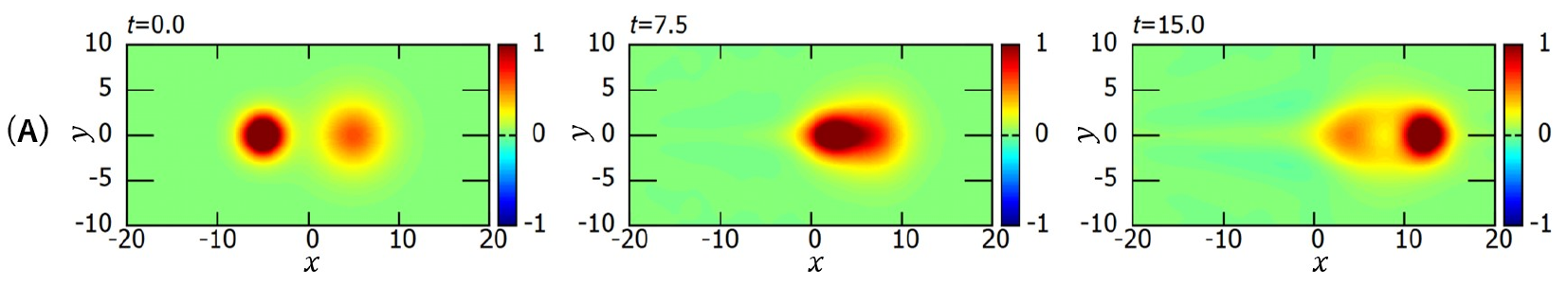}
	\includegraphics[width=1.0\linewidth]{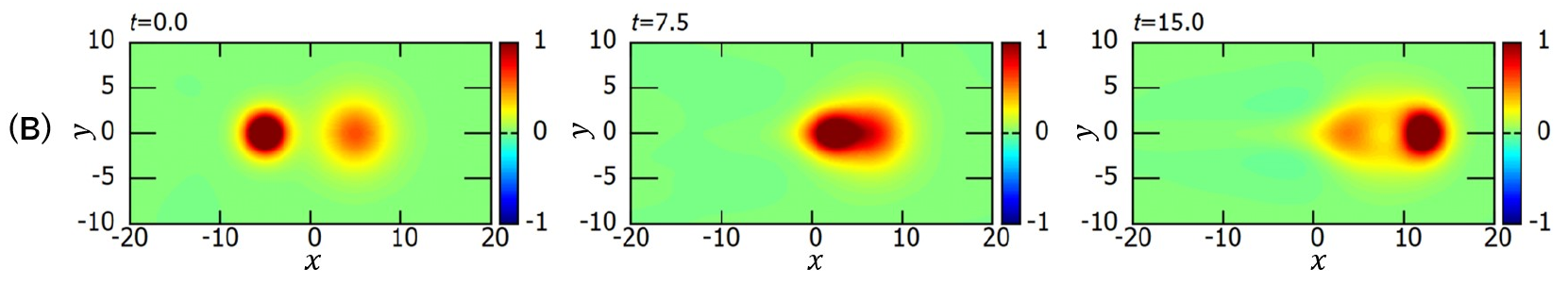}
	\includegraphics[width=1.0\linewidth]{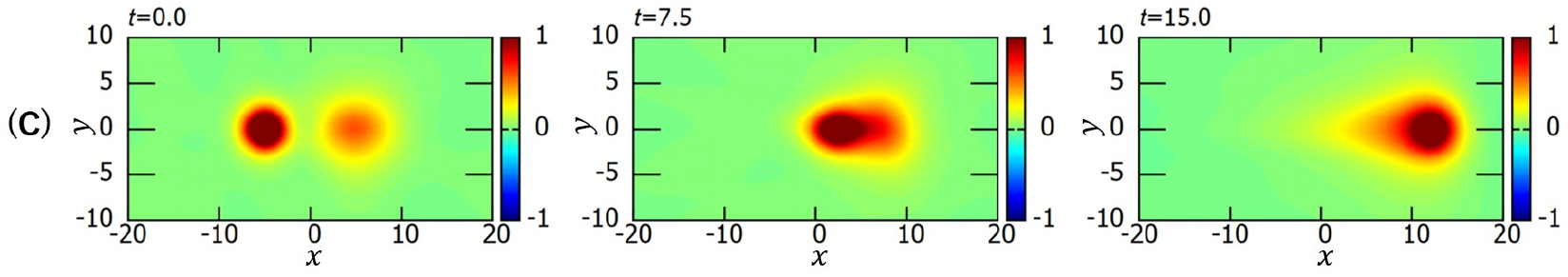}
	\caption{\label{fwd2on1}Onset collision of the solutions with $c=1.0$ and $0.25$ of 
	(A)~the exact numerical analysis, (B)~the PINNs, and (C)~the cPINNs
	at the time $t=0.0,7.5,15.0$.}

	\centering 
	\includegraphics[width=0.7\linewidth]{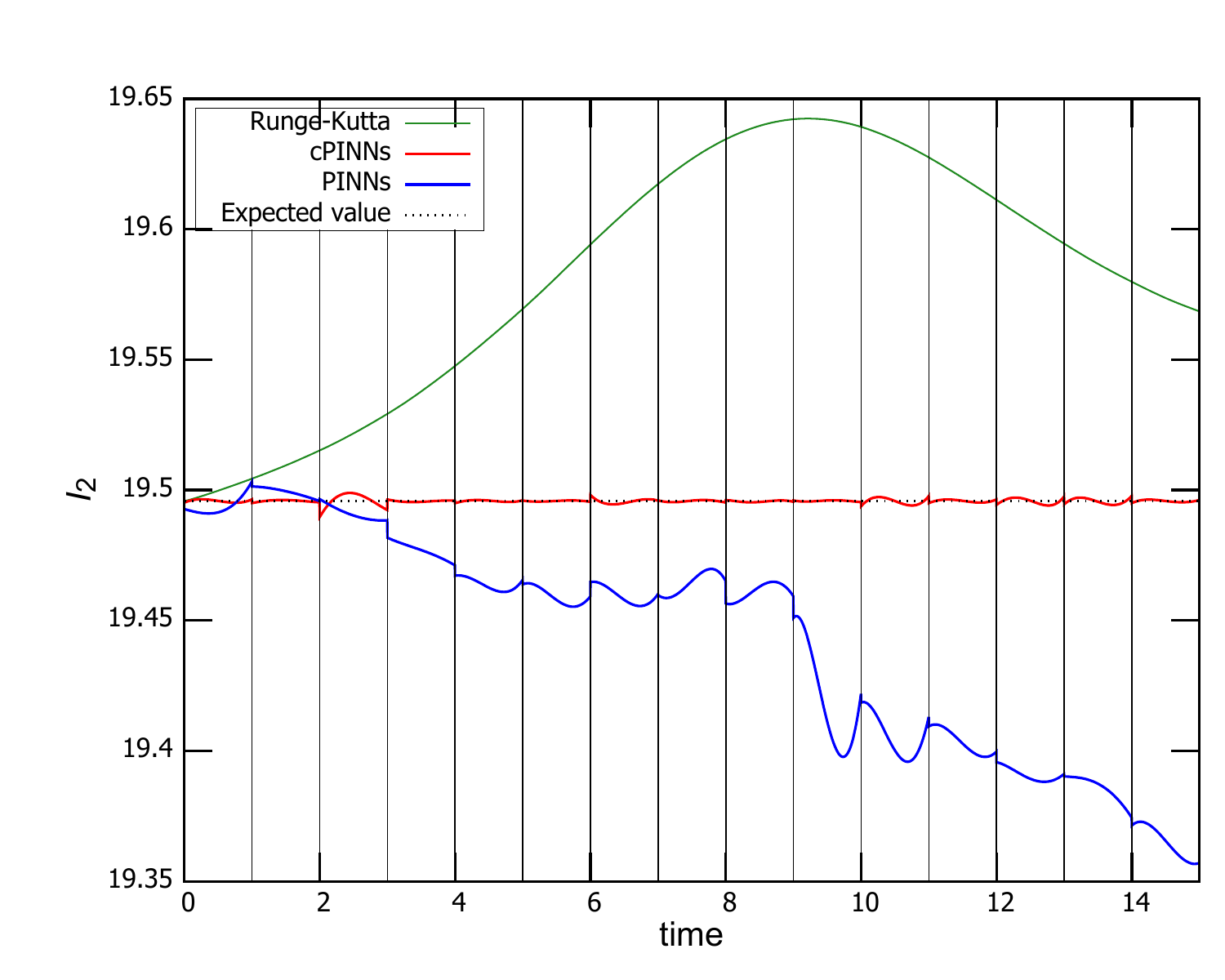}
	\caption{\label{conserv}Conserved quantity $I_2$ of the configurations of the onset collision 
    with $c=1.0, 0.25$ of Runge-Kutta, PINN and cPINN solutions, on
    a longer timescale of $0\leqq t\leqq 5$.
    For the PINNs, and cPINNs, we conducted the analysis with the short time segments 
    $[(s-1)\Delta t,s\Delta t],~~s=1,\cdots,15;~\Delta t=1.0$.}

\end{figure}

Therefore, it is worthwhile to apply cPINNs to the collision process, as this approach 
surely yields a novel solution.  
We show the convergence of the PINN and cPINN solutions on the first-time segment 
$[T_0=0,T_0+\Delta t]$. 
Fig.\ref{loss} shows the MSEs of the PINN and cPINN solutions plotted against the iteration number. 
As previously mentioned, 
convergence is attained if $|\nabla\mathit{MSE}|<\varepsilon$.  
Though the convergence of $MSE_\textrm{c}$ is superior in the cPINN as anticipated, 
the total MSE is much better in the PINN. This may seem strange because if the MSE 
had more conditions, the networks would provide a reasonable estimate of the actual 
solutions. 
The estimation of the conserved quantity $I_2$ using three distinct 
approaches might be used to explain the genesis of the unusual 
behavior (see Fig.\ref{cons0-1}). The results indicate that there is never an exact 
conserved quantity in the solution, not just in that obtained by 
the Runge-Kutta method.
The solution of the collision process is genuinely out of the submanifold of the 
conservative solution.

\begin{figure}[htbp]
	\centering 
	\includegraphics[width=1.0\linewidth]{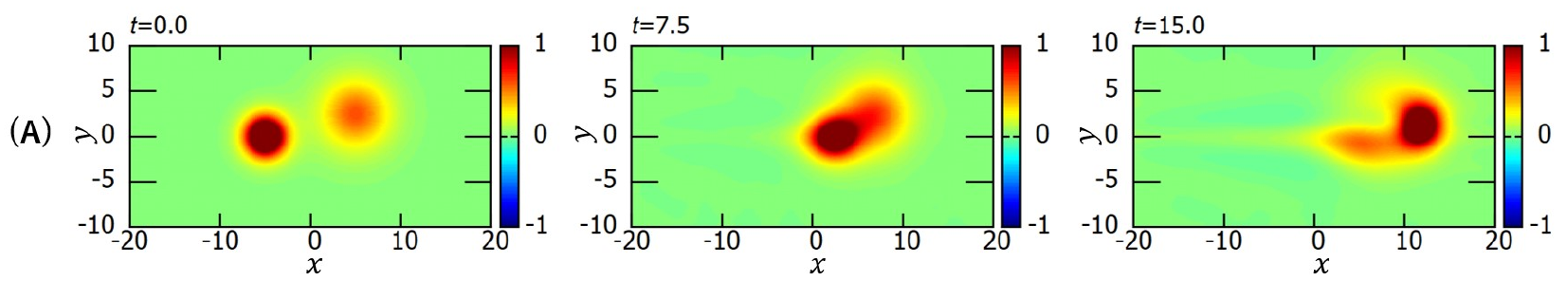}
	\includegraphics[width=1.0\linewidth]{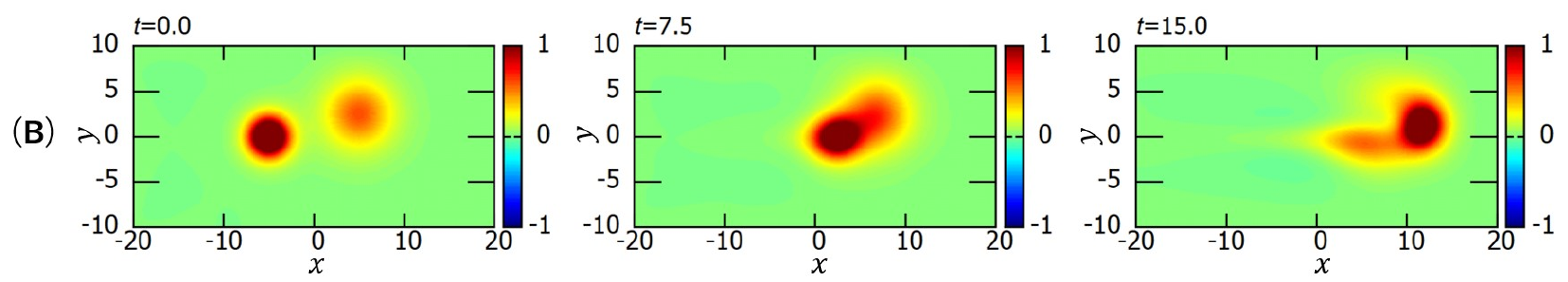}
	\includegraphics[width=1.0\linewidth]{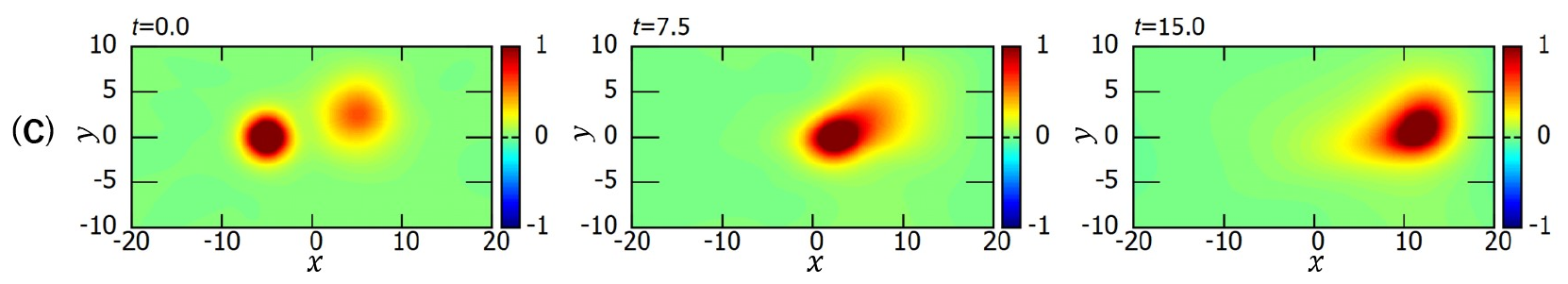}
	\caption{\label{fwd2off}Offset collision of the solutions with $c=1.0, 0.25$ 
	obtained with (A)~the exact numerical analysis, (B)~the PINN, and (C)~the cPINN
	at the time $t=0.0,7.5,15.0$.}

	\centering 
	\includegraphics[width=0.7\linewidth]{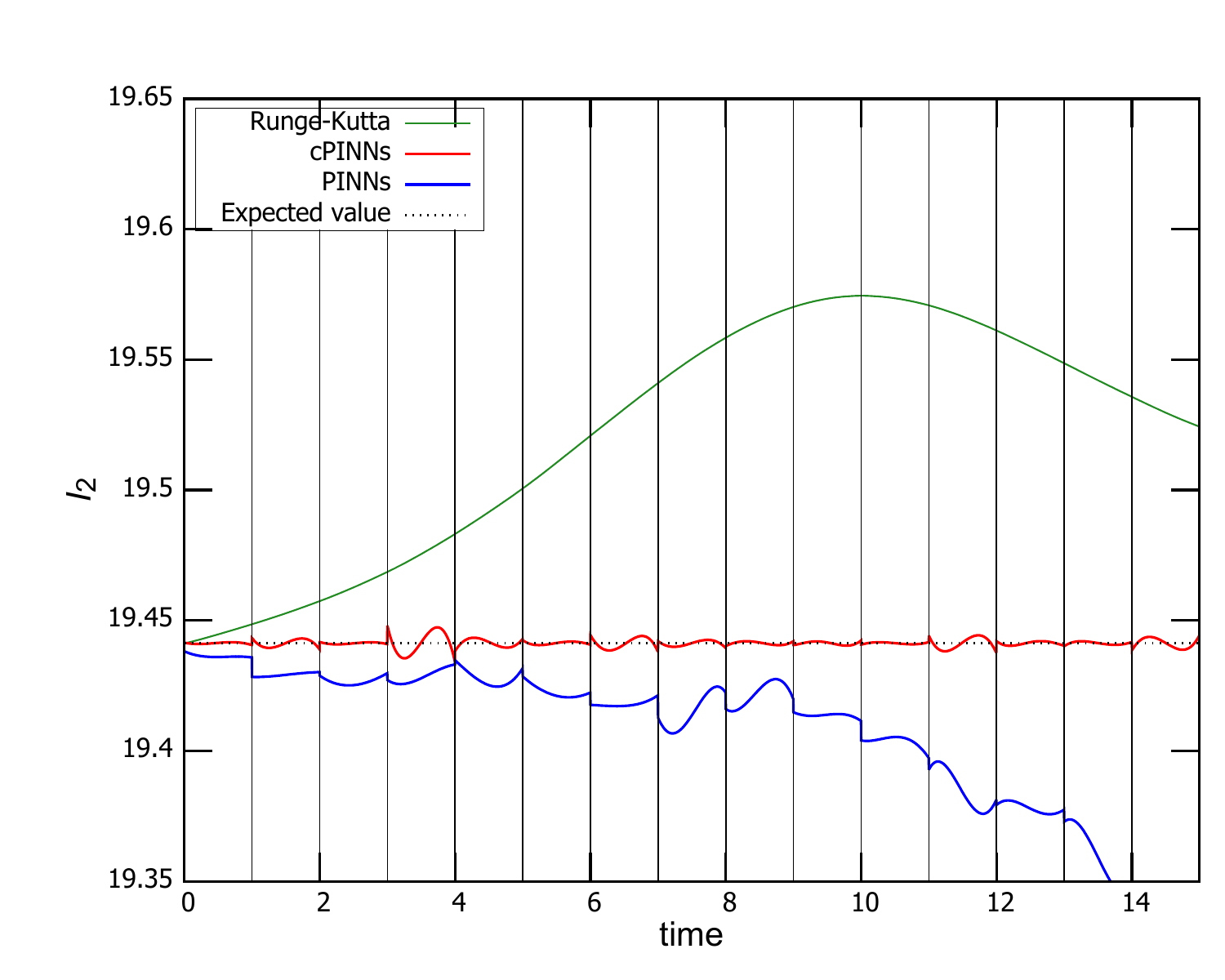}
	\caption{\label{conservoff}Conserved quantity $I_2$ of the configurations of 
	the offset collision (Fig.\ref{fwd2off}) with $c=1.0, 0.25$ 
	for the Runge--Kutta, PINN, and cPINN solutions on the longer timescale $0\leqq t\leqq 15$.}

\end{figure}

It is well known that 
some inelastic properties emerge when the heights of colliding solitons 
are significantly different: in such a collision, the taller soliton gains more height 
while the shorter tends to wane with the radiation~\cite{IWASAKI1990293}. 
Fig.\ref{fwd2on1} shows the collision of 2-solitons with 
the velocities $c=1.0$ and $0.25$. There is a slight difference between the numerical and the PINN
solutions,  
with the PINN solutions exhibiting a slight slowing, or subtle dissipation, of the solitons. 
Ref.\cite{IWASAKI1990293} also described a further unusual behavior of the collision process.  
In the case of an offset collision (a collision with a finite impact parameter), 
the smaller soliton is almost eliminated after impact. 
Fig.\ref{conserv} shows the conserved quantity $I_2$ plotted over a long time scale. 
This demonstrates that cPINN is 
efficient in identifying the solution that satisfies the conservation law. 
The 2-soliton results shown in Fig.\ref{fwd2off} more clearly highlight the discrepancy between the numerical
and PINN solutions. In Fig.\ref{conservoff}, we show the conserved quantity $I_2$ in the 
offset collision.

\subsection{Inverse analysis}

We define $\tilde{\mathcal{N}}$ for the inverse analysis of the ZK equation, 
using two unknown constants $\lambda_0,\lambda_1$
\begin{equation}
\tilde{\mathcal{N}}_\textrm{ZK}(u,u_x,u_{xxx},u_{xyy};\lambda_0,\lambda_1):=\lambda_0uu_x+\lambda_1\left(\nabla^2 u\right)_x\,. 
\label{inv2}
\end{equation}
We investigate the use of PINNs for inverse to find the parameter values ($\lambda_0,\lambda_1$) 
given training data derived from numerical analysis, PINNs and cPINNs. 
Table \ref{tab:Inv_ZK} gives the result of our 
inverse analysis with 1-soliton data. As shown in this table, all the coefficients of the equation 
are quite well-reproduced in the analysis.

\begin{table}[H]
	\centering
	\caption{\textit{Successful} parameterization of the 1-soliton solution}
	\label{tab:Inv_ZK}
	\centering
	\begin{tabular}{|c|c|c|}
		\hline
		& $\mathit{PDE}$ & $\mathit{MSE}$ $(\times 10^{-6})$ 
		\\ \hline
		correct  & $u_t+2uu_x+\left(\nabla^2 u\right)_x=0$ & $-$ 
		\\ \hline
		exact &$u_t+2.046uu_x+1.0048\left(\nabla^2 u\right)_x=0$ & 4.4 
		\\ \hline
		PINNs &$u_t+2.011uu_x+1.0033\left(\nabla^2 u\right)_x=0$ & 4.6
		\\ \hline
		cPINNs &$u_t+2.011uu_x+1.0033\left(\nabla^2 u\right)_x=0$ & 4.6
		\\ \hline
	\end{tabular}
\end{table}

\begin{figure}[t]
	\centering 
	\includegraphics[width=0.5\linewidth]{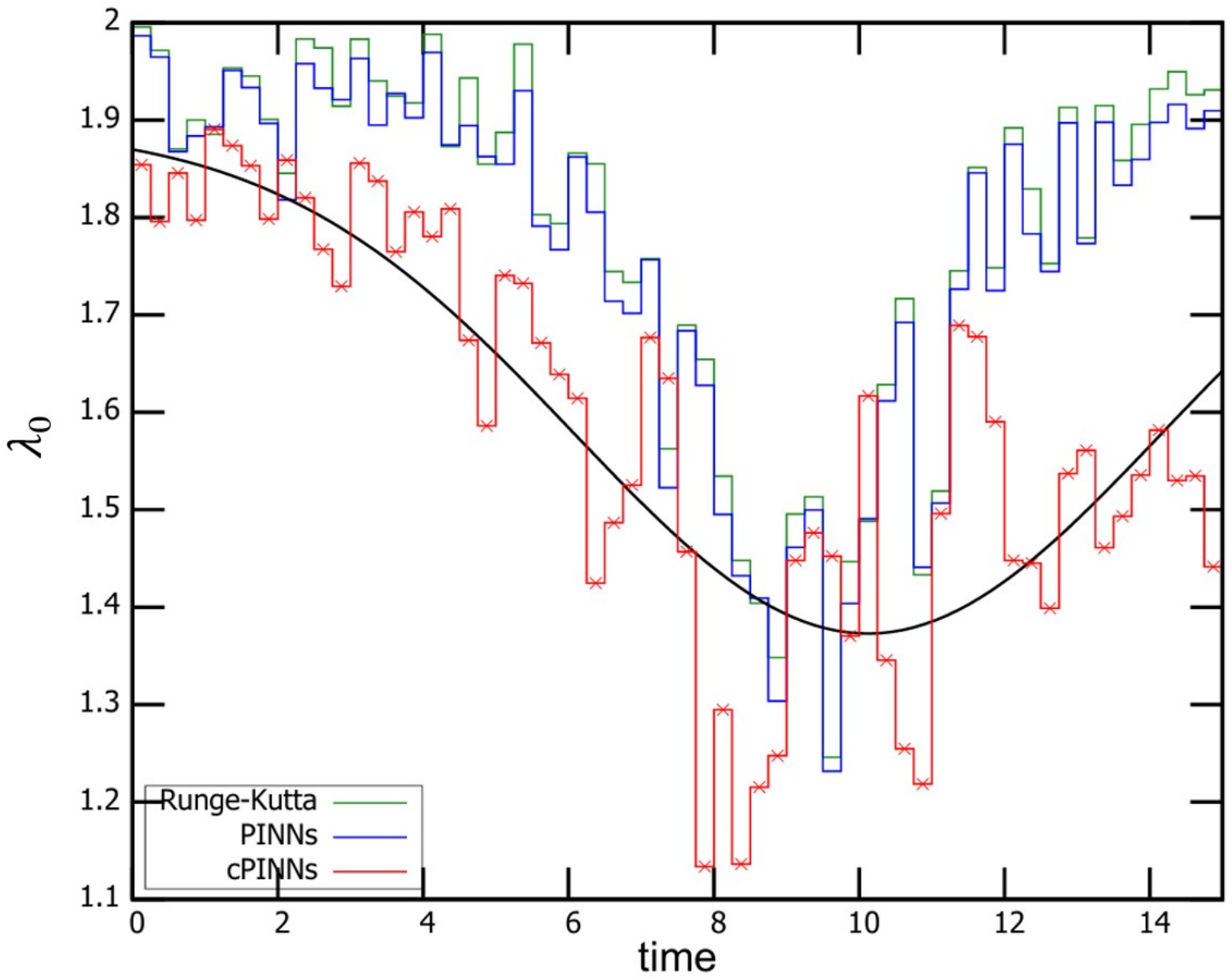}\hspace{-0.5cm}
	\includegraphics[width=0.5\linewidth]{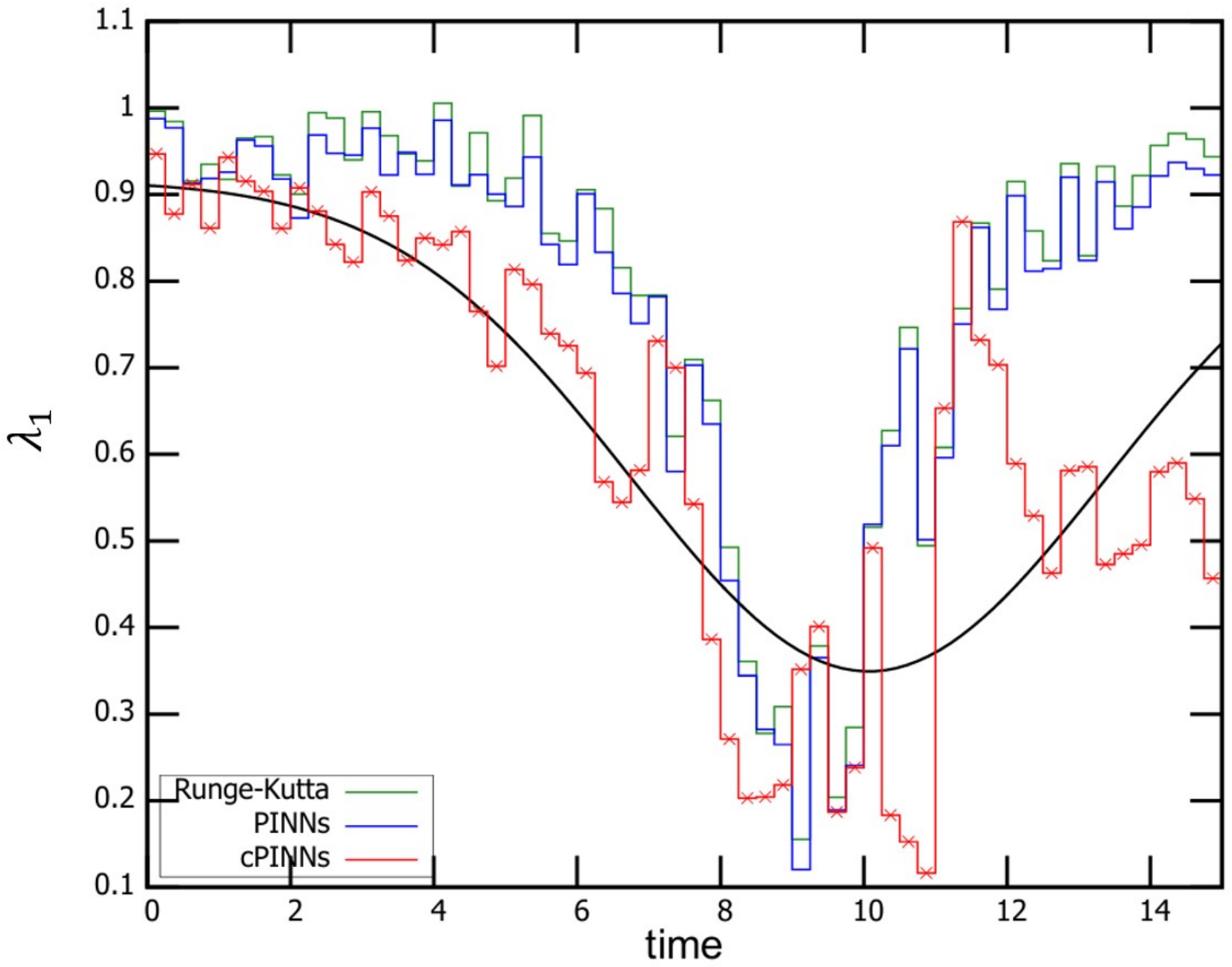}
	\caption{\label{invvalon}Inverse analysis with the PINNs. 
	The data are from the onset collision with $c=1.0$ and $0.25$ obtained by
	forward analysis using
	exact numerical analysis, PINNs, and cPINNs corresponding to 
	Fig.	\ref{fwd2on1}(A),(B), and  (C), respectively.
	The inverse analysis is realized by randomly sampling from 50000 data points. 
	The solid line shows an exponential fitting of the cPINN result 
	given in Eq.\eqref{modcoefon}.}

\end{figure}

\begin{figure}[t]
	\centering 
	\includegraphics[width=0.5\linewidth]{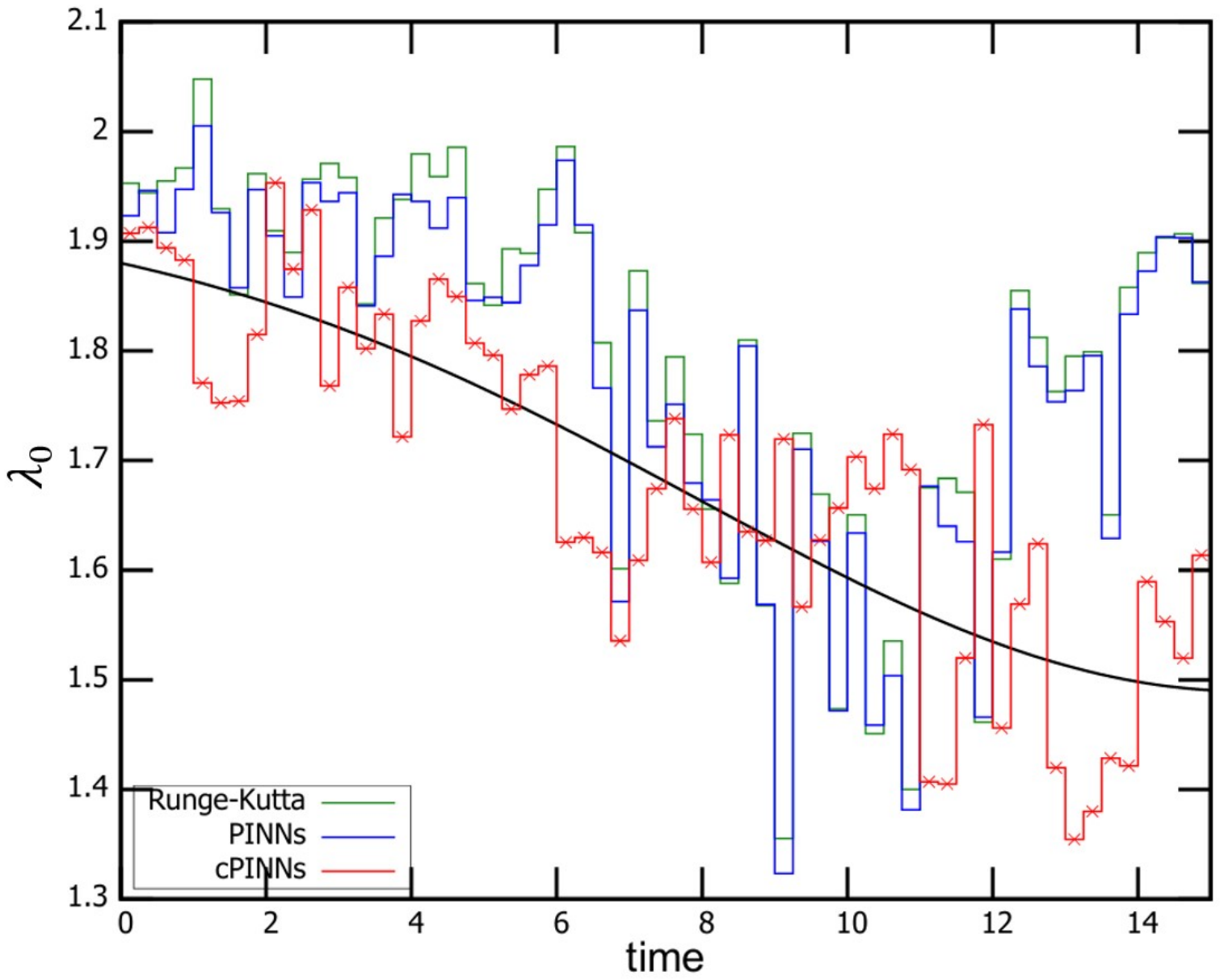}\hspace{-0.5cm}
	\includegraphics[width=0.5\linewidth]{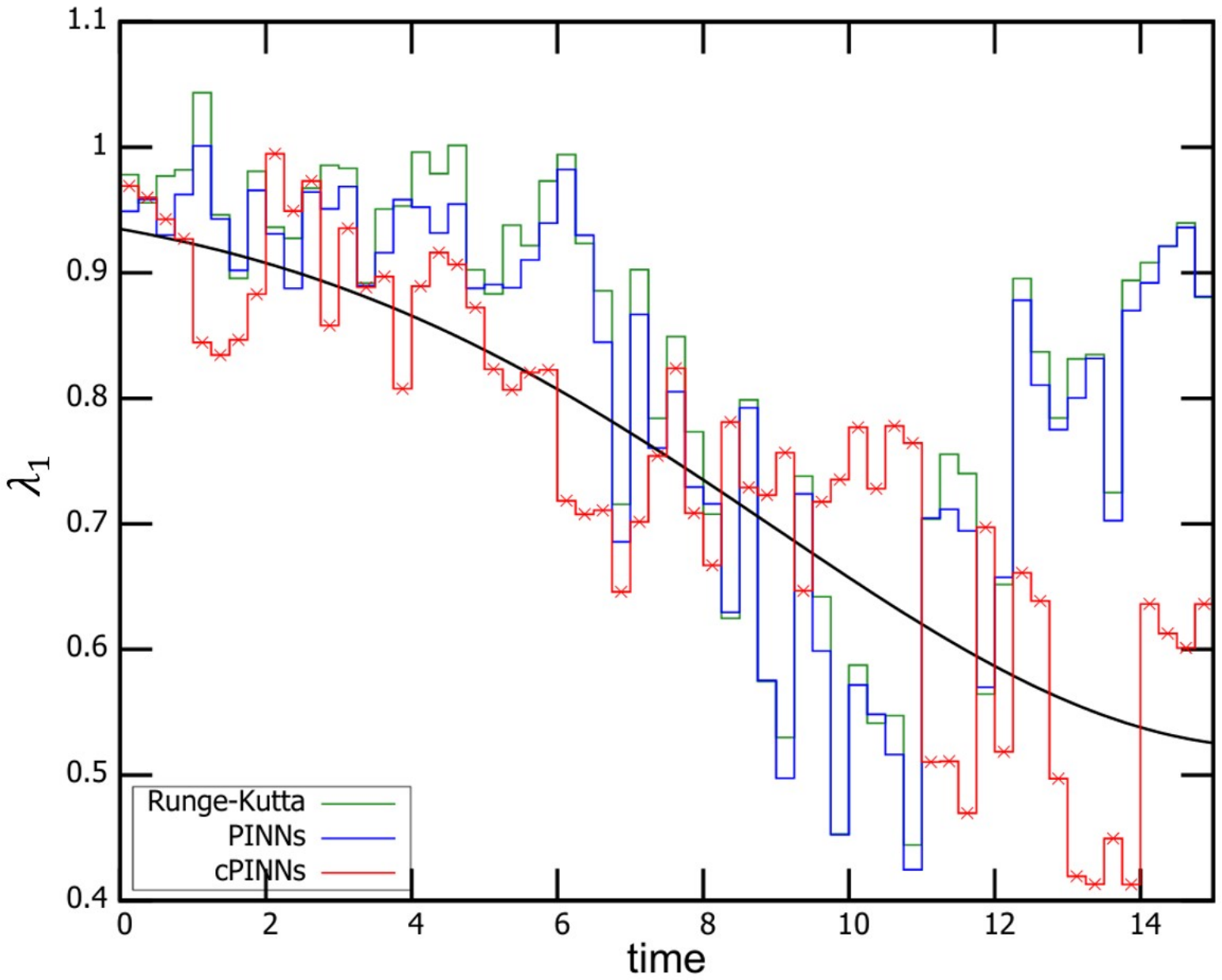}
	\caption{\label{invvaloff}Inverse analysis with PINNs. 
	The data are from the offset collision with $c=1.0$ and $0.25$ obtained by 
	forward analysis using
	exact numerical analysis, PINNs, and cPINNs, 
	corresponding to 
	Fig.\ref{fwd2off}(A),(B), and (C), respectively. 
	The inverse analysis is realized by randomly sampling from 50000 data points.
	The solid line shows an exponential fitting of the cPINN result 
	given in Eq.\eqref{modcoefoff}.}

\end{figure}

\subsection{Temporal coefficients analysis via inverse PINNs}

This section presents the inverse analysis~\eqref{inv2} of the 2-soliton training data, 
the results of which are shown in Figs.\ref{fwd2on1} and \ref{fwd2off}.
Instead of using the entire dataset for the inverse analysis, we considered only data from the short period of a single segment
at a time, 
allowing us to estimate the true value of the coefficients within each segment. 
Such a constraint was repeatedly used with several types of formulation to overcome
the issue of reducing the computational cost of training a larger network~\cite{JAGTAP2020113028,Jagtap2020CCP,YANG2024237}.  
Fig.\ref{invvalon} shows the results for the modulation of the coefficients $\lambda_0,\lambda_1$ over time $t$ 
in the onset collision process using forward analysis data 
obtained by the Runge-Kutta method, PINNs, and cPINNs 
with time segments of width $dt=0.25$. 
The coefficients depart from their initial values, indicating a special 
temporal effect that is not explicitly implemented in the equation. 
Of note is the significant modulation of the coefficients specifically their reduced values during 
impact and eventual recovery to their initial conditions. 
Because the data obtained with cPINNs are similar to the integrable data because of the method's conservative nature, 
it is natural to expect that the modulation of the coefficients would be suppressed. 
However, the result runs counter to this expectation, with the cPINNs showing the most drastic change. 
Fig.\ref{invvaloff} shows the same analysis in the offset collision case. 
A notable feature of the cPINNs that sets their result apart from the other two methods is that 
after the collision the coefficients do not 
return to their initial value (a process called mutation), 
suggesting that the equation decays into a different regime.

\begin{figure}[htbp]
	\centering 
	\vspace{-0.0cm}
	\includegraphics[width=0.9\linewidth]{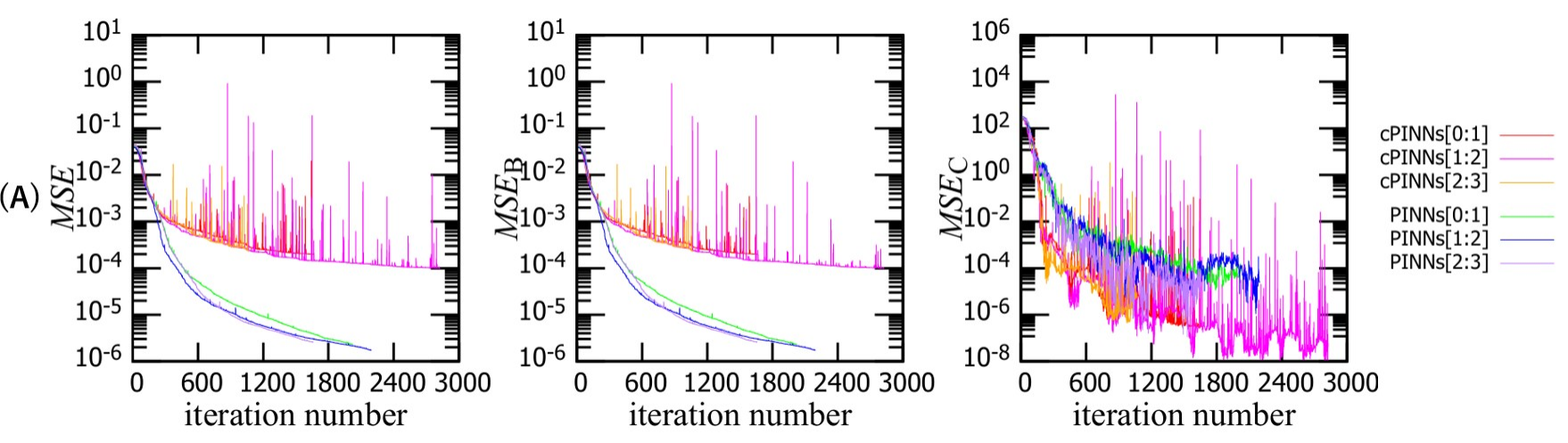}\\
	\vspace{-0.0cm}
	\includegraphics[width=0.9\linewidth]{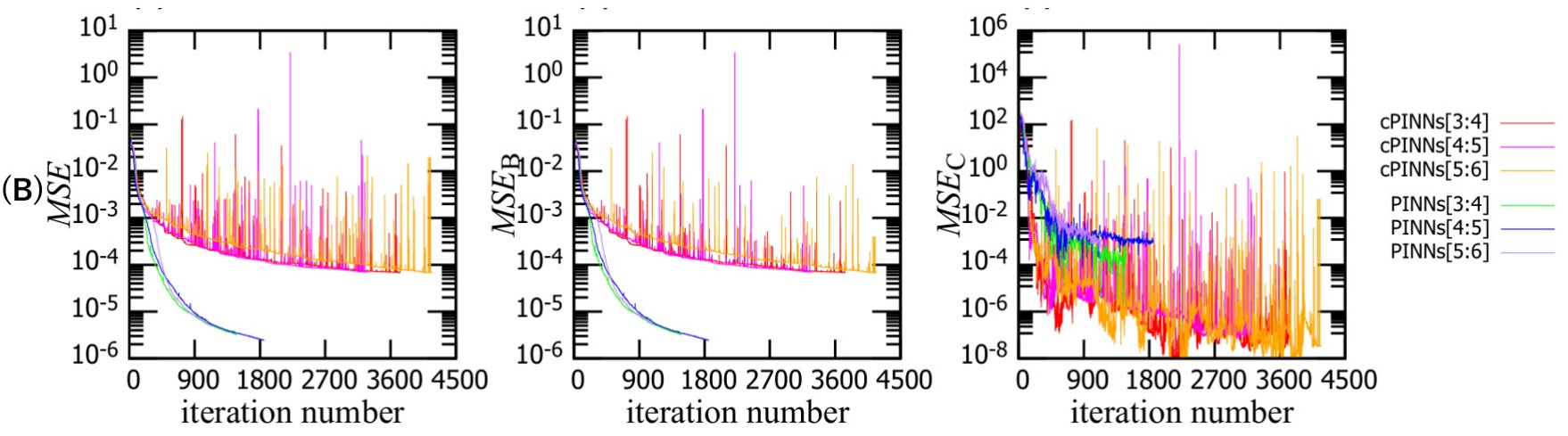}\\
	\vspace{-0.0cm}
	\includegraphics[width=0.9\linewidth]{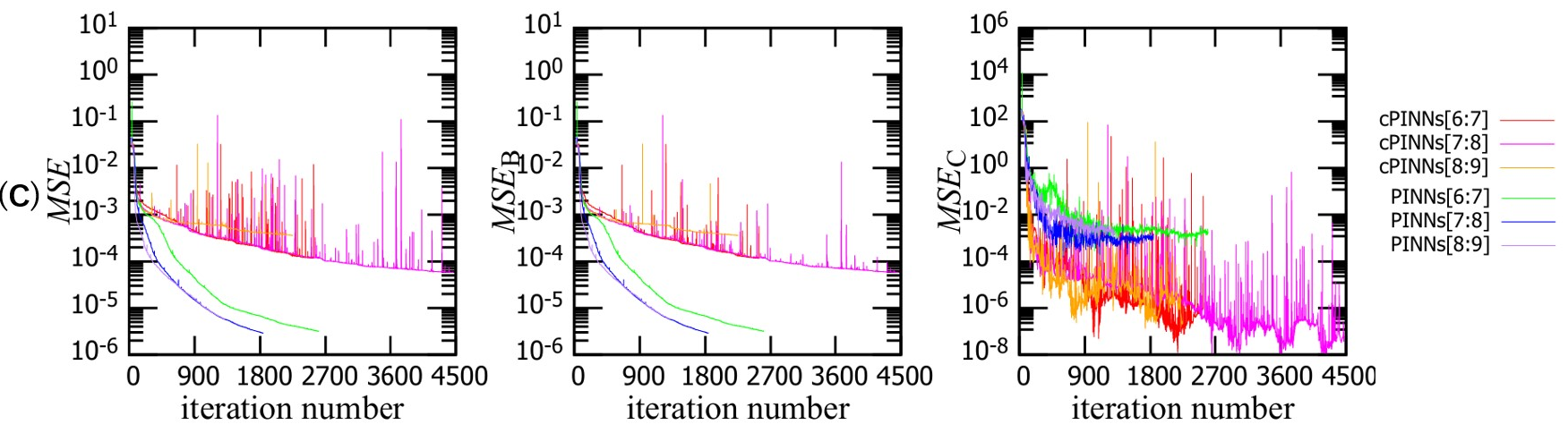}\\
	\vspace{-0.0cm}
	\includegraphics[width=0.9\linewidth]{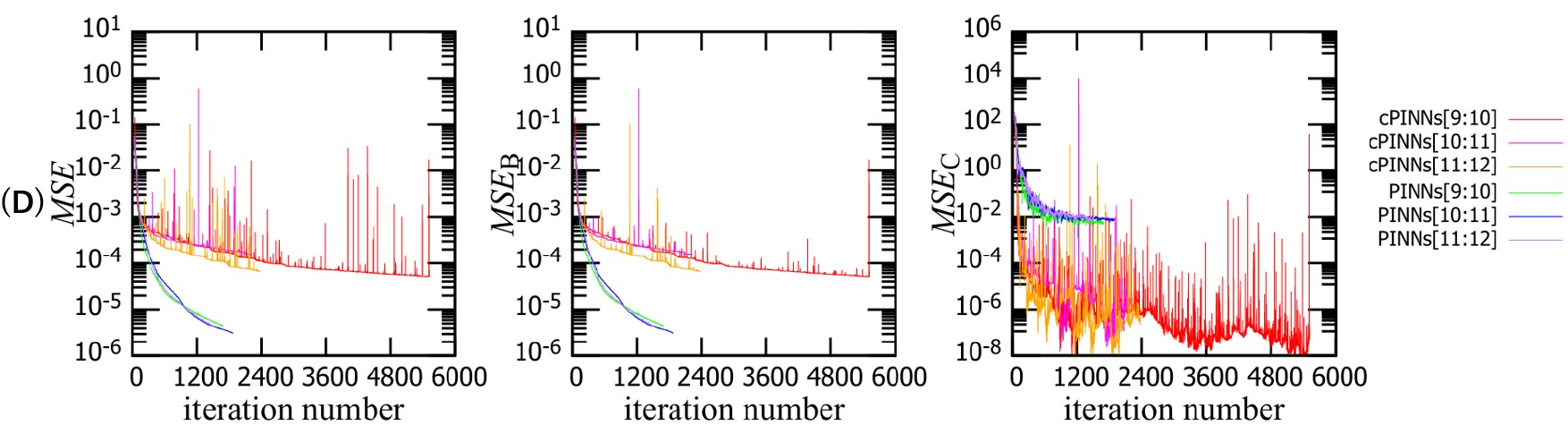}\\
	\vspace{-0.0cm}
	\includegraphics[width=0.9\linewidth]{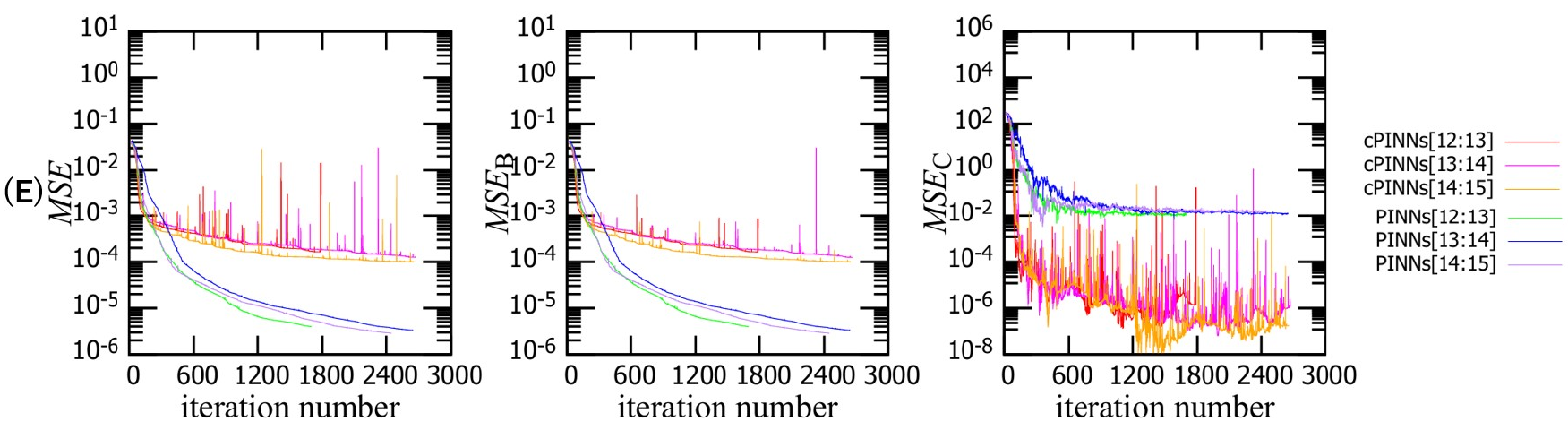}
	\caption{\label{loss_timings}
	MSE for the onset collision of the solutions with $c=1.0,0.25$. 
	For sake of visualization, we separate the data into five blocks with three time segments each: 
	(A)$[0,3]$,(B)$[3,6]$,(C)$[6,9]$,(D)$[9,12]$,(E)$[12,15]$. 
	Cool colors (green, blue, cyan) and warm (red, magenta, orange) colors represent PINN and cPINN results, 
	respectively. 
	The left, center, and right columns shows the total MSE, $\mathit{MSE}_\textrm{B}$, and 
	$\mathit{MSE}_\textrm{C}$, respectively. 
	}
\end{figure}

\begin{figure}[htbp]
	\centering 
	\vspace{-1.5cm}
	\includegraphics[width=0.9\linewidth]{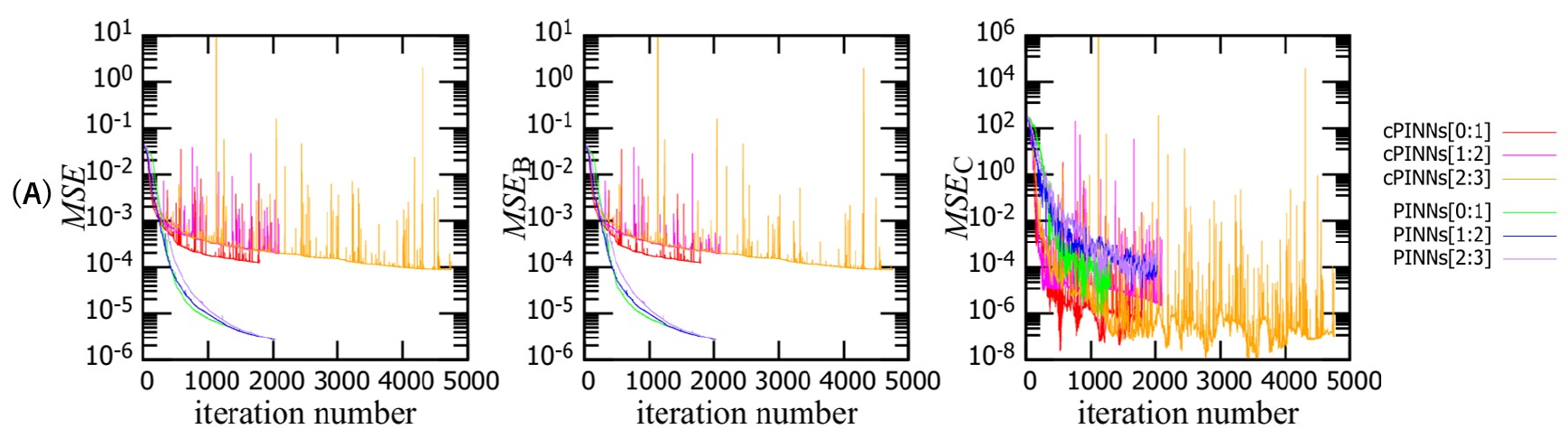}\\
	\vspace{-0.0cm}
	\includegraphics[width=0.9\linewidth]{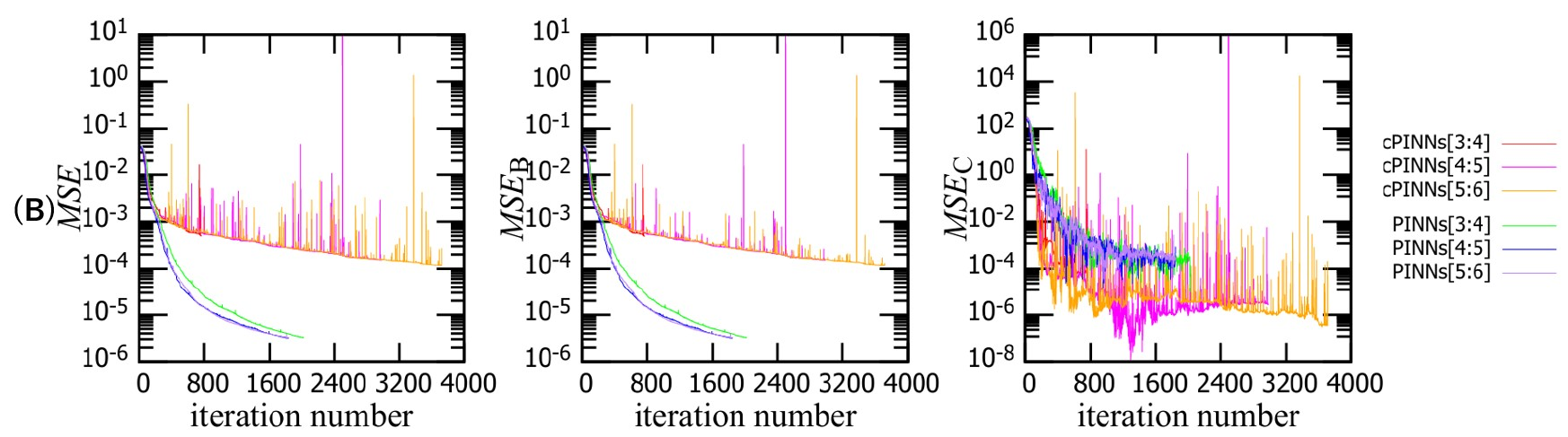}\\
	\vspace{-0.0cm}
	\includegraphics[width=0.9\linewidth]{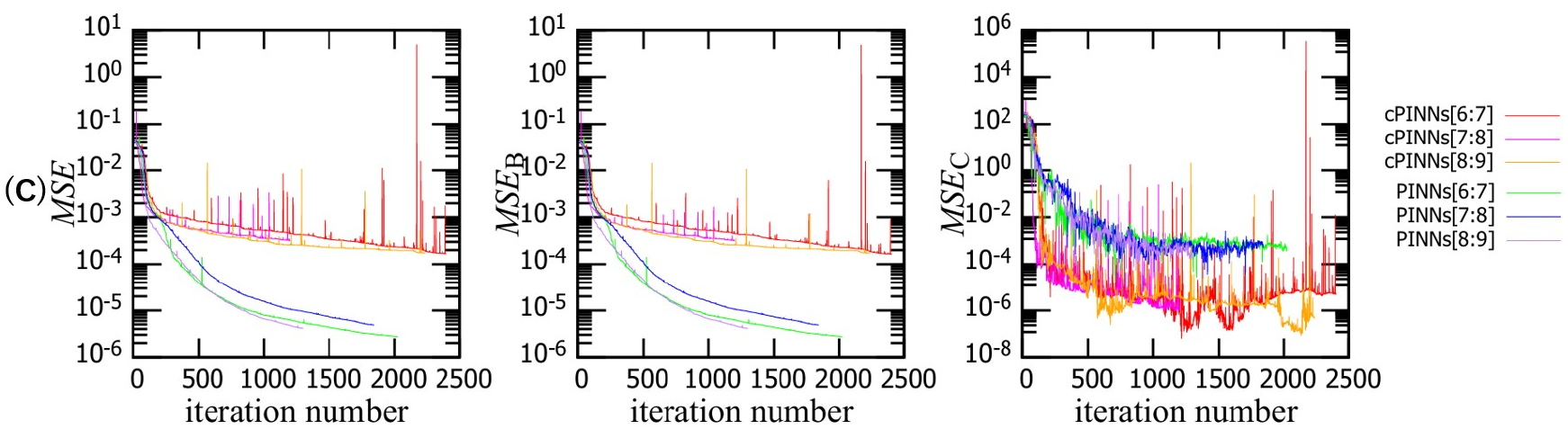}\\
	\vspace{-0.0cm}
	\includegraphics[width=0.9\linewidth]{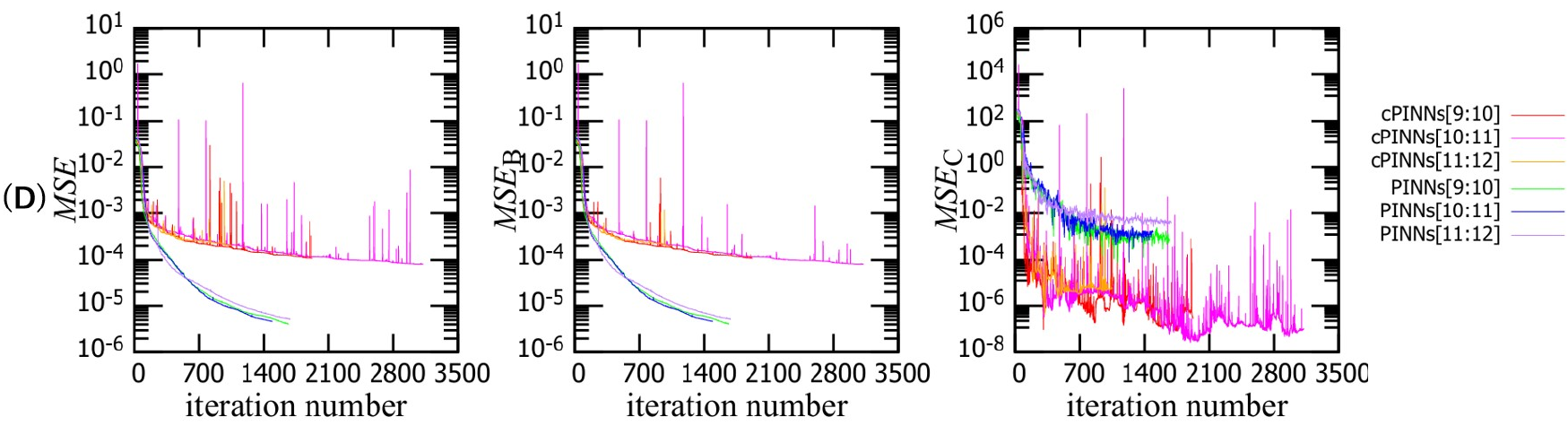}\\
	\vspace{-0.0cm}
	\includegraphics[width=0.9\linewidth]{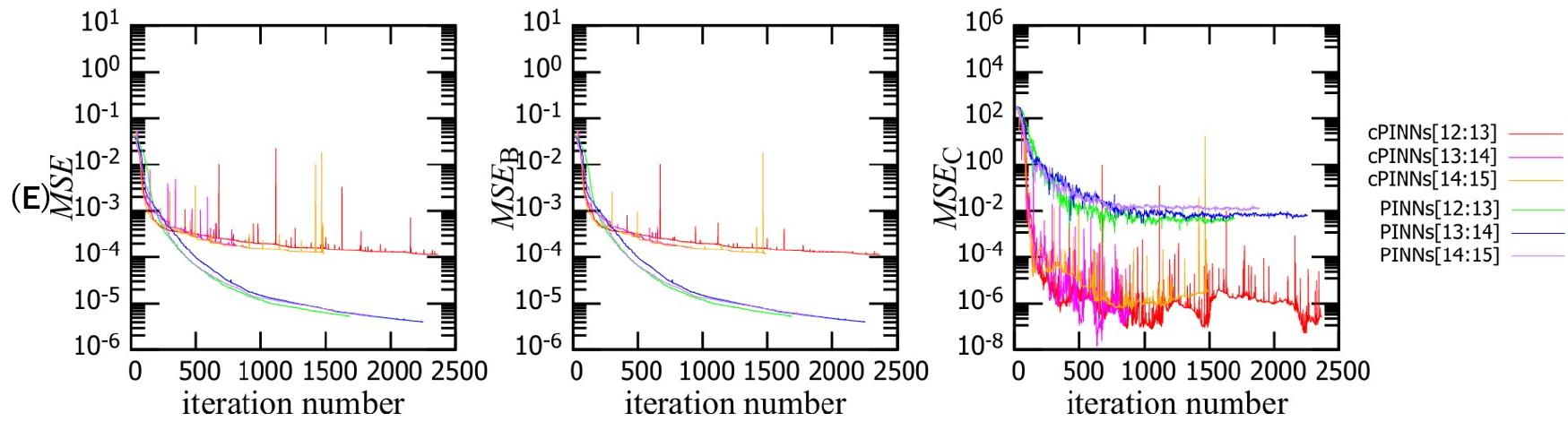}
	\caption{\label{loss_off_timings}
	MSE for the offset collision of the solutions with $c=1.0,0.25$. 
	For sake of visualization, we separate the data into five 
	blocks with three time segments each: 
	(A)$[0,3]$,(B)$[3,6]$,(C)$[6,9]$,(D)$[9,12]$,(E)$[12,15]$. 
	Cool (green, blue, cyan) colors and warm (red, magenta, orange) colors
	represent PINN and cPINN results, respectively. 
	The left, center, and right columns show the total MSE, 
	the $\mathit{MSE}_\textrm{B}$, and the $\mathit{MSE}_\textrm{C}$, 
	respectively. }
\end{figure}

\subsection{Interpretation of the coefficient modulation effect}

As discussed in the previous subsection, we have observed non-trivial modulations in
or mutations of the coefficients
derived from the inverse analysis of data from different configurations of the 
equation with constant coefficients. 
Our inverse analysis with PINNs have exposed this effect for the first time.
This phenomenon appears to be unique to quasi-integrable systems 
and does not evidently arise in
integrable equations such as the KdV and nonlinear Schr\"{o}dinger equations. 
In these cases, both the forward and the inverse analysis perfectly 
coincide in terms of the parameter mutations. 
The natural interpretation of this result is the temporal emergence of 
an effective interaction during the impact of the collision. It is difficult to evaluate such 
an interaction that was not included explicitly in the original equation. 
Therefore, we would like to 
approach in this question from a different point of view. 
One helpful way to distinguish between integrable 
and quasi-integrable equations is to look for their exact solutions. As is well-known the KdV equation
possesses an infinite number of exact solutions. 
However, to the best of our knowledge, there are no analytical multi-soliton solutions 
to the ZK equation. This naturally raises the question: 
Are the numerical 2-soliton ``solutions'' derived from PINNs or the Runge-Kutta method true solutions? 
The Runge-Kutta method and other numerical method are based on the finite difference formalism.
The solutions in discrete space-time (or discrete wave number--angular frequency space) 
are good approximations of the genuine ones.  
However, because it might not have a continuous limit counterpart, the collision process is a DIFC.
As previously stated, a distinguishing characteristic of PINNs is that 
they are based on a mesh-free algorithm. Therefore, the collision process does not 
directly solve the original equation, and, according to the inverse analysis 
the process is expressed using a modified equation. 
Concerning the forward PINNs or cPINNs, we have certain reservations. 
They are mesh-free but still exhibit the same type of behavior as the Runge-Kutta method. 
The hint is in the low convergence property of the result with PINNs. 
Fig.\ref{loss_timings} shows the 
MSE for PINNs and cPINNs in the collision process, corresponding to the solutions shown in Fig.\ref{fwd2on1}. 
For the sake of visibility, the data are divided into five blocks each consisting of three consecutive time 
segments. The warm color (red, magenta, and orange) show the cPINN results, 
and the cool color (green, blue, and cyan) show the PINN results. 
The PINN MSEs are already an order of magnitude worse than those for the corresponding 1-soliton solutions. 
The cPINN results show even worse performance; 
the MSEs are roughly $10^{-4}$, indicating that the original equation has no exact solution with 
a sufficiently low MSE.  
We may conclude that, in terms of PINN technology, we were successful in determining the correct 
equation for the collision of the quasi-integrable solitons. 
Fig.\ref{loss_off_timings} shows the 
PINN and cPINN MSEs for the offset collision, corresponding to the solutions in Fig.\ref{fwd2off}.

\begin{figure}[t]
	\centering 
	\includegraphics[width=1.0\linewidth]{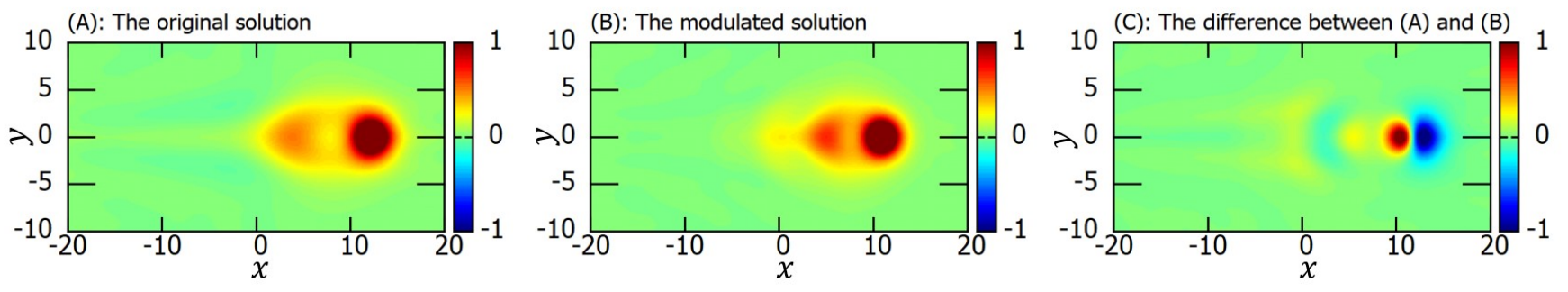}
	\caption{\label{Resolon}
	Snapshot at $t=15$ of the DIFC of the exact numerical analysis of the onset collision 
	with $c=1.0$ and $0.25$.  
	(A) Exact numerical analysis corresponding to Fig.\ref{fwd2on1}(A), 
	(B) the equation with the modulated coefficients~\eqref{modcoefon}, and 
	(C) the difference between (A) and (B).
	}

\end{figure}

\begin{figure}[t]
	\centering 
	\includegraphics[width=1.0\linewidth]{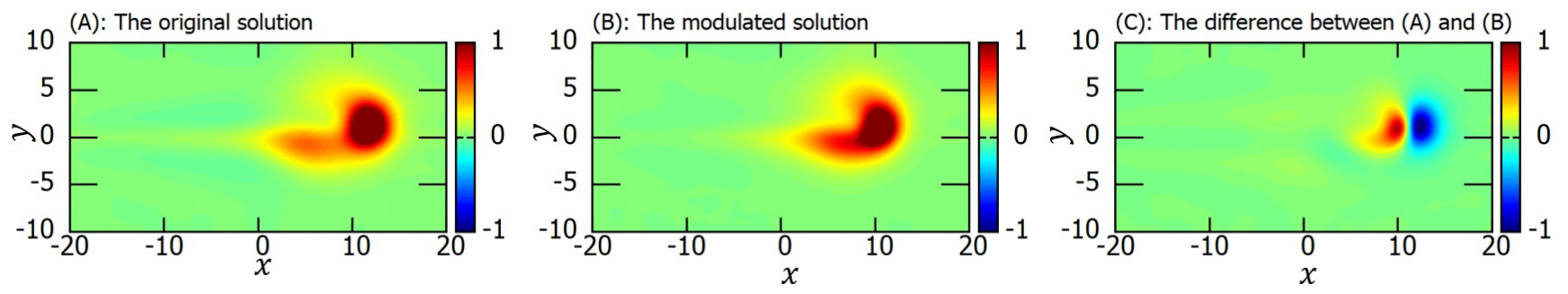}
	\caption{\label{Resoloff}
	Snapshot at $t=15$ of the DIFC of the exact numerical analysis of 
	the offset collision 
	with $c=1.0$ and $0.25$.  
	(A) Exact numerical analysis for corresponding to Fig.\ref{fwd2on1}(A), 
	(B) the equation with the modulated coefficients~\eqref{modcoefon}, and 
	(C) the difference between (A) and (B).
	}

\end{figure}

\begin{figure}[t]
	\centering 
	\includegraphics[width=0.5\linewidth]{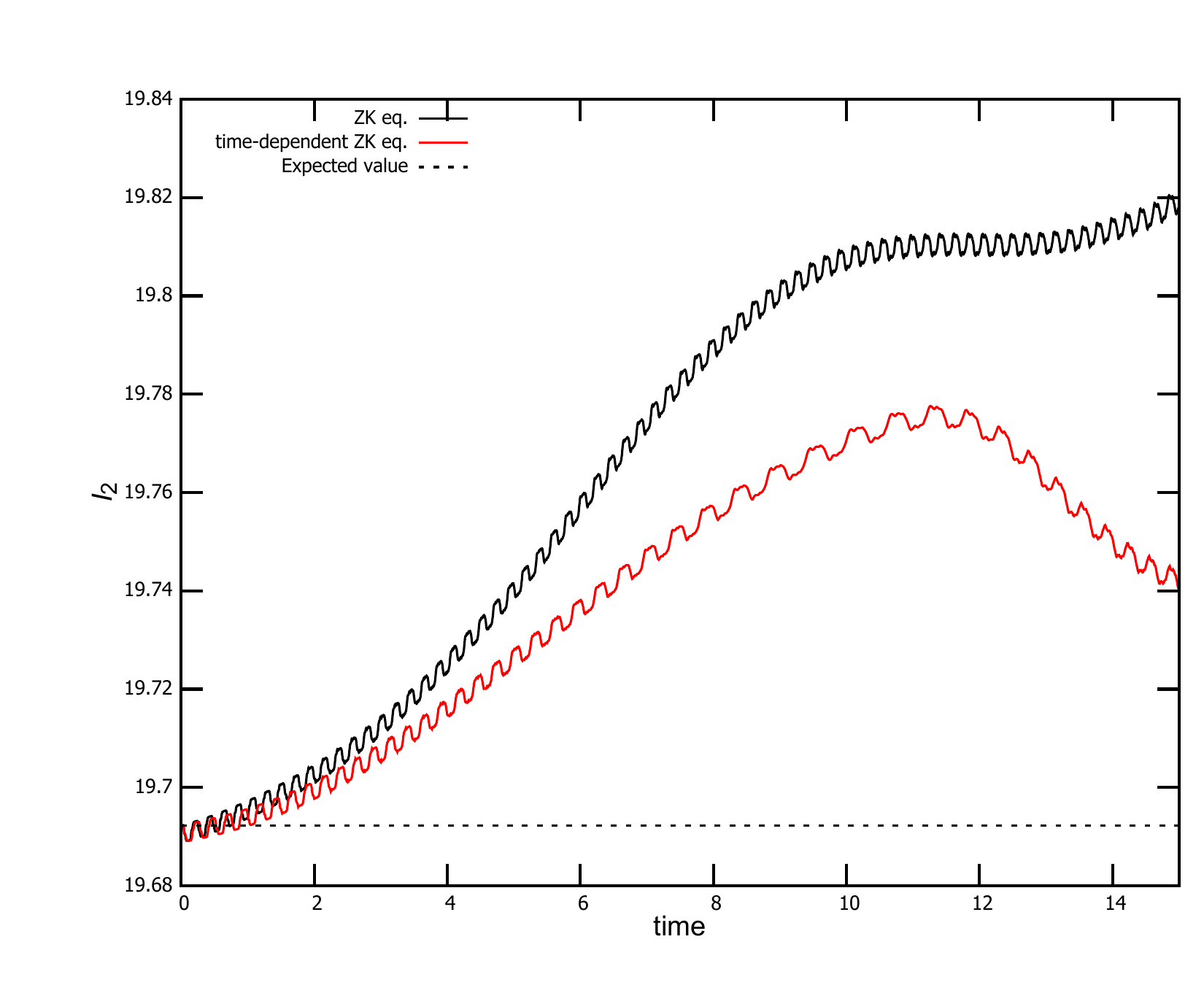}\hspace{-0.5cm}
	\includegraphics[width=0.5\linewidth]{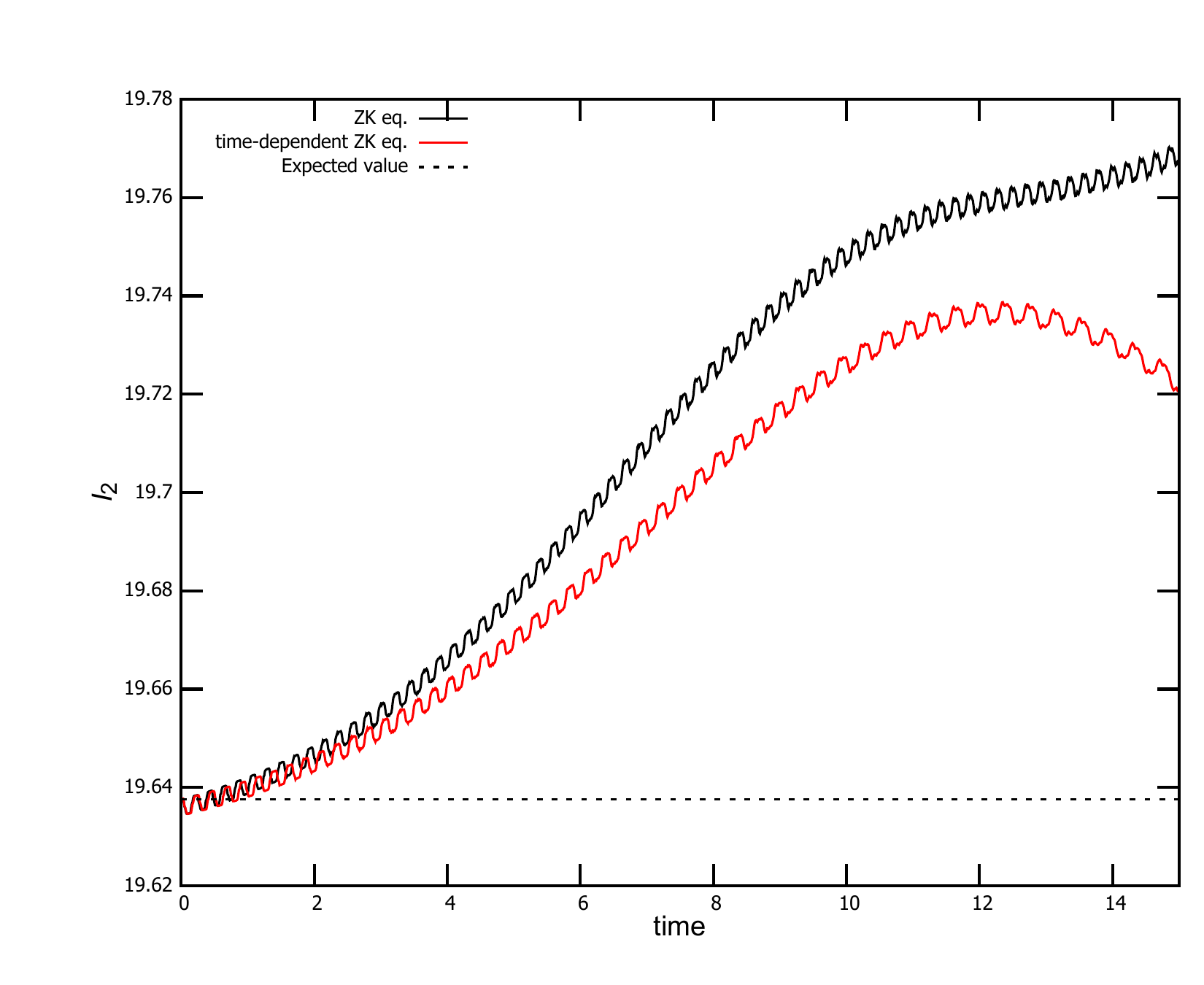}
	\caption{\label{Reconsoff}Conserved quantity $I_2$ of the DIFC of the 
	normal ZK equation and the time-dependent coefficients 
	for (A) the onset collision 
	with $c=1.0$ and $0.25$ in terms of the numerical analysis corresponding to 
	Fig.\ref{fwd2on1}(A),  
	and (B) the offset collision, corresponding to Fig.\ref{fwd2off}(A). 
	The inverse analysis is realized by the randomly sampling from 50000 data points.}

\end{figure}

\subsection{2-soliton solution of the equation with modulated coefficients}

As shown above, the PINN results clearly imply that the equation 
governing the system may vary over time, especially during the collision. 
Thus, it is natural to re-examine the new equation with modulated coefficients. 
Our procedure is as follows. 
We start by examining an interpolation function describing the temporal change 
in the coefficients in order to make the analysis tractable. 
Next,we solve the equation with that function by forward analysis (here, 
we use the Runge-Kutta method). 
Lastly, we analyze the conserved quantity $I_2$ with the obtained solution.  
For the form of the equation 
$\mathcal{N}_\textrm{ZK,mod}:=\lambda_0uu_x+\lambda_1(\nabla^2 u)_x$, 
we here employ a simple exponential fitting of the coefficients as
\begin{align}
\lambda_i=a_i-\exp(b_i+c_it+d_it^2),~~i=0,1
\end{align} 
with the eight fitting parameters $a_i,b_i,c_i,d_i$. 
We choose cPINN training data in this analysis 
because it exhibited a wider and greater shift in the coefficients
compared with the PINN or Runge--Kutta data,  
making the effect more noticeable. 
For the change in the coefficient at the onset collision obtained by cPINN (Fig.\ref{invvalon}), 
the interpolation function can be fixed approximately as
\begin{align}
&\lambda_0=\left(1.892\pm 0.069\right)-\exp\left[\left(-3.8\pm 1.1\right)+\left(0.62\pm 0.21\right)t+\left(-0.031\pm 0.010\right)t^2\right]\,,
\nonumber \\
&\lambda_1=\left(0.916\pm 0.054\right)-\exp\left[\left(-5.1\pm 1.3\right)+\left(0.91\pm 0.25\right)t+\left(-0.046\pm 0.013\right)t^2\right]\,,
\label{modcoefon}
\end{align}
which are also plotted as solid lines in Fig.\ref{invvalon}.
We solve the equation with Eqs.\eqref{modcoefon} by the Runge-Kutta method.   
In Fig.\ref{Resolon}, we show the DIFC of the snapshot at $t=15$ 
together with the original, DIFC from the unmodified equations 
reproduced from Fig.\ref{fwd2on1}. 
For the offset cPINN, the interpolation function becomes 
\begin{align}
&\lambda_0=\left(1.93\pm 0.12\right)-\exp\left[\left(-2.9\pm 1.7\right)+\left(0.27\pm 0.24\right)t+\left(-0.0088\pm 0.0098\right)t^2\right]\,,
\nonumber \\
&\lambda_1=\left(0.965\pm 0.076\right)-\exp\left[\left(-3.4\pm 1.5\right)+\left(0.33\pm 0.22\right)t+\left(-0.0106\pm 0.0093\right)t^2\right]\,,
\label{modcoefoff}
\end{align}
Again the behavior is shown in Fig.\ref{invvaloff}. 
The DIFC of the snapshot at $t=15$ is presented with the original DIFC 
(Fig.\ref{fwd2off}) in Fig.\ref{Resoloff}. 
The difference between the original equation and that with modulated 
coefficients is clear; the weaker solitons appear to be somewhat closer to the taller ones
following the collision, whereas the taller solitons of the latter are ahead of the former. 
As a result, they tend to maintain their shape after impact. 
This suggests that the integrable nature recovers when the influences are lessened by the 
weaker coefficients. 
Therefore, we anticipate that there may be a 2-soliton DIFC to 
a novel variable coefficients equation
~\cite{calogero1978exact,brugarino1980integration,joshi1987painleve,hlavaty1988painleve,brugarino1991painleve,
gao2001variable,kobayashi2005generalized,kobayashi2006painleve,gao2021optica,
gao2021varmodkdv,GAO20222707,Gao2022InNO,gao2022water}. 

\begin{figure}[t]
	\centering 
	\includegraphics[width=0.5\linewidth]{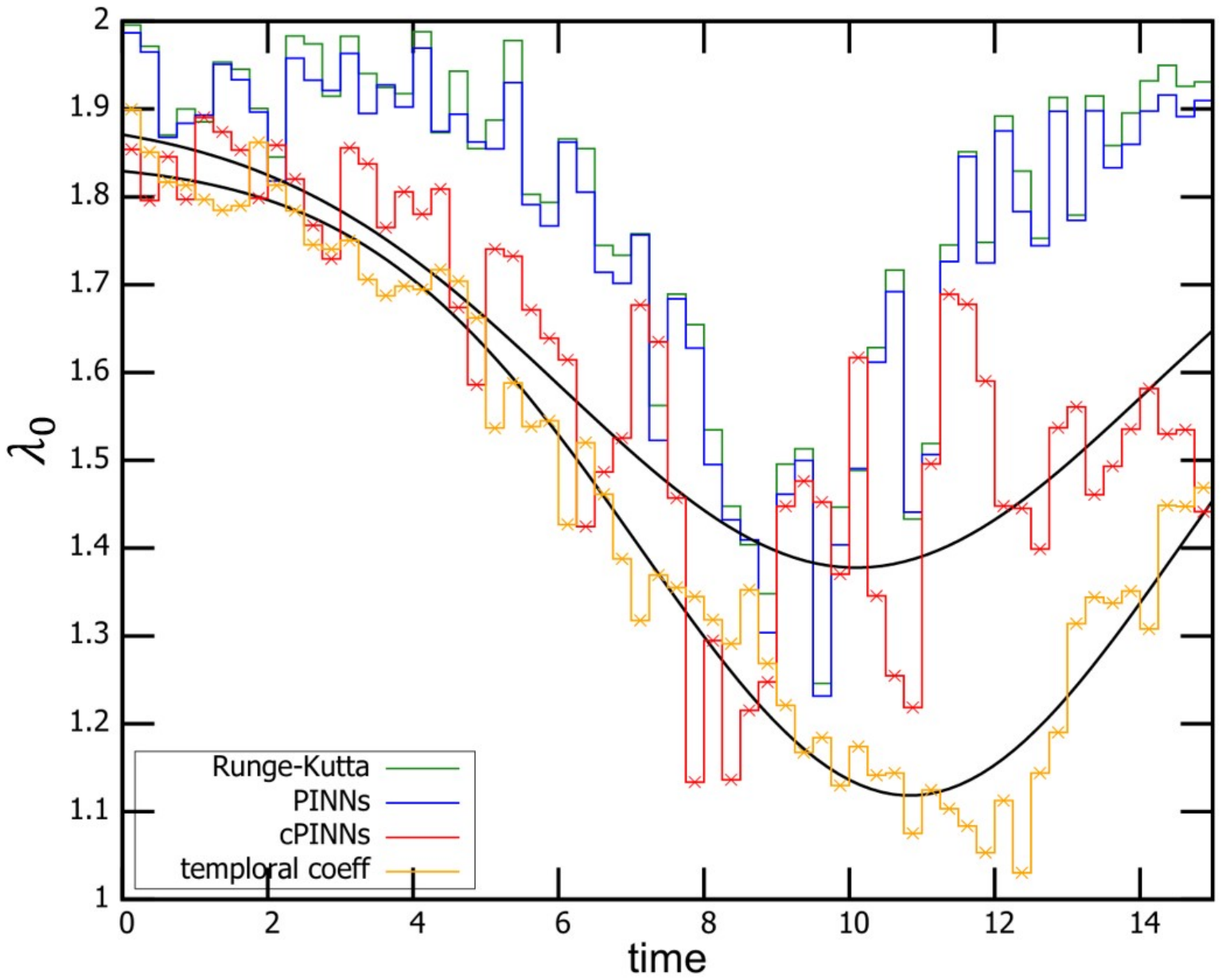}\hspace{-0.5cm}
	\includegraphics[width=0.5\linewidth]{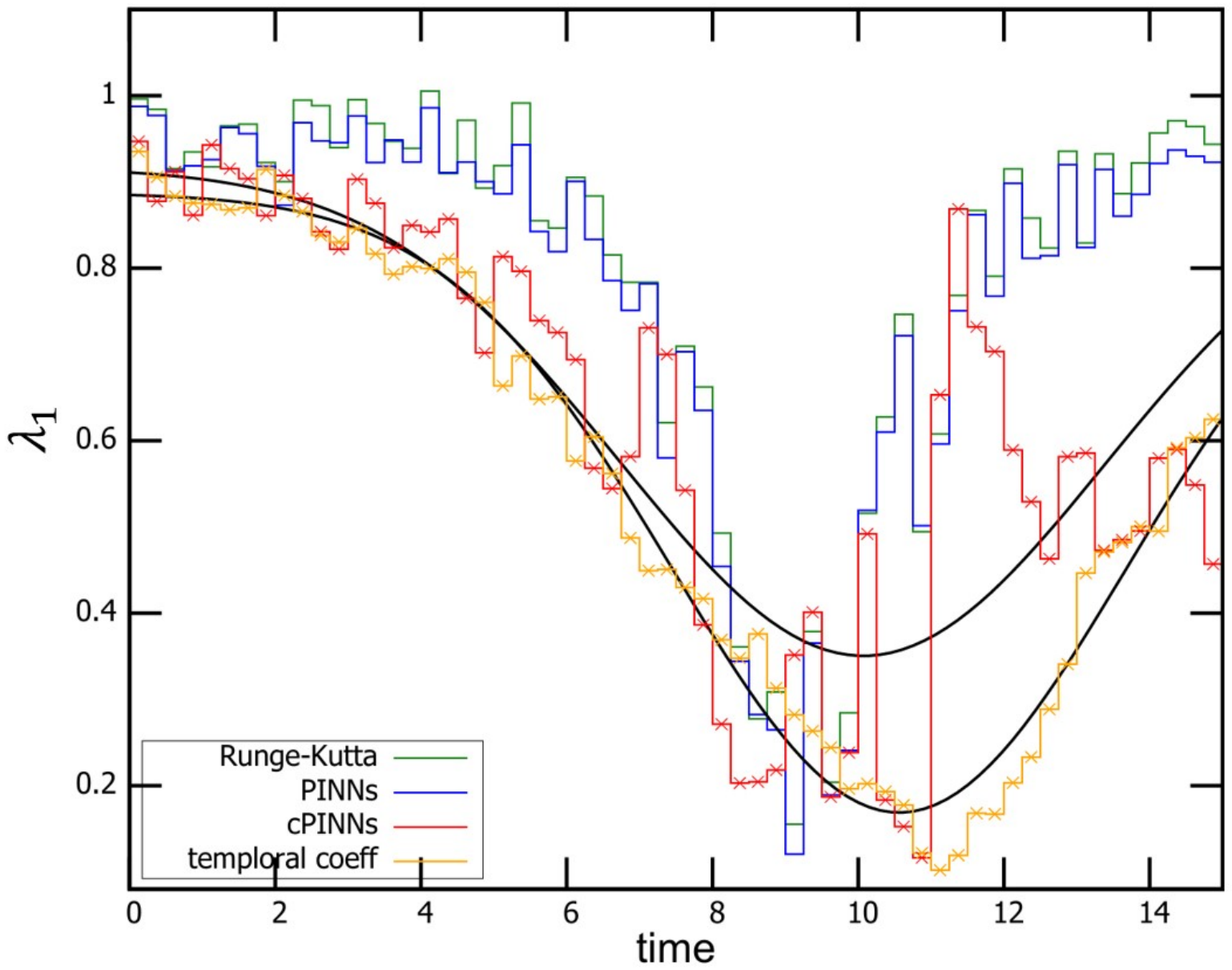}
	
	\includegraphics[width=0.6\linewidth]{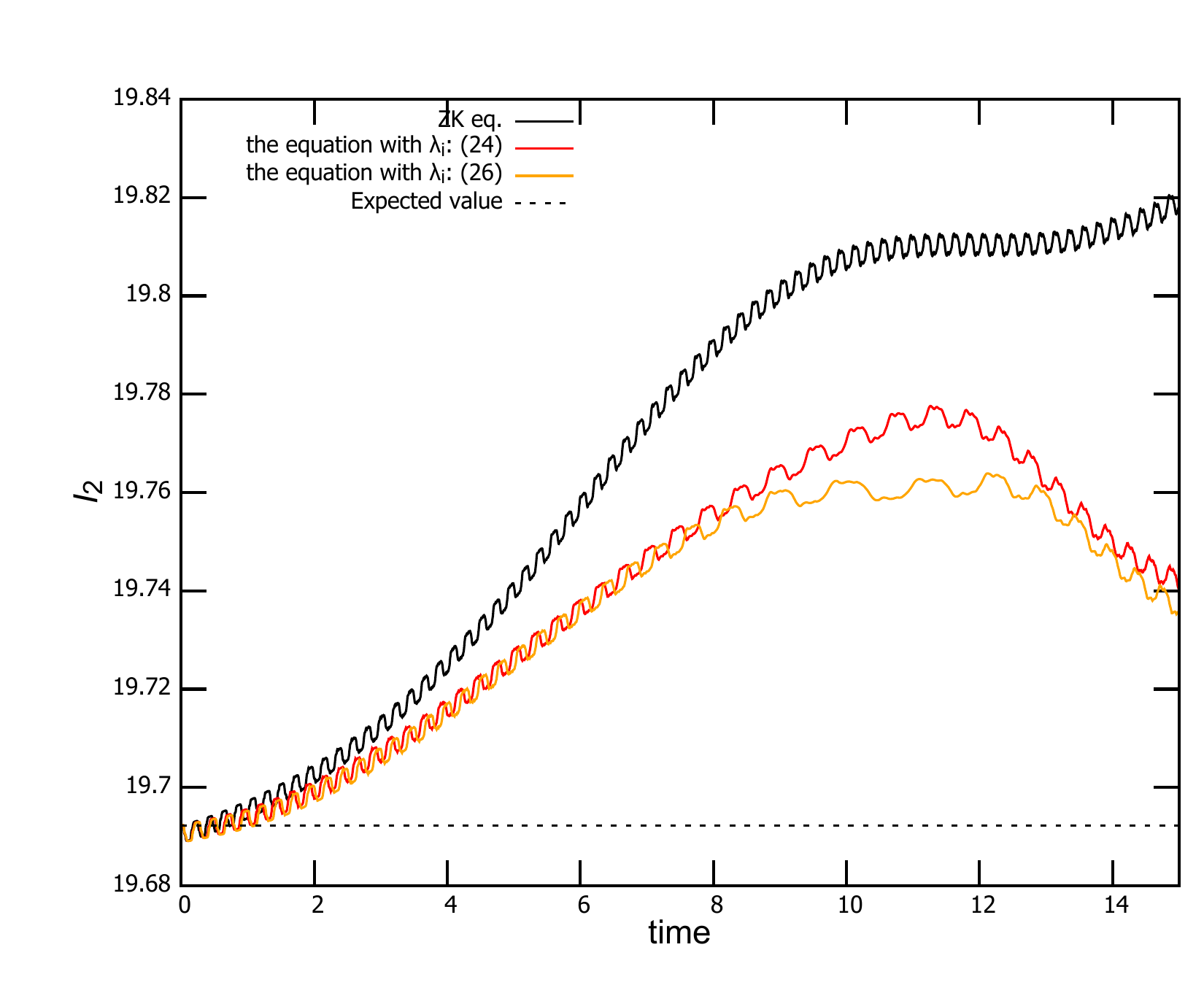}
	\caption{\label{Recons2}
	The top two figures show the coefficients $\lambda_0,\lambda_1$ by 
	the inverse analysis with the Runge-Kutta method, PINNs and cPINNs where the data are the 
	DIFC of the onset collision 	with $c=1.0$ and $0.25$ and of the temporal coefficients. 
	The inverse analysis is realized by the randomly sampling from 50000 data points.
	The solid lines show an exponential fitting to the cPINN result 
	defined by \eqref{modcoefon} and \eqref{modcoef2}.
	The bottom figure is the resulting conserved quantity $I_2$ of the DIFC of the 
	normal ZK equation and the time-dependent coefficients \eqref{modcoefon},		\eqref{modcoef2}. 
	}

\end{figure}

Using the obtained DIFC, we evaluate the 
conservation quantity $I_2$ and plot the time dependence in Fig.\ref{Reconsoff}. 
There is still a large discrepancy, but the quantity tends to revert to the 
expected exact value, unlike the result using the original equation (Fig.\ref{cons1}).
If we wish to find the exact solution of the collision with perfect conservation laws, 
we can recalculate the extra changing coefficients by computing the inverse PINNs 
once again using the data from Fig.\ref{Resolon} or Fig.\ref{Resoloff}, then 
repeat the procedure until self-consistency is attained.  
By performing one extra step, we demonstrate how the iteration analysis proceeds.
The interpolation function for the inverse PINNs result with the data from Fig.\ref{Resolon} is defined as
\begin{align}
\lambda_0=\left(1.839\pm 0.019\right)-\exp\left[\left(-4.55\pm 0.36\right)+\left(0.780\pm 0.065\right)t+\left(-0.0360\pm 0.0031\right)t^2\right]
\nonumber \\
\lambda_1=\left(0.887\pm 0.011\right)-\exp\left[\left(-6.08\pm 0.31\right)+\left(1.088\pm 0.059\right)t+\left(-0.0515\pm 0.0028\right)t^2\right]
\label{modcoef2}
\end{align}
We show the modulation of the coefficients and the conserved quantity $I_2$ 
in Fig.\ref{Recons2}. 

It appears that many iteration steps are required for complete convergence, 
which obviously necessitates a large amount of computation time.
In our typical analysis, the computations were carried out on four interconnected 
machines, each equipped with an Intel(R) UHD Graphics 730 GPU, 3.8 GB of memory, 
and an Intel Core i3-13100 CPU. 
The analysis takes roughly 75h per machine for one iteration step: 
15h for the forward analysis by cPINN and 60h for the inverse analysis. 
As a result, it takes a very long time to achieve solution with sufficient convergence, 
necessitating significant technological advancements such as developing 
algorithms with more comprehensive capabilities and using more powerful machine.

\section{Summary}

In the present paper, we have investigated the use of PINNs for the analysis of the inelastic collision process 
of solitons in the quasi-integrable Zakharov-Kuznetsov equation. 
It is well-known that the process exhibits odd behavior, i.e., 
the taller soliton gains more height 
while the shorter one tends to wane with the radiation. 
We confirmed that all the conserved quantities are broken during the impact. 
Therefore, we introduced the conservative PINN and obtained a solution that was completely distinct 
from the known solutions obtained by the Runge-Kutta method or the conventional PINN. 

With training data obtained by the Runge-Kutta method, PINNs and cPINNs, 
we examined the effectiveness of using inverse analysis to construct the equation. 
We observed that the coefficients in the resulting equations deviated from their initial values, which seems crucial in quasi-integrable systems.  
The natural interpretation of the effect is the temporal emergence of 
an effective interaction during the impact of the collision. 
We determined the inverse PINNs using the data once more to investigate the further mutation 
of the coefficients and more closely approach the precise equation for the collision. 
Apparently, this process requires numerous iterative steps that are highly 
computationally demanding before convergence is attained,  
requiring significant technological advancement to make this process practically feasible.  
Of course, the best way is to construct a huge NN for the complete process, 
but this is beyond scope of the present paper.
In future work, we will present the results of the aforementioned iterative 
method for a simpler, 1+1-dimensional case.  

\vspace{0.5cm}

\noindent {\bf Acknowledgments} 
The authors would like to thank Satoshi Horihata, Filip Blaschke, Sven Bjarke Gudnason, 
Luiz Agostinho Ferreira, Wojtek Zakrzewski
and Pawe\l~Klimas for their useful advice and comments. 
N.S. and K.S. would like to thank all the conference organizers of QTS12 and Prof. \v{C}estmir Burd\'{i}k for the hospitality and 
also kind consideration. K.S. is supported by Tokyo University of Science. 
A.N., N.S. and K.T. are supported in part by the Japan Society for the Promotion of Science (JSPS) KAKENHI Grant Number JP23K02794.
K.O. is partially supported by the JSPS through KAKENHI grant 21H05309.

\vspace{0.5cm}

\section*{References}

\bibliography{NN}

\begin{thebibliography}{75}%
\makeatletter
\providecommand \@ifxundefined [1]{%
 \@ifx{#1\undefined}
}%
\providecommand \@ifnum [1]{%
 \ifnum #1\expandafter \@firstoftwo
 \else \expandafter \@secondoftwo
 \fi
}%
\providecommand \@ifx [1]{%
 \ifx #1\expandafter \@firstoftwo
 \else \expandafter \@secondoftwo
 \fi
}%
\providecommand \natexlab [1]{#1}%
\providecommand \enquote  [1]{``#1''}%
\providecommand \bibnamefont  [1]{#1}%
\providecommand \bibfnamefont [1]{#1}%
\providecommand \citenamefont [1]{#1}%
\providecommand \href@noop [0]{\@secondoftwo}%
\providecommand \href [0]{\begingroup \@sanitize@url \@href}%
\providecommand \@href[1]{\@@startlink{#1}\@@href}%
\providecommand \@@href[1]{\endgroup#1\@@endlink}%
\providecommand \@sanitize@url [0]{\catcode `\\12\catcode `\$12\catcode
  `\&12\catcode `\#12\catcode `\^12\catcode `\_12\catcode `\%12\relax}%
\providecommand \@@startlink[1]{}%
\providecommand \@@endlink[0]{}%
\providecommand \url  [0]{\begingroup\@sanitize@url \@url }%
\providecommand \@url [1]{\endgroup\@href {#1}{\urlprefix }}%
\providecommand \urlprefix  [0]{URL }%
\providecommand \Eprint [0]{\href }%
\providecommand \doibase [0]{http://dx.doi.org/}%
\providecommand \selectlanguage [0]{\@gobble}%
\providecommand \bibinfo  [0]{\@secondoftwo}%
\providecommand \bibfield  [0]{\@secondoftwo}%
\providecommand \translation [1]{[#1]}%
\providecommand \BibitemOpen [0]{}%
\providecommand \bibitemStop [0]{}%
\providecommand \bibitemNoStop [0]{.\EOS\space}%
\providecommand \EOS [0]{\spacefactor3000\relax}%
\providecommand \BibitemShut  [1]{\csname bibitem#1\endcsname}%
\let\auto@bib@innerbib\@empty
\bibitem [{\citenamefont {Raissi}\ \emph {et~al.}(2020)\citenamefont {Raissi},
  \citenamefont {Yazdani},\ and\ \citenamefont {Karniadakis}}]{Raissi2020}%
  \BibitemOpen
  \bibfield  {author} {\bibinfo {author} {\bibfnamefont {Maziar}\ \bibnamefont
  {Raissi}}, \bibinfo {author} {\bibfnamefont {Alireza}\ \bibnamefont
  {Yazdani}}, \ and\ \bibinfo {author} {\bibfnamefont {George~Em}\ \bibnamefont
  {Karniadakis}},\ }\bibfield  {title} {\enquote {\bibinfo {title} {Hidden
  fluid mechanics: Learning velocity and pressure fields from flow
  visualizations},}\ }\href {\doibase 10.1126/science.aaw4741} {\bibfield
  {journal} {\bibinfo  {journal} {Science}\ }\textbf {\bibinfo {volume}
  {367}},\ \bibinfo {pages} {1026--1030} (\bibinfo {year} {2020})},\ \Eprint
  {http://arxiv.org/abs/https://www.science.org/doi/pdf/10.1126/science.aaw4741}
  {https://www.science.org/doi/pdf/10.1126/science.aaw4741} \BibitemShut
  {NoStop}%
\bibitem [{\citenamefont {Brunton}\ \emph {et~al.}(2020)\citenamefont
  {Brunton}, \citenamefont {Noack},\ and\ \citenamefont
  {Koumoutsakos}}]{Brunton2020}%
  \BibitemOpen
  \bibfield  {author} {\bibinfo {author} {\bibfnamefont {Steven~L.}\
  \bibnamefont {Brunton}}, \bibinfo {author} {\bibfnamefont {Bernd~R.}\
  \bibnamefont {Noack}}, \ and\ \bibinfo {author} {\bibfnamefont {Petros}\
  \bibnamefont {Koumoutsakos}},\ }\bibfield  {title} {\enquote {\bibinfo
  {title} {Machine learning for fluid mechanics},}\ }\href {\doibase
  https://doi.org/10.1146/annurev-fluid-010719-060214} {\bibfield  {journal}
  {\bibinfo  {journal} {Annual Review of Fluid Mechanics}\ }\textbf {\bibinfo
  {volume} {52}},\ \bibinfo {pages} {477--508} (\bibinfo {year}
  {2020})}\BibitemShut {NoStop}%
\bibitem [{\citenamefont {Kadeethum}\ \emph {et~al.}(2020)\citenamefont
  {Kadeethum}, \citenamefont {Jorgensen},\ and\ \citenamefont
  {Nick}}]{Kadeethum2020}%
  \BibitemOpen
  \bibfield  {author} {\bibinfo {author} {\bibfnamefont {Teeratorn}\
  \bibnamefont {Kadeethum}}, \bibinfo {author} {\bibfnamefont {Thomas~M.}\
  \bibnamefont {Jorgensen}}, \ and\ \bibinfo {author} {\bibfnamefont
  {Hamidreza~M.}\ \bibnamefont {Nick}},\ }\bibfield  {title} {\enquote
  {\bibinfo {title} {Physics-informed neural networks for solving nonlinear
  diffusivity and biot’s equations},}\ }\href {\doibase
  10.1371/journal.pone.0232683} {\bibfield  {journal} {\bibinfo  {journal}
  {PLOS ONE}\ }\textbf {\bibinfo {volume} {15}},\ \bibinfo {pages} {1--28}
  (\bibinfo {year} {2020})}\BibitemShut {NoStop}%
\bibitem [{\citenamefont {Cai}(2021)}]{cai2021}%
  \BibitemOpen
  \bibfield  {author} {\bibinfo {author} {\bibfnamefont {Mao Zhiping Wang
  Zhicheng Yin Minglang Karniadakis George~Em}\ \bibnamefont {Cai},
  \bibfnamefont {Shengze}},\ }\bibfield  {title} {\enquote {\bibinfo {title}
  {Physics-informed neural networks (pinns) for fluid mechanics: a review},}\
  }\href {\doibase 10.1007/s10409-021-01148-1} {\bibfield  {journal} {\bibinfo
  {journal} {Acta Mechanica Sinica}\ } (\bibinfo {year} {2021}),\
  10.1007/s10409-021-01148-1}\BibitemShut {NoStop}%
\bibitem [{\citenamefont {et~al.}(2021)}]{Kashinath2021}%
  \BibitemOpen
  \bibfield  {author} {\bibinfo {author} {\bibfnamefont {Kashinath~K}\
  \bibnamefont {et~al.}},\ }\bibfield  {title} {\enquote {\bibinfo {title}
  {Physics-informed machine learning: case studies for weather and climate
  modeling},}\ }\href {\doibase 10.1098/rsta.2020.0093} {\bibfield  {journal}
  {\bibinfo  {journal} {Phil.Trans.R.Soc.A}\ }\textbf {\bibinfo {volume}
  {379}},\ \bibinfo {pages} {20200093} (\bibinfo {year} {2021})}\BibitemShut
  {NoStop}%
\bibitem [{\citenamefont {Jin}(2021)}]{Jin2021}%
  \BibitemOpen
  \bibfield  {author} {\bibinfo {author} {\bibfnamefont {Cai S. Li H.
  Karniadakis~G.E}\ \bibnamefont {Jin}, \bibfnamefont {X.}},\ }\bibfield
  {title} {\enquote {\bibinfo {title} {{NSFnets (NavierStokes flow nets):
  Physics-informed neural networks for the incompressible Navier-Stokes
  equations}},}\ }\href@noop {} {\bibfield  {journal} {\bibinfo  {journal}
  {Journal of Computational Physics}\ }\textbf {\bibinfo {volume} {426}},\
  \bibinfo {pages} {109951} (\bibinfo {year} {2021})}\BibitemShut {NoStop}%
\bibitem [{\citenamefont {Linghu}\ \emph {et~al.}(2025)\citenamefont {Linghu},
  \citenamefont {Gao}, \citenamefont {Dong},\ and\ \citenamefont
  {Nie}}]{LINGHU2025116223}%
  \BibitemOpen
  \bibfield  {author} {\bibinfo {author} {\bibfnamefont {Jiale}\ \bibnamefont
  {Linghu}}, \bibinfo {author} {\bibfnamefont {Weifeng}\ \bibnamefont {Gao}},
  \bibinfo {author} {\bibfnamefont {Hao}\ \bibnamefont {Dong}}, \ and\ \bibinfo
  {author} {\bibfnamefont {Yufeng}\ \bibnamefont {Nie}},\ }\bibfield  {title}
  {\enquote {\bibinfo {title} {Higher-order multi-scale physics-informed neural
  network (homs-pinn) method and its convergence analysis for solving elastic
  problems of authentic composite materials},}\ }\href {\doibase
  https://doi.org/10.1016/j.cam.2024.116223} {\bibfield  {journal} {\bibinfo
  {journal} {Journal of Computational and Applied Mathematics}\ }\textbf
  {\bibinfo {volume} {456}},\ \bibinfo {pages} {116223} (\bibinfo {year}
  {2025})}\BibitemShut {NoStop}%
\bibitem [{\citenamefont {Alhubail}\ \emph {et~al.}(2024)\citenamefont
  {Alhubail}, \citenamefont {Fahs}, \citenamefont {Lehmann},\ and\
  \citenamefont {Hoteit}}]{ALHUBAIL2024104797}%
  \BibitemOpen
  \bibfield  {author} {\bibinfo {author} {\bibfnamefont {Ali}\ \bibnamefont
  {Alhubail}}, \bibinfo {author} {\bibfnamefont {Marwan}\ \bibnamefont {Fahs}},
  \bibinfo {author} {\bibfnamefont {François}\ \bibnamefont {Lehmann}}, \ and\
  \bibinfo {author} {\bibfnamefont {Hussein}\ \bibnamefont {Hoteit}},\
  }\bibfield  {title} {\enquote {\bibinfo {title} {Modeling fluid flow in
  heterogeneous porous media with physics-informed neural networks: Weighting
  strategies for the mixed pressure head-velocity formulation},}\ }\href
  {\doibase https://doi.org/10.1016/j.advwatres.2024.104797} {\bibfield
  {journal} {\bibinfo  {journal} {Advances in Water Resources}\ }\textbf
  {\bibinfo {volume} {193}},\ \bibinfo {pages} {104797} (\bibinfo {year}
  {2024})}\BibitemShut {NoStop}%
\bibitem [{\citenamefont {Eshkofti}\ and\ \citenamefont
  {Hosseini}(2024)}]{ESHKOFTI2024118485}%
  \BibitemOpen
  \bibfield  {author} {\bibinfo {author} {\bibfnamefont {Katayoun}\
  \bibnamefont {Eshkofti}}\ and\ \bibinfo {author} {\bibfnamefont
  {Seyed~Mahmoud}\ \bibnamefont {Hosseini}},\ }\bibfield  {title} {\enquote
  {\bibinfo {title} {The modified physics-informed neural network (pinn) method
  for the thermoelastic wave propagation analysis based on the
  moore-gibson-thompson theory in porous materials},}\ }\href {\doibase
  https://doi.org/10.1016/j.compstruct.2024.118485} {\bibfield  {journal}
  {\bibinfo  {journal} {Composite Structures}\ }\textbf {\bibinfo {volume}
  {348}},\ \bibinfo {pages} {118485} (\bibinfo {year} {2024})}\BibitemShut
  {NoStop}%
\bibitem [{\citenamefont {Sun}\ \emph {et~al.}(2024)\citenamefont {Sun},
  \citenamefont {Jeong}, \citenamefont {Zhao}, \citenamefont {Gou},
  \citenamefont {Sauret}, \citenamefont {Li},\ and\ \citenamefont
  {Gu}}]{SUN2024106421}%
  \BibitemOpen
  \bibfield  {author} {\bibinfo {author} {\bibfnamefont {Runze}\ \bibnamefont
  {Sun}}, \bibinfo {author} {\bibfnamefont {Hyogu}\ \bibnamefont {Jeong}},
  \bibinfo {author} {\bibfnamefont {Jiachen}\ \bibnamefont {Zhao}}, \bibinfo
  {author} {\bibfnamefont {Yixing}\ \bibnamefont {Gou}}, \bibinfo {author}
  {\bibfnamefont {Emilie}\ \bibnamefont {Sauret}}, \bibinfo {author}
  {\bibfnamefont {Zirui}\ \bibnamefont {Li}}, \ and\ \bibinfo {author}
  {\bibfnamefont {Yuantong}\ \bibnamefont {Gu}},\ }\bibfield  {title} {\enquote
  {\bibinfo {title} {A physics-informed neural network framework for
  multi-physics coupling microfluidic problems},}\ }\href {\doibase
  https://doi.org/10.1016/j.compfluid.2024.106421} {\bibfield  {journal}
  {\bibinfo  {journal} {Computers and Fluids}\ }\textbf {\bibinfo {volume}
  {284}},\ \bibinfo {pages} {106421} (\bibinfo {year} {2024})}\BibitemShut
  {NoStop}%
\bibitem [{\citenamefont {Rai}\ and\ \citenamefont {Sahu}(2020)}]{9064519}%
  \BibitemOpen
  \bibfield  {author} {\bibinfo {author} {\bibfnamefont {Rahul}\ \bibnamefont
  {Rai}}\ and\ \bibinfo {author} {\bibfnamefont {Chandan~K.}\ \bibnamefont
  {Sahu}},\ }\bibfield  {title} {\enquote {\bibinfo {title} {Driven by data or
  derived through physics? a review of hybrid physics guided machine learning
  techniques with cyber-physical system (cps) focus},}\ }\href {\doibase
  10.1109/ACCESS.2020.2987324} {\bibfield  {journal} {\bibinfo  {journal} {IEEE
  Access}\ }\textbf {\bibinfo {volume} {8}},\ \bibinfo {pages} {71050--71073}
  (\bibinfo {year} {2020})}\BibitemShut {NoStop}%
\bibitem [{\citenamefont {Wu}\ \emph {et~al.}(2017)\citenamefont {Wu},
  \citenamefont {Nüske}, \citenamefont {Paul}, \citenamefont {Klus},
  \citenamefont {Koltai},\ and\ \citenamefont {Noé}}]{Wu2017}%
  \BibitemOpen
  \bibfield  {author} {\bibinfo {author} {\bibfnamefont {Hao}\ \bibnamefont
  {Wu}}, \bibinfo {author} {\bibfnamefont {Feliks}\ \bibnamefont {Nüske}},
  \bibinfo {author} {\bibfnamefont {Fabian}\ \bibnamefont {Paul}}, \bibinfo
  {author} {\bibfnamefont {Stefan}\ \bibnamefont {Klus}}, \bibinfo {author}
  {\bibfnamefont {Péter}\ \bibnamefont {Koltai}}, \ and\ \bibinfo {author}
  {\bibfnamefont {Frank}\ \bibnamefont {Noé}},\ }\bibfield  {title} {\enquote
  {\bibinfo {title} {{Variational Koopman models: Slow collective variables and
  molecular kinetics from short off-equilibrium simulations}},}\ }\href
  {\doibase 10.1063/1.4979344} {\bibfield  {journal} {\bibinfo  {journal} {The
  Journal of Chemical Physics}\ }\textbf {\bibinfo {volume} {146}},\ \bibinfo
  {pages} {154104} (\bibinfo {year} {2017})},\ \Eprint
  {http://arxiv.org/abs/https://pubs.aip.org/aip/jcp/article-pdf/doi/10.1063/1.4979344/14899047/154104\_1\_online.pdf}
  {https://pubs.aip.org/aip/jcp/article-pdf/doi/10.1063/1.4979344/14899047/154104\_1\_online.pdf}
  \BibitemShut {NoStop}%
\bibitem [{\citenamefont {Kissas}\ \emph {et~al.}(2020)\citenamefont {Kissas},
  \citenamefont {Yang}, \citenamefont {Hwuang}, \citenamefont {Witschey},
  \citenamefont {Detre},\ and\ \citenamefont {Perdikaris}}]{KISSAS2020112623}%
  \BibitemOpen
  \bibfield  {author} {\bibinfo {author} {\bibfnamefont {Georgios}\
  \bibnamefont {Kissas}}, \bibinfo {author} {\bibfnamefont {Yibo}\ \bibnamefont
  {Yang}}, \bibinfo {author} {\bibfnamefont {Eileen}\ \bibnamefont {Hwuang}},
  \bibinfo {author} {\bibfnamefont {Walter~R.}\ \bibnamefont {Witschey}},
  \bibinfo {author} {\bibfnamefont {John~A.}\ \bibnamefont {Detre}}, \ and\
  \bibinfo {author} {\bibfnamefont {Paris}\ \bibnamefont {Perdikaris}},\
  }\bibfield  {title} {\enquote {\bibinfo {title} {Machine learning in
  cardiovascular flows modeling: Predicting arterial blood pressure from
  non-invasive 4d flow mri data using physics-informed neural networks},}\
  }\href {\doibase https://doi.org/10.1016/j.cma.2019.112623} {\bibfield
  {journal} {\bibinfo  {journal} {Computer Methods in Applied Mechanics and
  Engineering}\ }\textbf {\bibinfo {volume} {358}},\ \bibinfo {pages} {112623}
  (\bibinfo {year} {2020})}\BibitemShut {NoStop}%
\bibitem [{\citenamefont {Ruiz~Herrera}\ \emph {et~al.}(2022)\citenamefont
  {Ruiz~Herrera}, \citenamefont {Grandits}, \citenamefont {Plank},
  \citenamefont {Perdikaris},\ and\ \citenamefont
  {Sahli~Costabal}}]{Ruizherrera2021}%
  \BibitemOpen
  \bibfield  {author} {\bibinfo {author} {\bibfnamefont {Carlos}\ \bibnamefont
  {Ruiz~Herrera}}, \bibinfo {author} {\bibfnamefont {Thomas}\ \bibnamefont
  {Grandits}}, \bibinfo {author} {\bibfnamefont {Gernot}\ \bibnamefont
  {Plank}}, \bibinfo {author} {\bibfnamefont {Paris}\ \bibnamefont
  {Perdikaris}}, \ and\ \bibinfo {author} {\bibfnamefont {Simone}\ \bibnamefont
  {Sahli~Costabal}, \bibfnamefont {Francisco~andPezzuto}},\ }\bibfield  {title}
  {\enquote {\bibinfo {title} {Physics-informed neural networks to learn
  cardiac fiber orientation from multiple electroanatomical maps},}\ }\href
  {\doibase 10.1007/s00366-022-01709-3} {\bibfield  {journal} {\bibinfo
  {journal} {Engineering with Computers}\ }\textbf {\bibinfo {volume} {38}},\
  \bibinfo {pages} {3957--3973} (\bibinfo {year} {2022})}\BibitemShut {NoStop}%
\bibitem [{\citenamefont {Sel}(2023)}]{SEL2023}%
  \BibitemOpen
  \bibfield  {author} {\bibinfo {author} {\bibfnamefont {Mohammadi Amirmohammad
  Pettigrew Roderic I. Jafari~Roozbeh}\ \bibnamefont {Sel}, \bibfnamefont
  {Kaan}},\ }\bibfield  {title} {\enquote {\bibinfo {title} {Physics-informed
  neural networks for modeling physiological time series for cuffless blood
  pressure estimation},}\ }\href {\doibase 10.1038/s41746-023-00853-4}
  {\bibfield  {journal} {\bibinfo  {journal} {npj Digital Medicine}\ }\textbf
  {\bibinfo {volume} {6}},\ \bibinfo {pages} {110} (\bibinfo {year}
  {2023})}\BibitemShut {NoStop}%
\bibitem [{\citenamefont {Raissi}\ \emph
  {et~al.}(2017{\natexlab{a}})\citenamefont {Raissi}, \citenamefont
  {Perdikaris},\ and\ \citenamefont {Karniadakis}}]{RaissarxivI}%
  \BibitemOpen
  \bibfield  {author} {\bibinfo {author} {\bibfnamefont {Maziar}\ \bibnamefont
  {Raissi}}, \bibinfo {author} {\bibfnamefont {Paris}\ \bibnamefont
  {Perdikaris}}, \ and\ \bibinfo {author} {\bibfnamefont {George}\ \bibnamefont
  {Karniadakis}},\ }\bibfield  {title} {\enquote {\bibinfo {title} {Physics
  informed deep learning (part i): Data-driven solutions of nonlinear partial
  differential equations},}\ }\href {\doibase 10.48550/arXiv.1711.10561} {\
  (\bibinfo {year} {2017}{\natexlab{a}}),\
  10.48550/arXiv.1711.10561}\BibitemShut {NoStop}%
\bibitem [{\citenamefont {Raissi}\ \emph
  {et~al.}(2017{\natexlab{b}})\citenamefont {Raissi}, \citenamefont
  {Perdikaris},\ and\ \citenamefont {Karniadakis}}]{RaissarxivII}%
  \BibitemOpen
  \bibfield  {author} {\bibinfo {author} {\bibfnamefont {Maziar}\ \bibnamefont
  {Raissi}}, \bibinfo {author} {\bibfnamefont {Paris}\ \bibnamefont
  {Perdikaris}}, \ and\ \bibinfo {author} {\bibfnamefont {George}\ \bibnamefont
  {Karniadakis}},\ }\bibfield  {title} {\enquote {\bibinfo {title} {Physics
  informed deep learning (part ii): Data-driven discovery of nonlinear partial
  differential equations},}\ }\href {\doibase 10.48550/arXiv.1711.10566} {\
  (\bibinfo {year} {2017}{\natexlab{b}}),\
  10.48550/arXiv.1711.10566}\BibitemShut {NoStop}%
\bibitem [{\citenamefont {Raissi}\ \emph {et~al.}(2019)\citenamefont {Raissi},
  \citenamefont {Perdikaris},\ and\ \citenamefont
  {Karniadakis}}]{RAISSI2019686}%
  \BibitemOpen
  \bibfield  {author} {\bibinfo {author} {\bibfnamefont {M.}~\bibnamefont
  {Raissi}}, \bibinfo {author} {\bibfnamefont {P.}~\bibnamefont {Perdikaris}},
  \ and\ \bibinfo {author} {\bibfnamefont {G.E.}\ \bibnamefont {Karniadakis}},\
  }\bibfield  {title} {\enquote {\bibinfo {title} {Physics-informed neural
  networks: A deep learning framework for solving forward and inverse problems
  involving nonlinear partial differential equations},}\ }\href {\doibase
  https://doi.org/10.1016/j.jcp.2018.10.045} {\bibfield  {journal} {\bibinfo
  {journal} {Journal of Computational Physics}\ }\textbf {\bibinfo {volume}
  {378}},\ \bibinfo {pages} {686--707} (\bibinfo {year} {2019})}\BibitemShut
  {NoStop}%
\bibitem [{\citenamefont {Fang}\ and\ \citenamefont {Zhan}(2020)}]{Fang2020}%
  \BibitemOpen
  \bibfield  {author} {\bibinfo {author} {\bibfnamefont {Zhiwei}\ \bibnamefont
  {Fang}}\ and\ \bibinfo {author} {\bibfnamefont {Justin}\ \bibnamefont
  {Zhan}},\ }\bibfield  {title} {\enquote {\bibinfo {title} {A physics-informed
  neural network framework for pdes on 3d surfaces: Time independent
  problems},}\ }\href {\doibase 10.1109/ACCESS.2019.2963390} {\bibfield
  {journal} {\bibinfo  {journal} {IEEE Access}\ }\textbf {\bibinfo {volume}
  {8}},\ \bibinfo {pages} {26328--26335} (\bibinfo {year} {2020})}\BibitemShut
  {NoStop}%
\bibitem [{\citenamefont {{Sahli Costabal}}\ \emph {et~al.}(2024)\citenamefont
  {{Sahli Costabal}}, \citenamefont {Pezzuto},\ and\ \citenamefont
  {Perdikaris}}]{SAHLICOSTABAL2024107324}%
  \BibitemOpen
  \bibfield  {author} {\bibinfo {author} {\bibfnamefont {Francisco}\
  \bibnamefont {{Sahli Costabal}}}, \bibinfo {author} {\bibfnamefont {Simone}\
  \bibnamefont {Pezzuto}}, \ and\ \bibinfo {author} {\bibfnamefont {Paris}\
  \bibnamefont {Perdikaris}},\ }\bibfield  {title} {\enquote {\bibinfo {title}
  {Delta-pinns: Physics-informed neural networks on complex geometries},}\
  }\href {\doibase https://doi.org/10.1016/j.engappai.2023.107324} {\bibfield
  {journal} {\bibinfo  {journal} {Engineering Applications of Artificial
  Intelligence}\ }\textbf {\bibinfo {volume} {127}},\ \bibinfo {pages} {107324}
  (\bibinfo {year} {2024})}\BibitemShut {NoStop}%
\bibitem [{\citenamefont {Wu}\ \emph {et~al.}(2025)\citenamefont {Wu},
  \citenamefont {Duan}, \citenamefont {Sun}, \citenamefont {Yu}, \citenamefont
  {Liu},\ and\ \citenamefont {Peng}}]{WU2025106750}%
  \BibitemOpen
  \bibfield  {author} {\bibinfo {author} {\bibfnamefont {Wenyuan}\ \bibnamefont
  {Wu}}, \bibinfo {author} {\bibfnamefont {Siyuan}\ \bibnamefont {Duan}},
  \bibinfo {author} {\bibfnamefont {Yuan}\ \bibnamefont {Sun}}, \bibinfo
  {author} {\bibfnamefont {Yang}\ \bibnamefont {Yu}}, \bibinfo {author}
  {\bibfnamefont {Dong}\ \bibnamefont {Liu}}, \ and\ \bibinfo {author}
  {\bibfnamefont {Dezhong}\ \bibnamefont {Peng}},\ }\bibfield  {title}
  {\enquote {\bibinfo {title} {Deep fuzzy physics-informed neural networks for
  forward and inverse pde problems},}\ }\href {\doibase
  https://doi.org/10.1016/j.neunet.2024.106750} {\bibfield  {journal} {\bibinfo
   {journal} {Neural Networks}\ }\textbf {\bibinfo {volume} {181}},\ \bibinfo
  {pages} {106750} (\bibinfo {year} {2025})}\BibitemShut {NoStop}%
\bibitem [{\citenamefont {Jagtap}\ \emph {et~al.}(2020)\citenamefont {Jagtap},
  \citenamefont {Kharazmi},\ and\ \citenamefont
  {Karniadakis}}]{JAGTAP2020113028}%
  \BibitemOpen
  \bibfield  {author} {\bibinfo {author} {\bibfnamefont {Ameya~D.}\
  \bibnamefont {Jagtap}}, \bibinfo {author} {\bibfnamefont {Ehsan}\
  \bibnamefont {Kharazmi}}, \ and\ \bibinfo {author} {\bibfnamefont
  {George~Em}\ \bibnamefont {Karniadakis}},\ }\bibfield  {title} {\enquote
  {\bibinfo {title} {Conservative physics-informed neural networks on discrete
  domains for conservation laws: Applications to forward and inverse
  problems},}\ }\href {\doibase https://doi.org/10.1016/j.cma.2020.113028}
  {\bibfield  {journal} {\bibinfo  {journal} {Computer Methods in Applied
  Mechanics and Engineering}\ }\textbf {\bibinfo {volume} {365}},\ \bibinfo
  {pages} {113028} (\bibinfo {year} {2020})}\BibitemShut {NoStop}%
\bibitem [{\citenamefont {Mishra}\ and\ \citenamefont
  {Molinaro}(2021)}]{MISHRA2021107705}%
  \BibitemOpen
  \bibfield  {author} {\bibinfo {author} {\bibfnamefont {Siddhartha}\
  \bibnamefont {Mishra}}\ and\ \bibinfo {author} {\bibfnamefont {Roberto}\
  \bibnamefont {Molinaro}},\ }\bibfield  {title} {\enquote {\bibinfo {title}
  {Physics informed neural networks for simulating radiative transfer},}\
  }\href {\doibase https://doi.org/10.1016/j.jqsrt.2021.107705} {\bibfield
  {journal} {\bibinfo  {journal} {Journal of Quantitative Spectroscopy and
  Radiative Transfer}\ }\textbf {\bibinfo {volume} {270}},\ \bibinfo {pages}
  {107705} (\bibinfo {year} {2021})}\BibitemShut {NoStop}%
\bibitem [{\citenamefont {Chen}\ \emph {et~al.}(2022)\citenamefont {Chen},
  \citenamefont {Jeffery}, \citenamefont {Zhong}, \citenamefont {McClenny},
  \citenamefont {Braga-Neto},\ and\ \citenamefont {Wang}}]{Chen2022UsingPI}%
  \BibitemOpen
  \bibfield  {author} {\bibinfo {author} {\bibfnamefont {Xingzhuo}\
  \bibnamefont {Chen}}, \bibinfo {author} {\bibfnamefont {David~J.}\
  \bibnamefont {Jeffery}}, \bibinfo {author} {\bibfnamefont {Ming}\
  \bibnamefont {Zhong}}, \bibinfo {author} {\bibfnamefont {Levi~D.}\
  \bibnamefont {McClenny}}, \bibinfo {author} {\bibfnamefont {Ulisses~M.}\
  \bibnamefont {Braga-Neto}}, \ and\ \bibinfo {author} {\bibfnamefont {Lifan}\
  \bibnamefont {Wang}},\ }\bibfield  {title} {\enquote {\bibinfo {title} {Using
  physics informed neural networks for supernova radiative transfer
  simulation},}\ }\href {https://api.semanticscholar.org/CorpusID:253446818} {\
   (\bibinfo {year} {2022})}\BibitemShut {NoStop}%
\bibitem [{\citenamefont {Yang}\ \emph {et~al.}(2023)\citenamefont {Yang},
  \citenamefont {Gong}, \citenamefont {Zhang}, \citenamefont {Yang},
  \citenamefont {Chen}, \citenamefont {He},\ and\ \citenamefont
  {Li}}]{YANG2023109656}%
  \BibitemOpen
  \bibfield  {author} {\bibinfo {author} {\bibfnamefont {Yu}~\bibnamefont
  {Yang}}, \bibinfo {author} {\bibfnamefont {Helin}\ \bibnamefont {Gong}},
  \bibinfo {author} {\bibfnamefont {Shiquan}\ \bibnamefont {Zhang}}, \bibinfo
  {author} {\bibfnamefont {Qihong}\ \bibnamefont {Yang}}, \bibinfo {author}
  {\bibfnamefont {Zhang}\ \bibnamefont {Chen}}, \bibinfo {author}
  {\bibfnamefont {Qiaolin}\ \bibnamefont {He}}, \ and\ \bibinfo {author}
  {\bibfnamefont {Qing}\ \bibnamefont {Li}},\ }\bibfield  {title} {\enquote
  {\bibinfo {title} {A data-enabled physics-informed neural network with
  comprehensive numerical study on solving neutron diffusion eigenvalue
  problems},}\ }\href {\doibase https://doi.org/10.1016/j.anucene.2022.109656}
  {\bibfield  {journal} {\bibinfo  {journal} {Annals of Nuclear Energy}\
  }\textbf {\bibinfo {volume} {183}},\ \bibinfo {pages} {109656} (\bibinfo
  {year} {2023})}\BibitemShut {NoStop}%
\bibitem [{\citenamefont {Sedykh}\ \emph {et~al.}(2024)\citenamefont {Sedykh},
  \citenamefont {Podapaka}, \citenamefont {Sagingalieva}, \citenamefont
  {Pinto}, \citenamefont {Pflitsch},\ and\ \citenamefont
  {Melnikov}}]{Sedykh_2024}%
  \BibitemOpen
  \bibfield  {author} {\bibinfo {author} {\bibfnamefont {Alexandr}\
  \bibnamefont {Sedykh}}, \bibinfo {author} {\bibfnamefont {Maninadh}\
  \bibnamefont {Podapaka}}, \bibinfo {author} {\bibfnamefont {Asel}\
  \bibnamefont {Sagingalieva}}, \bibinfo {author} {\bibfnamefont {Karan}\
  \bibnamefont {Pinto}}, \bibinfo {author} {\bibfnamefont {Markus}\
  \bibnamefont {Pflitsch}}, \ and\ \bibinfo {author} {\bibfnamefont {Alexey}\
  \bibnamefont {Melnikov}},\ }\bibfield  {title} {\enquote {\bibinfo {title}
  {Hybrid quantum physics-informed neural networks for simulating computational
  fluid dynamics in complex shapes},}\ }\href {\doibase
  10.1088/2632-2153/ad43b2} {\bibfield  {journal} {\bibinfo  {journal} {Machine
  Learning: Science and Technology}\ }\textbf {\bibinfo {volume} {5}},\
  \bibinfo {pages} {025045} (\bibinfo {year} {2024})}\BibitemShut {NoStop}%
\bibitem [{\citenamefont {Lin}\ and\ \citenamefont
  {Chen}(2022)}]{LIN2022111053}%
  \BibitemOpen
  \bibfield  {author} {\bibinfo {author} {\bibfnamefont {Shuning}\ \bibnamefont
  {Lin}}\ and\ \bibinfo {author} {\bibfnamefont {Yong}\ \bibnamefont {Chen}},\
  }\bibfield  {title} {\enquote {\bibinfo {title} {A two-stage physics-informed
  neural network method based on conserved quantities and applications in
  localized wave solutions},}\ }\href {\doibase
  https://doi.org/10.1016/j.jcp.2022.111053} {\bibfield  {journal} {\bibinfo
  {journal} {Journal of Computational Physics}\ }\textbf {\bibinfo {volume}
  {457}},\ \bibinfo {pages} {111053} (\bibinfo {year} {2022})}\BibitemShut
  {NoStop}%
\bibitem [{\citenamefont {Fang}\ \emph {et~al.}(2022)\citenamefont {Fang},
  \citenamefont {Wu}, \citenamefont {Kudryashov}, \citenamefont {Wang},\ and\
  \citenamefont {Dai}}]{FANG2022112118}%
  \BibitemOpen
  \bibfield  {author} {\bibinfo {author} {\bibfnamefont {Yin}\ \bibnamefont
  {Fang}}, \bibinfo {author} {\bibfnamefont {Gang-Zhou}\ \bibnamefont {Wu}},
  \bibinfo {author} {\bibfnamefont {Nikolay~A.}\ \bibnamefont {Kudryashov}},
  \bibinfo {author} {\bibfnamefont {Yue-Yue}\ \bibnamefont {Wang}}, \ and\
  \bibinfo {author} {\bibfnamefont {Chao-Qing}\ \bibnamefont {Dai}},\
  }\bibfield  {title} {\enquote {\bibinfo {title} {Data-driven soliton
  solutions and model parameters of nonlinear wave models via the
  conservation-law constrained neural network method},}\ }\href {\doibase
  https://doi.org/10.1016/j.chaos.2022.112118} {\bibfield  {journal} {\bibinfo
  {journal} {Chaos, Solitons and Fractals}\ }\textbf {\bibinfo {volume}
  {158}},\ \bibinfo {pages} {112118} (\bibinfo {year} {2022})}\BibitemShut
  {NoStop}%
\bibitem [{\citenamefont {Wu}\ \emph {et~al.}(2022)\citenamefont {Wu},
  \citenamefont {Fang}, \citenamefont {Kudryashov}, \citenamefont {Wang},\ and\
  \citenamefont {Dai}}]{WU2022112143}%
  \BibitemOpen
  \bibfield  {author} {\bibinfo {author} {\bibfnamefont {Gang-Zhou}\
  \bibnamefont {Wu}}, \bibinfo {author} {\bibfnamefont {Yin}\ \bibnamefont
  {Fang}}, \bibinfo {author} {\bibfnamefont {Nikolay~A.}\ \bibnamefont
  {Kudryashov}}, \bibinfo {author} {\bibfnamefont {Yue-Yue}\ \bibnamefont
  {Wang}}, \ and\ \bibinfo {author} {\bibfnamefont {Chao-Qing}\ \bibnamefont
  {Dai}},\ }\bibfield  {title} {\enquote {\bibinfo {title} {Prediction of
  optical solitons using an improved physics-informed neural network method
  with the conservation law constraint},}\ }\href {\doibase
  https://doi.org/10.1016/j.chaos.2022.112143} {\bibfield  {journal} {\bibinfo
  {journal} {Chaos, Solitons and Fractals}\ }\textbf {\bibinfo {volume}
  {159}},\ \bibinfo {pages} {112143} (\bibinfo {year} {2022})}\BibitemShut
  {NoStop}%
\bibitem [{\citenamefont {Cardoso-Bihlo}\ and\ \citenamefont
  {Bihlo}(2025)}]{cardosobihlo2025}%
  \BibitemOpen
  \bibfield  {author} {\bibinfo {author} {\bibfnamefont {Elsa}\ \bibnamefont
  {Cardoso-Bihlo}}\ and\ \bibinfo {author} {\bibfnamefont {Alex}\ \bibnamefont
  {Bihlo}},\ }\bibfield  {title} {\enquote {\bibinfo {title} {Exactly
  conservative physics-informed neural networks and deep operator networks for
  dynamical systems},}\ }\href {\doibase
  https://doi.org/10.1016/j.neunet.2024.106826} {\bibfield  {journal} {\bibinfo
   {journal} {Neural Networks}\ }\textbf {\bibinfo {volume} {181}},\ \bibinfo
  {pages} {106826} (\bibinfo {year} {2025})}\BibitemShut {NoStop}%
\bibitem [{\citenamefont {Nakamula}\ \emph {et~al.}(2024)\citenamefont
  {Nakamula}, \citenamefont {Obuse}, \citenamefont {Sawado}, \citenamefont
  {Shimasaki}, \citenamefont {Shimazaki}, \citenamefont {Suzuki},\ and\
  \citenamefont {Toda}}]{Nakamula:2024cmx}%
  \BibitemOpen
  \bibfield  {author} {\bibinfo {author} {\bibfnamefont {A.}~\bibnamefont
  {Nakamula}}, \bibinfo {author} {\bibfnamefont {N.}~\bibnamefont {Obuse}},
  \bibinfo {author} {\bibfnamefont {N.}~\bibnamefont {Sawado}}, \bibinfo
  {author} {\bibfnamefont {K.}~\bibnamefont {Shimasaki}}, \bibinfo {author}
  {\bibfnamefont {Y.}~\bibnamefont {Shimazaki}}, \bibinfo {author}
  {\bibfnamefont {Y.}~\bibnamefont {Suzuki}}, \ and\ \bibinfo {author}
  {\bibfnamefont {K.}~\bibnamefont {Toda}},\ }\bibfield  {title} {\enquote
  {\bibinfo {title} {{Discovery of Quasi-Integrable Equations from
  traveling-wave data using the Physics-Informed Neural Networks}},}\
  }\href@noop {} {\  (\bibinfo {year} {2024})},\ \Eprint
  {http://arxiv.org/abs/2410.19014} {arXiv:2410.19014 [physics.flu-dyn]}
  \BibitemShut {NoStop}%
\bibitem [{\citenamefont {Li}\ and\ \citenamefont {Chen}(2020)}]{Li_2020}%
  \BibitemOpen
  \bibfield  {author} {\bibinfo {author} {\bibfnamefont {Jun}\ \bibnamefont
  {Li}}\ and\ \bibinfo {author} {\bibfnamefont {Yong}\ \bibnamefont {Chen}},\
  }\bibfield  {title} {\enquote {\bibinfo {title} {A deep learning method for
  solving third-order nonlinear evolution equations},}\ }\href {\doibase
  10.1088/1572-9494/abb7c8} {\bibfield  {journal} {\bibinfo  {journal}
  {Communications in Theoretical Physics}\ }\textbf {\bibinfo {volume} {72}},\
  \bibinfo {pages} {115003} (\bibinfo {year} {2020})}\BibitemShut {NoStop}%
\bibitem [{\citenamefont {J.C.Pu}\ and\ \citenamefont
  {Y.Chen}(2024)}]{Junkai2024}%
  \BibitemOpen
  \bibfield  {author} {\bibinfo {author} {\bibnamefont {J.C.Pu}}\ and\ \bibinfo
  {author} {\bibnamefont {Y.Chen}},\ }\bibfield  {title} {\enquote {\bibinfo
  {title} {{Lax pairs informed neural networks solving integrable systems}},}\
  }\href@noop {} {\bibfield  {journal} {\bibinfo  {journal} {J. Comput. Phys.}\
  }\textbf {\bibinfo {volume} {510}},\ \bibinfo {pages} {113090} (\bibinfo
  {year} {2024})}\BibitemShut {NoStop}%
\bibitem [{\citenamefont {Huijuan}(2024)}]{Zhou2024}%
  \BibitemOpen
  \bibfield  {author} {\bibinfo {author} {\bibfnamefont {Zhou}\ \bibnamefont
  {Huijuan}},\ }\bibfield  {title} {\enquote {\bibinfo {title} {{Parallel
  Physics-Informed Neural Networks Method with Regularization Strategies for
  the Forward-Inverse Problems of the Variable Coefficient Modified KdV
  Equation}},}\ }\href@noop {} {\bibfield  {journal} {\bibinfo  {journal} {J.
  Syst. Sci. Complex.}\ }\textbf {\bibinfo {volume} {37}},\ \bibinfo {pages}
  {511--544} (\bibinfo {year} {2024})}\BibitemShut {NoStop}%
\bibitem [{\citenamefont {Lin}\ and\ \citenamefont
  {Chen}(2023)}]{LIN2023133629}%
  \BibitemOpen
  \bibfield  {author} {\bibinfo {author} {\bibfnamefont {Shuning}\ \bibnamefont
  {Lin}}\ and\ \bibinfo {author} {\bibfnamefont {Yong}\ \bibnamefont {Chen}},\
  }\bibfield  {title} {\enquote {\bibinfo {title} {Physics-informed neural
  network methods based on miura transformations and discovery of new localized
  wave solutions},}\ }\href {\doibase
  https://doi.org/10.1016/j.physd.2022.133629} {\bibfield  {journal} {\bibinfo
  {journal} {Physica D: Nonlinear Phenomena}\ }\textbf {\bibinfo {volume}
  {445}},\ \bibinfo {pages} {133629} (\bibinfo {year} {2023})}\BibitemShut
  {NoStop}%
\bibitem [{\citenamefont {Zhou}\ and\ \citenamefont
  {Yan}(2021)}]{ZHOU2021127010}%
  \BibitemOpen
  \bibfield  {author} {\bibinfo {author} {\bibfnamefont {Zijian}\ \bibnamefont
  {Zhou}}\ and\ \bibinfo {author} {\bibfnamefont {Zhenya}\ \bibnamefont
  {Yan}},\ }\bibfield  {title} {\enquote {\bibinfo {title} {Solving forward and
  inverse problems of the logarithmic nonlinear schrödinger equation with
  pt-symmetric harmonic potential via deep learning},}\ }\href {\doibase
  https://doi.org/10.1016/j.physleta.2020.127010} {\bibfield  {journal}
  {\bibinfo  {journal} {Physics Letters A}\ }\textbf {\bibinfo {volume}
  {387}},\ \bibinfo {pages} {127010} (\bibinfo {year} {2021})}\BibitemShut
  {NoStop}%
\bibitem [{\citenamefont {Pu}\ \emph {et~al.}(2021)\citenamefont {Pu},
  \citenamefont {Li},\ and\ \citenamefont {Chen}}]{Pu2021}%
  \BibitemOpen
  \bibfield  {author} {\bibinfo {author} {\bibfnamefont {Juncai}\ \bibnamefont
  {Pu}}, \bibinfo {author} {\bibfnamefont {Jun}\ \bibnamefont {Li}}, \ and\
  \bibinfo {author} {\bibfnamefont {Yong~and}\ \bibnamefont {Chen}},\
  }\bibfield  {title} {\enquote {\bibinfo {title} {Solving localized wave
  solutions of the derivative nonlinear schrödinger equation using an improved
  pinn method},}\ }\href {\doibase 10.1007/s11071-021-06554-5} {\bibfield
  {journal} {\bibinfo  {journal} {Nonlinear Dynamics}\ }\textbf {\bibinfo
  {volume} {105}},\ \bibinfo {pages} {1723--1739} (\bibinfo {year}
  {2021})}\BibitemShut {NoStop}%
\bibitem [{\citenamefont {Zhang}\ \emph {et~al.}(2024)\citenamefont {Zhang},
  \citenamefont {Qiu}, \citenamefont {Hou},\ and\ \citenamefont
  {Yan}}]{ZHANG2024108229}%
  \BibitemOpen
  \bibfield  {author} {\bibinfo {author} {\bibfnamefont {Qiongni}\ \bibnamefont
  {Zhang}}, \bibinfo {author} {\bibfnamefont {Changxin}\ \bibnamefont {Qiu}},
  \bibinfo {author} {\bibfnamefont {Jiangyong}\ \bibnamefont {Hou}}, \ and\
  \bibinfo {author} {\bibfnamefont {Wenjing}\ \bibnamefont {Yan}},\ }\bibfield
  {title} {\enquote {\bibinfo {title} {Advanced physics-informed neural
  networks for numerical approximation of the coupled schrödinger–kdv
  equation},}\ }\href {\doibase https://doi.org/10.1016/j.cnsns.2024.108229}
  {\bibfield  {journal} {\bibinfo  {journal} {Communications in Nonlinear
  Science and Numerical Simulation}\ }\textbf {\bibinfo {volume} {138}},\
  \bibinfo {pages} {108229} (\bibinfo {year} {2024})}\BibitemShut {NoStop}%
\bibitem [{\citenamefont {Z.W.Miao}\ and\ \citenamefont
  {Y.Chen}(2022)}]{Zhengwu2022}%
  \BibitemOpen
  \bibfield  {author} {\bibinfo {author} {\bibnamefont {Z.W.Miao}}\ and\
  \bibinfo {author} {\bibnamefont {Y.Chen}},\ }\bibfield  {title} {\enquote
  {\bibinfo {title} {{Physics-informed neural networks method in
  high-dimensional integrable systems}},}\ }\href@noop {} {\bibfield  {journal}
  {\bibinfo  {journal} {Mod.Phys.Lett.B.}\ }\textbf {\bibinfo {volume}
  {36(1)}},\ \bibinfo {pages} {2150531} (\bibinfo {year} {2022})}\BibitemShut
  {NoStop}%
\bibitem [{\citenamefont {Zhou}\ \emph {et~al.}(2023)\citenamefont {Zhou},
  \citenamefont {Wang},\ and\ \citenamefont {Yan}}]{ZHOU2023164}%
  \BibitemOpen
  \bibfield  {author} {\bibinfo {author} {\bibfnamefont {Zijian}\ \bibnamefont
  {Zhou}}, \bibinfo {author} {\bibfnamefont {Li}~\bibnamefont {Wang}}, \ and\
  \bibinfo {author} {\bibfnamefont {Zhenya}\ \bibnamefont {Yan}},\ }\bibfield
  {title} {\enquote {\bibinfo {title} {Deep neural networks learning forward
  and inverse problems of two-dimensional nonlinear wave equations with
  rational solitons},}\ }\href {\doibase
  https://doi.org/10.1016/j.camwa.2023.09.047} {\bibfield  {journal} {\bibinfo
  {journal} {Computers and Mathematics with Applications}\ }\textbf {\bibinfo
  {volume} {151}},\ \bibinfo {pages} {164--171} (\bibinfo {year}
  {2023})}\BibitemShut {NoStop}%
\bibitem [{\citenamefont {Wang}\ \emph {et~al.}(2023)\citenamefont {Wang},
  \citenamefont {Zhou},\ and\ \citenamefont {Yan}}]{WANG202317}%
  \BibitemOpen
  \bibfield  {author} {\bibinfo {author} {\bibfnamefont {Li}~\bibnamefont
  {Wang}}, \bibinfo {author} {\bibfnamefont {Zijian}\ \bibnamefont {Zhou}}, \
  and\ \bibinfo {author} {\bibfnamefont {Zhenya}\ \bibnamefont {Yan}},\
  }\bibfield  {title} {\enquote {\bibinfo {title} {Data-driven vortex solitons
  and parameter discovery of 2d generalized nonlinear schrödinger equations
  with a pt-symmetric optical lattice},}\ }\href {\doibase
  https://doi.org/10.1016/j.camwa.2023.03.015} {\bibfield  {journal} {\bibinfo
  {journal} {Computers and Mathematics with Applications}\ }\textbf {\bibinfo
  {volume} {140}},\ \bibinfo {pages} {17--23} (\bibinfo {year}
  {2023})}\BibitemShut {NoStop}%
\bibitem [{\citenamefont {Bai}\ and\ \citenamefont {Molinaro}(2021)}]{Bai2021}%
  \BibitemOpen
  \bibfield  {author} {\bibinfo {author} {\bibfnamefont
  {UjjwalMishra~Siddhartha}\ \bibnamefont {Bai}, \bibfnamefont {GenmingKoley}}\
  and\ \bibinfo {author} {\bibfnamefont {Roberto}\ \bibnamefont {Molinaro}},\
  }\bibfield  {title} {\enquote {\bibinfo {title} {Physics informed neural
  networks (pinns) for approximating nonlinear dispersive pdes},}\ }\href
  {\doibase https://doi.org/10.4208/jcm.2101-m2020-0342} {\bibfield  {journal}
  {\bibinfo  {journal} {Journal of Computational Mathematics}\ }\textbf
  {\bibinfo {volume} {39}},\ \bibinfo {pages} {816--847} (\bibinfo {year}
  {2021})}\BibitemShut {NoStop}%
\bibitem [{\citenamefont {Z.J.~Zhou}(2023)}]{Zijian2023}%
  \BibitemOpen
  \bibfield  {author} {\bibinfo {author} {\bibfnamefont {Z.Y.~Yan}\
  \bibnamefont {Z.J.~Zhou}, \bibfnamefont {L.~Wang}},\ }\bibfield  {title}
  {\enquote {\bibinfo {title} {{Deep neural networks learning forward and
  inverse problems of two-dimensional nonlinear wave equations with rational
  solitons}},}\ }\href@noop {} {\bibfield  {journal} {\bibinfo  {journal}
  {Comput. Math. Appl.}\ }\textbf {\bibinfo {volume} {151}},\ \bibinfo {pages}
  {164--171} (\bibinfo {year} {2023})}\BibitemShut {NoStop}%
\bibitem [{\citenamefont {Zakharov}\ and\ \citenamefont
  {Kuznetsov}(1974)}]{Zakharov74}%
  \BibitemOpen
  \bibfield  {author} {\bibinfo {author} {\bibfnamefont {V.}~\bibnamefont
  {Zakharov}}\ and\ \bibinfo {author} {\bibfnamefont {E~A.}\ \bibnamefont
  {Kuznetsov}},\ }\bibfield  {title} {\enquote {\bibinfo {title}
  {{Three-dimensional solitons}},}\ }\href@noop {} {\bibfield  {journal}
  {\bibinfo  {journal} {Soviet Physics JETP}\ }\textbf {\bibinfo {volume}
  {29}},\ \bibinfo {pages} {594--597} (\bibinfo {year} {1974})}\BibitemShut
  {NoStop}%
\bibitem [{\citenamefont {Iwasaki}\ \emph {et~al.}(1990)\citenamefont
  {Iwasaki}, \citenamefont {Toh},\ and\ \citenamefont
  {Kawahara}}]{IWASAKI1990293}%
  \BibitemOpen
  \bibfield  {author} {\bibinfo {author} {\bibfnamefont {Hiroshi}\ \bibnamefont
  {Iwasaki}}, \bibinfo {author} {\bibfnamefont {Sadayoshi}\ \bibnamefont
  {Toh}}, \ and\ \bibinfo {author} {\bibfnamefont {Takuji}\ \bibnamefont
  {Kawahara}},\ }\bibfield  {title} {\enquote {\bibinfo {title} {{Cylindrical
  quasi-solitons of the Zakharov-Kuznetsov equation}},}\ }\href {\doibase
  https://doi.org/10.1016/0167-2789(90)90138-F} {\bibfield  {journal} {\bibinfo
   {journal} {Physica D: Nonlinear Phenomena}\ }\textbf {\bibinfo {volume}
  {43}},\ \bibinfo {pages} {293--303} (\bibinfo {year} {1990})}\BibitemShut
  {NoStop}%
\bibitem [{\citenamefont {Petviashvili}\ and\ \citenamefont
  {Yan'kov}(1982)}]{PetYan82}%
  \BibitemOpen
  \bibfield  {author} {\bibinfo {author} {\bibfnamefont {V.~I.}\ \bibnamefont
  {Petviashvili}}\ and\ \bibinfo {author} {\bibfnamefont {V.~V.}\ \bibnamefont
  {Yan'kov}},\ }\bibfield  {title} {\enquote {\bibinfo {title} {{Bilayer
  vortices in rotating stratified fluid}},}\ }\href@noop {} {\bibfield
  {journal} {\bibinfo  {journal} {Dokl. Akad. Nauk SSSR}\ }\textbf {\bibinfo
  {volume} {267}},\ \bibinfo {pages} {825--828} (\bibinfo {year}
  {1982})}\BibitemShut {NoStop}%
\bibitem [{\citenamefont {Klein}\ \emph {et~al.}(2021)\citenamefont {Klein},
  \citenamefont {Roudenko},\ and\ \citenamefont {Stoilov}}]{Klein21}%
  \BibitemOpen
  \bibfield  {author} {\bibinfo {author} {\bibfnamefont {Christian}\
  \bibnamefont {Klein}}, \bibinfo {author} {\bibfnamefont {Svetlana}\
  \bibnamefont {Roudenko}}, \ and\ \bibinfo {author} {\bibfnamefont {Nikola}\
  \bibnamefont {Stoilov}},\ }\bibfield  {title} {\enquote {\bibinfo {title}
  {{Numerical study of Zakhavor-Kuznetsov equations in two dimensions}},}\
  }\href@noop {} {\bibfield  {journal} {\bibinfo  {journal} {Journal of
  Nonlinear Science}\ }\textbf {\bibinfo {volume} {31}},\ \bibinfo {pages}
  {1--28} (\bibinfo {year} {2021})}\BibitemShut {NoStop}%
\bibitem [{\citenamefont {Koike}\ \emph {et~al.}(2022)\citenamefont {Koike},
  \citenamefont {Nakamula}, \citenamefont {Nishie}, \citenamefont {Obuse},
  \citenamefont {Sawado}, \citenamefont {Suda},\ and\ \citenamefont
  {Toda}}]{Koike:2022gfq}%
  \BibitemOpen
  \bibfield  {author} {\bibinfo {author} {\bibfnamefont {Yukito}\ \bibnamefont
  {Koike}}, \bibinfo {author} {\bibfnamefont {Atsushi}\ \bibnamefont
  {Nakamula}}, \bibinfo {author} {\bibfnamefont {Akihiro}\ \bibnamefont
  {Nishie}}, \bibinfo {author} {\bibfnamefont {Kiori}\ \bibnamefont {Obuse}},
  \bibinfo {author} {\bibfnamefont {Nobuyuki}\ \bibnamefont {Sawado}}, \bibinfo
  {author} {\bibfnamefont {Yamato}\ \bibnamefont {Suda}}, \ and\ \bibinfo
  {author} {\bibfnamefont {Kouichi}\ \bibnamefont {Toda}},\ }\bibfield  {title}
  {\enquote {\bibinfo {title} {{Mock-integrability and stable solitary
  vortices}},}\ }\href {\doibase 10.1016/j.chaos.2022.112782} {\bibfield
  {journal} {\bibinfo  {journal} {Chaos Solitons and Fractals: the
  interdisciplinary journal of Nonlinear Science and Nonequilibrium and Complex
  Phenomena}\ }\textbf {\bibinfo {volume} {165}},\ \bibinfo {pages} {112782}
  (\bibinfo {year} {2022})},\ \Eprint {http://arxiv.org/abs/2204.01985}
  {arXiv:2204.01985 [math-ph]} \BibitemShut {NoStop}%
\bibitem [{\citenamefont {Abdulloev}\ \emph {et~al.}(1976)\citenamefont
  {Abdulloev}, \citenamefont {Bogolubsky},\ and\ \citenamefont
  {Makhankov}}]{ABDULLOEV1976427}%
  \BibitemOpen
  \bibfield  {author} {\bibinfo {author} {\bibfnamefont {Kh.O.}\ \bibnamefont
  {Abdulloev}}, \bibinfo {author} {\bibfnamefont {I.L.}\ \bibnamefont
  {Bogolubsky}}, \ and\ \bibinfo {author} {\bibfnamefont {V.G.}\ \bibnamefont
  {Makhankov}},\ }\bibfield  {title} {\enquote {\bibinfo {title} {One more
  example of inelastic soliton interaction},}\ }\href {\doibase
  https://doi.org/10.1016/0375-9601(76)90714-3} {\bibfield  {journal} {\bibinfo
   {journal} {Physics Letters A}\ }\textbf {\bibinfo {volume} {56}},\ \bibinfo
  {pages} {427--428} (\bibinfo {year} {1976})}\BibitemShut {NoStop}%
\bibitem [{\citenamefont {Lewis}\ and\ \citenamefont
  {Tjon}(1979)}]{COURTENAYLEWIS1979275}%
  \BibitemOpen
  \bibfield  {author} {\bibinfo {author} {\bibfnamefont {J.~Courtenay}\
  \bibnamefont {Lewis}}\ and\ \bibinfo {author} {\bibfnamefont {J.A.}\
  \bibnamefont {Tjon}},\ }\bibfield  {title} {\enquote {\bibinfo {title}
  {Resonant production of solitons in the rlw equation},}\ }\href {\doibase
  https://doi.org/10.1016/0375-9601(79)90532-2} {\bibfield  {journal} {\bibinfo
   {journal} {Physics Letters A}\ }\textbf {\bibinfo {volume} {73}},\ \bibinfo
  {pages} {275--279} (\bibinfo {year} {1979})}\BibitemShut {NoStop}%
\bibitem [{\citenamefont {ter Braak}\ \emph {et~al.}(2019)\citenamefont {ter
  Braak}, \citenamefont {Ferreira},\ and\ \citenamefont
  {Zakrzewski}}]{terBraak:2017jpe}%
  \BibitemOpen
  \bibfield  {author} {\bibinfo {author} {\bibfnamefont {F.}~\bibnamefont {ter
  Braak}}, \bibinfo {author} {\bibfnamefont {L.~A.}\ \bibnamefont {Ferreira}},
  \ and\ \bibinfo {author} {\bibfnamefont {W.~J.}\ \bibnamefont {Zakrzewski}},\
  }\bibfield  {title} {\enquote {\bibinfo {title} {{Quasi-integrability of
  deformations of the KdV equation}},}\ }\href {\doibase
  10.1016/j.nuclphysb.2018.12.004} {\bibfield  {journal} {\bibinfo  {journal}
  {Nucl. Phys. B}\ }\textbf {\bibinfo {volume} {939}},\ \bibinfo {pages}
  {49--94} (\bibinfo {year} {2019})},\ \Eprint
  {http://arxiv.org/abs/1710.00918} {arXiv:1710.00918 [hep-th]} \BibitemShut
  {NoStop}%
\bibitem [{\citenamefont {Kawahara}\ \emph {et~al.}(1992)\citenamefont
  {Kawahara}, \citenamefont {Araki},\ and\ \citenamefont
  {Toh}}]{KAWAHARA199279}%
  \BibitemOpen
  \bibfield  {author} {\bibinfo {author} {\bibfnamefont {Takuji}\ \bibnamefont
  {Kawahara}}, \bibinfo {author} {\bibfnamefont {Keisuke}\ \bibnamefont
  {Araki}}, \ and\ \bibinfo {author} {\bibfnamefont {Sadayoshi}\ \bibnamefont
  {Toh}},\ }\bibfield  {title} {\enquote {\bibinfo {title} {Interactions of
  two-dimensionally localized pulses of the regularized-long-wave equation},}\
  }\href {\doibase https://doi.org/10.1016/0167-2789(92)90207-4} {\bibfield
  {journal} {\bibinfo  {journal} {Physica D: Nonlinear Phenomena}\ }\textbf
  {\bibinfo {volume} {59}},\ \bibinfo {pages} {79--89} (\bibinfo {year}
  {1992})}\BibitemShut {NoStop}%
\bibitem [{\citenamefont {Benjamin}\ \emph {et~al.}(1972)\citenamefont
  {Benjamin}, \citenamefont {Bona},\ and\ \citenamefont {Mahony}}]{Benjamin72}%
  \BibitemOpen
  \bibfield  {author} {\bibinfo {author} {\bibfnamefont {Thomas~Brooke}\
  \bibnamefont {Benjamin}}, \bibinfo {author} {\bibfnamefont {J.~L.}\
  \bibnamefont {Bona}}, \ and\ \bibinfo {author} {\bibfnamefont {J.~J.}\
  \bibnamefont {Mahony}},\ }\bibfield  {title} {\enquote {\bibinfo {title}
  {Model equations for long waves in nonlinear dispersive systems},}\ }\href
  {\doibase 10.1098/rsta.1972.0032} {\bibfield  {journal} {\bibinfo  {journal}
  {Philosophical Transactions of the Royal Society of London. Series A,
  Mathematical and Physical Sciences}\ }\textbf {\bibinfo {volume} {272}},\
  \bibinfo {pages} {47--78} (\bibinfo {year} {1972})}\BibitemShut {NoStop}%
\bibitem [{\citenamefont {Kuznetsov}\ \emph {et~al.}(1986)\citenamefont
  {Kuznetsov}, \citenamefont {Rubenchik},\ and\ \citenamefont
  {Zakharov}}]{KUZNETSOV1986103}%
  \BibitemOpen
  \bibfield  {author} {\bibinfo {author} {\bibfnamefont {E.A.}\ \bibnamefont
  {Kuznetsov}}, \bibinfo {author} {\bibfnamefont {A.M.}\ \bibnamefont
  {Rubenchik}}, \ and\ \bibinfo {author} {\bibfnamefont {V.E.}\ \bibnamefont
  {Zakharov}},\ }\bibfield  {title} {\enquote {\bibinfo {title} {{Soliton
  stability in plasmas and hydrodynamics}},}\ }\href@noop {} {\bibfield
  {journal} {\bibinfo  {journal} {Physics Reports}\ }\textbf {\bibinfo {volume}
  {142}},\ \bibinfo {pages} {103--165} (\bibinfo {year} {1986})}\BibitemShut
  {NoStop}%
\bibitem [{\citenamefont {Liu}\ and\ \citenamefont {Nocedal}(1989)}]{LIU1989}%
  \BibitemOpen
  \bibfield  {author} {\bibinfo {author} {\bibfnamefont {Dong~C.}\ \bibnamefont
  {Liu}}\ and\ \bibinfo {author} {\bibfnamefont {Jorge}\ \bibnamefont
  {Nocedal}},\ }\bibfield  {title} {\enquote {\bibinfo {title} {On the limited
  memory bfgs method for large scale optimization},}\ }\href {\doibase
  https://doi.org/10.1007/BF01589116} {\bibfield  {journal} {\bibinfo
  {journal} {Mathematical Programming}\ }\textbf {\bibinfo {volume} {45}},\
  \bibinfo {pages} {503--528} (\bibinfo {year} {1989})}\BibitemShut {NoStop}%
\bibitem [{\citenamefont {Bihlo}\ and\ \citenamefont
  {Popovych}(2022)}]{Bihlo2022}%
  \BibitemOpen
  \bibfield  {author} {\bibinfo {author} {\bibfnamefont {Alex}\ \bibnamefont
  {Bihlo}}\ and\ \bibinfo {author} {\bibfnamefont {Roman~O.}\ \bibnamefont
  {Popovych}},\ }\bibfield  {title} {\enquote {\bibinfo {title}
  {Physics-informed neural networks for the shallow-water equations on the
  sphere},}\ }\href {\doibase https://doi.org/10.1016/j.jcp.2022.111024}
  {\bibfield  {journal} {\bibinfo  {journal} {Journal of Computational
  Physics}\ }\textbf {\bibinfo {volume} {456}},\ \bibinfo {pages} {111024}
  (\bibinfo {year} {2022})}\BibitemShut {NoStop}%
\bibitem [{\citenamefont {Krishnapriyan}\ \emph {et~al.}(2024)\citenamefont
  {Krishnapriyan}, \citenamefont {Gholami}, \citenamefont {Zhe}, \citenamefont
  {Kirby},\ and\ \citenamefont {Mahoney}}]{Krishnapriyan2024}%
  \BibitemOpen
  \bibfield  {author} {\bibinfo {author} {\bibfnamefont {Aditi~S.}\
  \bibnamefont {Krishnapriyan}}, \bibinfo {author} {\bibfnamefont {Amir}\
  \bibnamefont {Gholami}}, \bibinfo {author} {\bibfnamefont {Shandian}\
  \bibnamefont {Zhe}}, \bibinfo {author} {\bibfnamefont {Robert~M.}\
  \bibnamefont {Kirby}}, \ and\ \bibinfo {author} {\bibfnamefont {Michael~W.}\
  \bibnamefont {Mahoney}},\ }\bibfield  {title} {\enquote {\bibinfo {title}
  {Characterizing possible failure modes in physics-informed neural
  networks},}\ }\bibfield  {booktitle} {\emph {\bibinfo {booktitle}
  {Proceedings of the 35th International Conference on Neural Information
  Processing Systems}},\ }\href@noop {} {\ \bibinfo {series} {NIPS '21}
  (\bibinfo {year} {2024})}\BibitemShut {NoStop}%
\bibitem [{\citenamefont {Goda}\ and\ \citenamefont {Fukui}(1980)}]{Goda1980}%
  \BibitemOpen
  \bibfield  {author} {\bibinfo {author} {\bibfnamefont {Katuhiko}\
  \bibnamefont {Goda}}\ and\ \bibinfo {author} {\bibfnamefont {Yoshinari}\
  \bibnamefont {Fukui}},\ }\bibfield  {title} {\enquote {\bibinfo {title}
  {Numerical studies of the regularized long wave equation},}\ }\href {\doibase
  10.1143/JPSJ.48.623} {\bibfield  {journal} {\bibinfo  {journal} {Journal of
  the Physical Society of Japan}\ }\textbf {\bibinfo {volume} {48}},\ \bibinfo
  {pages} {623--630} (\bibinfo {year} {1980})},\ \Eprint
  {http://arxiv.org/abs/https://doi.org/10.1143/JPSJ.48.623}
  {https://doi.org/10.1143/JPSJ.48.623} \BibitemShut {NoStop}%
\bibitem [{\citenamefont {Makino}\ \emph {et~al.}(1981)\citenamefont {Makino},
  \citenamefont {Kamimura},\ and\ \citenamefont {Taniuti}}]{Makino81}%
  \BibitemOpen
  \bibfield  {author} {\bibinfo {author} {\bibfnamefont {Mitsuhiro}\
  \bibnamefont {Makino}}, \bibinfo {author} {\bibfnamefont {Tetsuo}\
  \bibnamefont {Kamimura}}, \ and\ \bibinfo {author} {\bibfnamefont {Tosiya}\
  \bibnamefont {Taniuti}},\ }\bibfield  {title} {\enquote {\bibinfo {title}
  {Dynamics of two-dimensional solitary vortices in a low- beta plasma with
  convective motion},}\ }\href {\doibase 10.1143/JPSJ.50.980} {\bibfield
  {journal} {\bibinfo  {journal} {Journal of the Physical Society of Japan}\
  }\textbf {\bibinfo {volume} {50}},\ \bibinfo {pages} {980--989} (\bibinfo
  {year} {1981})},\ \Eprint
  {http://arxiv.org/abs/https://doi.org/10.1143/JPSJ.50.980}
  {https://doi.org/10.1143/JPSJ.50.980} \BibitemShut {NoStop}%
\bibitem [{\citenamefont {Ferreira}\ \emph {et~al.}(2013)\citenamefont
  {Ferreira}, \citenamefont {Luchini},\ and\ \citenamefont
  {Zakrzewski}}]{Ferreira:2013nda}%
  \BibitemOpen
  \bibfield  {author} {\bibinfo {author} {\bibfnamefont {L.~A.}\ \bibnamefont
  {Ferreira}}, \bibinfo {author} {\bibfnamefont {G.}~\bibnamefont {Luchini}}, \
  and\ \bibinfo {author} {\bibfnamefont {Wojtek~J.}\ \bibnamefont
  {Zakrzewski}},\ }\bibfield  {title} {\enquote {\bibinfo {title} {{The concept
  of quasi-integrability}},}\ }\href {\doibase 10.1063/1.4828681} {\bibfield
  {journal} {\bibinfo  {journal} {AIP Conf. Proc.}\ }\textbf {\bibinfo {volume}
  {1562}},\ \bibinfo {pages} {43--49} (\bibinfo {year} {2013})},\ \Eprint
  {http://arxiv.org/abs/1307.7722} {arXiv:1307.7722 [hep-th]} \BibitemShut
  {NoStop}%
\bibitem [{\citenamefont {D.~Jagtap}\ and\ \citenamefont
  {Em~Karniadakis}(2020)}]{Jagtap2020CCP}%
  \BibitemOpen
  \bibfield  {author} {\bibinfo {author} {\bibfnamefont {Ameya}\ \bibnamefont
  {D.~Jagtap}}\ and\ \bibinfo {author} {\bibfnamefont {George}\ \bibnamefont
  {Em~Karniadakis}},\ }\bibfield  {title} {\enquote {\bibinfo {title} {Extended
  physics-informed neural networks (xpinns): A generalized space-time domain
  decomposition based deep learning framework for nonlinear partial
  differential equations},}\ }\href {\doibase
  https://doi.org/10.4208/cicp.OA-2020-0164} {\bibfield  {journal} {\bibinfo
  {journal} {Communications in Computational Physics}\ }\textbf {\bibinfo
  {volume} {28}},\ \bibinfo {pages} {2002--2041} (\bibinfo {year}
  {2020})}\BibitemShut {NoStop}%
\bibitem [{\citenamefont {Yang}\ and\ \citenamefont {Kim}(2024)}]{YANG2024237}%
  \BibitemOpen
  \bibfield  {author} {\bibinfo {author} {\bibfnamefont {Hee~Jun}\ \bibnamefont
  {Yang}}\ and\ \bibinfo {author} {\bibfnamefont {Hyea~Hyun}\ \bibnamefont
  {Kim}},\ }\bibfield  {title} {\enquote {\bibinfo {title} {Iterative
  algorithms for partitioned neural network approximation to partial
  differential equations},}\ }\href {\doibase
  https://doi.org/10.1016/j.camwa.2024.07.007} {\bibfield  {journal} {\bibinfo
  {journal} {Computers and Mathematics with Applications}\ }\textbf {\bibinfo
  {volume} {170}},\ \bibinfo {pages} {237--259} (\bibinfo {year}
  {2024})}\BibitemShut {NoStop}%
\bibitem [{\citenamefont {Calogero}\ and\ \citenamefont
  {Degasperis}(1978)}]{calogero1978exact}%
  \BibitemOpen
  \bibfield  {author} {\bibinfo {author} {\bibfnamefont {F.}~\bibnamefont
  {Calogero}}\ and\ \bibinfo {author} {\bibfnamefont {Ao}~\bibnamefont
  {Degasperis}},\ }\bibfield  {title} {\enquote {\bibinfo {title} {Exact
  solution via the spectral transform of a generalization with linearly
  x-dependent coefficients of the modified korteweg-de-vries equation},}\
  }\href@noop {} {\bibfield  {journal} {\bibinfo  {journal} {Lett. Nuovo Cim}\
  }\textbf {\bibinfo {volume} {22}},\ \bibinfo {pages} {270--273} (\bibinfo
  {year} {1978})}\BibitemShut {NoStop}%
\bibitem [{\citenamefont {Brugarino}\ and\ \citenamefont
  {Pantano}(1980)}]{brugarino1980integration}%
  \BibitemOpen
  \bibfield  {author} {\bibinfo {author} {\bibfnamefont {T.}~\bibnamefont
  {Brugarino}}\ and\ \bibinfo {author} {\bibfnamefont {P.}~\bibnamefont
  {Pantano}},\ }\bibfield  {title} {\enquote {\bibinfo {title} {The integration
  of burgers and korteweg-de vries equations with nonuniformities},}\
  }\href@noop {} {\bibfield  {journal} {\bibinfo  {journal} {Physics Letters
  A}\ }\textbf {\bibinfo {volume} {80}},\ \bibinfo {pages} {223--224} (\bibinfo
  {year} {1980})}\BibitemShut {NoStop}%
\bibitem [{\citenamefont {Joshi}(1987)}]{joshi1987painleve}%
  \BibitemOpen
  \bibfield  {author} {\bibinfo {author} {\bibfnamefont {Nalini}\ \bibnamefont
  {Joshi}},\ }\bibfield  {title} {\enquote {\bibinfo {title} {Painlev{\'e}
  property of general variable-coefficient versions of the korteweg-de vries
  and non-linear schr{\"o}dinger equations},}\ }\href@noop {} {\bibfield
  {journal} {\bibinfo  {journal} {Physics Letters A}\ }\textbf {\bibinfo
  {volume} {125}},\ \bibinfo {pages} {456--460} (\bibinfo {year}
  {1987})}\BibitemShut {NoStop}%
\bibitem [{\citenamefont {Hlavat{\`y}}(1988)}]{hlavaty1988painleve}%
  \BibitemOpen
  \bibfield  {author} {\bibinfo {author} {\bibfnamefont {Ladislav}\
  \bibnamefont {Hlavat{\`y}}},\ }\bibfield  {title} {\enquote {\bibinfo {title}
  {Painlev{\'e} analysis of nonautonomous evolution equations},}\ }\href@noop
  {} {\bibfield  {journal} {\bibinfo  {journal} {Physics Letters A}\ }\textbf
  {\bibinfo {volume} {128}},\ \bibinfo {pages} {335--338} (\bibinfo {year}
  {1988})}\BibitemShut {NoStop}%
\bibitem [{\citenamefont {Brugarino}\ and\ \citenamefont
  {Greco}(1991)}]{brugarino1991painleve}%
  \BibitemOpen
  \bibfield  {author} {\bibinfo {author} {\bibfnamefont {Tommaso}\ \bibnamefont
  {Brugarino}}\ and\ \bibinfo {author} {\bibfnamefont {Antonio~M.}\
  \bibnamefont {Greco}},\ }\bibfield  {title} {\enquote {\bibinfo {title}
  {Painlev{\'e} analysis and reducibility to the canonical form for the
  generalized kadomtsev--petviashvili equation},}\ }\href@noop {} {\bibfield
  {journal} {\bibinfo  {journal} {Journal of mathematical physics}\ }\textbf
  {\bibinfo {volume} {32}},\ \bibinfo {pages} {69--71} (\bibinfo {year}
  {1991})}\BibitemShut {NoStop}%
\bibitem [{\citenamefont {Gao}\ and\ \citenamefont
  {Tian}(2001)}]{gao2001variable}%
  \BibitemOpen
  \bibfield  {author} {\bibinfo {author} {\bibfnamefont {Yi-Tian}\ \bibnamefont
  {Gao}}\ and\ \bibinfo {author} {\bibfnamefont {Bo}~\bibnamefont {Tian}},\
  }\bibfield  {title} {\enquote {\bibinfo {title} {Variable-coefficient
  balancing-act algorithm extended to a variable-coefficient mkp model for the
  rotating fluids},}\ }\href@noop {} {\bibfield  {journal} {\bibinfo  {journal}
  {International Journal of Modern Physics C}\ }\textbf {\bibinfo {volume}
  {12}},\ \bibinfo {pages} {1383--1389} (\bibinfo {year} {2001})}\BibitemShut
  {NoStop}%
\bibitem [{\citenamefont {Kobayashi}\ and\ \citenamefont
  {Toda}(2005)}]{kobayashi2005generalized}%
  \BibitemOpen
  \bibfield  {author} {\bibinfo {author} {\bibfnamefont {Tadashi}\ \bibnamefont
  {Kobayashi}}\ and\ \bibinfo {author} {\bibfnamefont {Kouichi}\ \bibnamefont
  {Toda}},\ }\bibfield  {title} {\enquote {\bibinfo {title} {A generalized
  kdv-family with variable coefficients in (2+ 1) dimensions},}\ }\href@noop {}
  {\bibfield  {journal} {\bibinfo  {journal} {IEICE Transactions on
  Fundamentals of Electronics, Communications and Computer Sciences}\ }\textbf
  {\bibinfo {volume} {88}},\ \bibinfo {pages} {2548--2553} (\bibinfo {year}
  {2005})}\BibitemShut {NoStop}%
\bibitem [{\citenamefont {Kobayashi}\ and\ \citenamefont
  {Toda}(2006)}]{kobayashi2006painleve}%
  \BibitemOpen
  \bibfield  {author} {\bibinfo {author} {\bibfnamefont {Tadashi}\ \bibnamefont
  {Kobayashi}}\ and\ \bibinfo {author} {\bibfnamefont {Kouichi}\ \bibnamefont
  {Toda}},\ }\bibfield  {title} {\enquote {\bibinfo {title} {The painlev{\'e}
  test and reducibility to the canonical forms for higher-dimensional soliton
  equations with variable-coefficients},}\ }\href@noop {} {\bibfield  {journal}
  {\bibinfo  {journal} {SIGMA. Symmetry, Integrability and Geometry: Methods
  and Applications}\ }\textbf {\bibinfo {volume} {2}},\ \bibinfo {pages} {063}
  (\bibinfo {year} {2006})}\BibitemShut {NoStop}%
\bibitem [{\citenamefont {Gao}\ \emph {et~al.}(2021{\natexlab{a}})\citenamefont
  {Gao}, \citenamefont {Guo},\ and\ \citenamefont {Shan}}]{gao2021optica}%
  \BibitemOpen
  \bibfield  {author} {\bibinfo {author} {\bibfnamefont {Xin-Yi}\ \bibnamefont
  {Gao}}, \bibinfo {author} {\bibfnamefont {Yong-Jiang}\ \bibnamefont {Guo}}, \
  and\ \bibinfo {author} {\bibfnamefont {Wen-Rui}\ \bibnamefont {Shan}},\
  }\bibfield  {title} {\enquote {\bibinfo {title} {Optical waves/modes in a
  multicomponent inhomogeneous optical fiber via a three-coupled
  variable-coefficient nonlinear schr{\"o}dinger system},}\ }\href@noop {}
  {\bibfield  {journal} {\bibinfo  {journal} {Applied Mathematics Letters}\
  }\textbf {\bibinfo {volume} {120}},\ \bibinfo {pages} {107161} (\bibinfo
  {year} {2021}{\natexlab{a}})}\BibitemShut {NoStop}%
\bibitem [{\citenamefont {Gao}\ \emph {et~al.}(2021{\natexlab{b}})\citenamefont
  {Gao}, \citenamefont {Guo}, \citenamefont {Shan}, \citenamefont {Zhou},
  \citenamefont {Wang},\ and\ \citenamefont {Yang}}]{gao2021varmodkdv}%
  \BibitemOpen
  \bibfield  {author} {\bibinfo {author} {\bibfnamefont {Xin-yi}\ \bibnamefont
  {Gao}}, \bibinfo {author} {\bibfnamefont {Yong-jiang}\ \bibnamefont {Guo}},
  \bibinfo {author} {\bibfnamefont {Wen-rui}\ \bibnamefont {Shan}}, \bibinfo
  {author} {\bibfnamefont {Tian-yu}\ \bibnamefont {Zhou}}, \bibinfo {author}
  {\bibfnamefont {Meng}\ \bibnamefont {Wang}}, \ and\ \bibinfo {author}
  {\bibfnamefont {Dan-yu}\ \bibnamefont {Yang}},\ }\bibfield  {title} {\enquote
  {\bibinfo {title} {In the atmosphere and oceanic fluids: Scaling
  transformations, bilinear forms, b{\"a}cklund transformations and solitons
  for a generalized variable-coefficient korteweg-de vries-modified korteweg-de
  vries equation},}\ }\href@noop {} {\bibfield  {journal} {\bibinfo  {journal}
  {China Ocean Engineering}\ }\textbf {\bibinfo {volume} {35}},\ \bibinfo
  {pages} {518} (\bibinfo {year} {2021}{\natexlab{b}})}\BibitemShut {NoStop}%
\bibitem [{\citenamefont {Gao}\ \emph {et~al.}(2022{\natexlab{a}})\citenamefont
  {Gao}, \citenamefont {Guo},\ and\ \citenamefont {Shan}}]{GAO20222707}%
  \BibitemOpen
  \bibfield  {author} {\bibinfo {author} {\bibfnamefont {Xin-Yi}\ \bibnamefont
  {Gao}}, \bibinfo {author} {\bibfnamefont {Yong-Jiang}\ \bibnamefont {Guo}}, \
  and\ \bibinfo {author} {\bibfnamefont {Wen-Rui}\ \bibnamefont {Shan}},\
  }\bibfield  {title} {\enquote {\bibinfo {title} {Similarity reductions for a
  generalized (3+1)-dimensional variable-coefficient b-type
  kadomtsev-petviashvili equation in fluid dynamics},}\ }\href {\doibase
  https://doi.org/10.1016/j.cjph.2022.04.014} {\bibfield  {journal} {\bibinfo
  {journal} {Chinese Journal of Physics}\ }\textbf {\bibinfo {volume} {77}},\
  \bibinfo {pages} {2707--2712} (\bibinfo {year}
  {2022}{\natexlab{a}})}\BibitemShut {NoStop}%
\bibitem [{\citenamefont {Gao}\ \emph {et~al.}(2022{\natexlab{b}})\citenamefont
  {Gao}, \citenamefont {Guo},\ and\ \citenamefont {Shan}}]{Gao2022InNO}%
  \BibitemOpen
  \bibfield  {author} {\bibinfo {author} {\bibfnamefont {Xin-Yi}\ \bibnamefont
  {Gao}}, \bibinfo {author} {\bibfnamefont {Yong}\ \bibnamefont {Guo}}, \ and\
  \bibinfo {author} {\bibfnamefont {Wen-Rui}\ \bibnamefont {Shan}},\ }\bibfield
   {title} {\enquote {\bibinfo {title} {In nonlinear optics, fluid mechanics,
  plasma physics or atmospheric science: symbolic computation on a generalized
  variable-coefficient korteweg-de vries equation},}\ }\href@noop {} {\bibfield
   {journal} {\bibinfo  {journal} {Acta Mathematica Sinica, English Series}\ }
  (\bibinfo {year} {2022}{\natexlab{b}})}\BibitemShut {NoStop}%
\bibitem [{\citenamefont {Gao}\ and\ \citenamefont
  {Tian}(2022)}]{gao2022water}%
  \BibitemOpen
  \bibfield  {author} {\bibinfo {author} {\bibfnamefont {Xiao-Tian}\
  \bibnamefont {Gao}}\ and\ \bibinfo {author} {\bibfnamefont {Bo}~\bibnamefont
  {Tian}},\ }\bibfield  {title} {\enquote {\bibinfo {title} {Water-wave studies
  on a (2+ 1)-dimensional generalized variable-coefficient
  boiti--leon--pempinelli system},}\ }\href@noop {} {\bibfield  {journal}
  {\bibinfo  {journal} {Applied Mathematics Letters}\ }\textbf {\bibinfo
  {volume} {128}},\ \bibinfo {pages} {107858} (\bibinfo {year}
  {2022})}\BibitemShut {NoStop}%
\end{thebibliography}%

\end{document}